\documentstyle[12pt,epsf,eqsection,latexsym]{article}

 \large\normalsize

\newcommand{\be}{\begin{equation}}
\newcommand{\ee}{\end{equation}}
\newcommand{\ba}{\begin{eqnarray}}
\newcommand{\ea}{\end{eqnarray}}
\newcommand{\Tr}{\mbox{$\mathrm{Tr}$}}
\def\reff#1{(\ref{#1})}
\newcommand{\1}{1\!\!\!\bot}
\def\gtapprox{\raisebox{0.45ex}{$\,\,>$}
         \raisebox{-0.7ex}{$\!\!\!\!\!\sim\,\,$}}
\def\ltapprox{\raisebox{0.45ex}{$\,\,<$}
         \raisebox{-0.7ex}{$\!\!\!\!\!\sim\,\,$}}

\def\bg{\mbox{\protect\boldmath $g$}}
\def\bx{x}
\def\by{y}
\def\bun{e}
\def\bk{k}
\def\bsig{\vec{\mbox{$\sigma$}}}
\def\bfv{\vec{\mbox{$f$}}}
\def\bT{T}
\def\bvg{\vec{\mbox{$g$}}}
\def\bvw{\vec{\mbox{${\mathrm w}$}}}
\def\bvu{\vec{\mbox{$u$}}}

\begin{document}


\title{Critical Slowing-Down in $SU(2)$ \\
       Landau-Gauge-Fixing Algorithms at $\beta = \infty$}

\author{
  \small  Attilio Cucchieri and Tereza Mendes \\[-0.2cm]
  \small\it IFSC-USP, Caixa Postal 369       \\[-0.2cm]
  \small\it 13560-970 S\~ao Carlos, SP, Brazil  \\[-0.2cm]
  \small E-mail:  {\tt attilio@if.sc.usp.br},
                  {\tt mendes@if.sc.usp.br} \\[-0.2cm]
  {\protect\makebox[5in]{\quad}} \\
}

\vspace{0.2cm}
\maketitle
\thispagestyle{empty}   


\begin{abstract}
We evaluate numerically and analytically the dynamic critical
exponent $z$ for five gauge-fixing algorithms in $SU(2)$ lattice
Landau-gauge theory by considering the case $\beta = \infty$.
Numerical data are obtained in two, three and four dimensions.
Results are in agreement with those obtained previously at finite
$\beta$ in two dimensions. The theoretical analysis, valid for
any dimension $d$, helps us clarify the tuning of these algorithms.
We also study generalizations of the overrelaxation algorithm
and of the stochastic overrelaxation algorithm and verify that we
cannot have a dynamic critical exponent $z$ smaller than 1 with
these local algorithms. Finally, the analytic approach is applied
to the so-called $\lambda$-gauges, again at $\beta = \infty$, and
verified numerically for the two-dimensional case.
\end{abstract}

\clearpage


\section{Introduction}
\label{Intro}

Lattice gauge fixing is a necessary step in our understanding
of the relationship between continuum and lattice models.
In fact, the continuum limit is the weak-coupling limit and a
weak-coupling expansion requires gauge fixing. Thus, even though
gauge fixing is in principle not needed for the lattice formulation
of QCD, one is led to consider gauge-dependent quantities
on the lattice as well, such as gluon, ghost and quark propagators,
vertices, etc.\ \cite{review}.
It is therefore important to devise numerical
algorithms to gauge-fix a lattice configuration efficiently. 
The main issue regarding the efficiency of these algorithms
is the problem of {\em critical slowing-down} (CSD), which
occurs when the relaxation time of an algorithm diverges as
the lattice volume is increased (see for example \cite{A,Wo2+So}).
Besides the problem of CSD, we are also interested in
understanding which quantities should be used to test the
convergence of the gauge fixing and in finding prescriptions
for the tuning of parameters in the algorithms, when tuning is needed.

In Ref.\ \cite{CM} we have studied the problem of critical
slowing-down for five gauge-fixing algorithms
in $SU(2)$ lattice Landau-gauge theory on two-dimensional lattices with
periodic boundary conditions. We obtained that
the local method called {\em Los Alamos}
has {\em dynamic critical exponent} $z \approx 2$, the three
improved local methods we considered --- the {\em 
overrelaxation} method, the {\em stochastic overrelaxation} method
and the so-called {\em Cornell} method --- have critical exponent
$z \approx 1$, and the global method of {\em Fourier acceleration}
completely eliminates critical slowing-down. All these methods,
except for the Los Alamos method, involve tuning.
In that reference we also
reported a detailed discussion and analysis of the tuning for the
overrelaxation, the stochastic overrelaxation and the Cornell methods,
and we made a comparative study of several quantities usually employed
in the literature to check the convergence of the gauge fixing.

Here we redo that analysis for the case $\beta = \infty$, which can
be studied analytically, and we include test runs in two, three and
four dimensions\footnote{~Partial results can be found in
\cite{CMLat96}.}.
Let us recall the numerical problem we want to study:
for a  given (i.e.\ fixed) lattice configuration $\{ U_{\mu}(\bx) \}$,
we look for a gauge transformation $\{ g (\bx) \}$ that brings the
functional
\ba
{\cal E}_{U}\left[ \, g\, \right] & \equiv &
 1 \,-\, \frac{1}{d\,V} \sum_{\mu = 1}^{d} \sum_{\bx} \,
         \frac{\Tr}{2} \left[ \;
               U_{\mu}^{\left( g \right)}(\bx)
                        \; \right]
\label{eq:minfun} \\
 & \equiv &
   1 \,-\, \frac{1}{d\,V} \sum_{\mu = 1}^{d} \sum_{\bx} \,
         \frac{\Tr}{2} \left[ \;
             g(\bx) \; U_{\mu}(\bx) \; g^{\dagger}(\bx + \bun_{\mu})
                        \; \right]
\label{eq:spingl} \\
 & = &
   1 \,-\, \frac{1}{d\,V} \sum_{\mu = 1}^{d} \sum_{\bx} \,
         \frac{\Tr}{2} \left[ \; U_{\mu}(\bx)
                              \; g^{\dagger}(\bx + \bun_{\mu})
                              \; g(\bx) \; \right]
\label{eq:spingl2}
\ea
to a local minimum, starting from a randomly chosen configuration
$\{g(\bx)\}$. Here, $\bx$ (with coordinates $x_{\mu} = 1\mbox{,}\,
2\mbox{,}\,\ldots\mbox{,}\,N$)
are sites on a $d$-dimensional lattice
with periodic boundary conditions,
$V = N^d$ is the lattice volume\footnote{~In order
to simplify the notation we don't consider asymmetric lattices.}
and $U_{\mu}(\bx)$ and $g(\bx)$ are $SU(2)$ matrices.
In order to analyze the CSD of an algorithm we have
to  investigate if, and with what exponent, its {\em relaxation
time} $\tau$ diverges as the lattice size increases.
To this end, we have to evaluate $\tau$ for different lattice sides
$N$ at ``constant physics'', namely as the lattice size is increased,
the {\em physical size} of the lattice should remain fixed. This is
done by introducing a {\em correlation length} $\xi$ and by keeping
the ratio $N / \xi$ constant. The lattice configuration $\{ U_{\mu}(\bx)
\}$ is usually chosen in equilibrium with the Gibbs weight 
$\exp\left(- S\left[\beta, U \right]\right)$, where $S\left[\beta,
U \right]$ is the standard Wilson action in $d$ dimensions. Since
$S\left[\beta, U \right]$ depends only on the coupling $\beta$ and on the
lattice side $N$, we must change $\beta$ and $N$ in such a way that the
pairs $(\beta\mbox{,}\,N)$ correspond to the same value of $N / \xi$.
For example --- in two dimensions and for $\xi \equiv 1 / \sqrt{\kappa}$,
where $\kappa$ is the {\em string tension} --- $\,\xi(\beta\mbox{,}\,N)$ is
known in the infinite-volume limit \cite{DM} and for large $\beta$
we have the simple relation $\xi(\beta\mbox{,}\,\infty) \propto
\beta^{1/2}\,$. [In Ref.\ \cite{CM} we considered different pairs
$(\beta\mbox{,}\,N)$ with the ratio $N^2 / \beta$ fixed and equal to $32$,
which corresponds to $N / \xi \approx 7$.] Of course, if the function
$\xi(\beta\mbox{,}\,N)$ is not available, one should  evaluate directly
(i.e.\ numerically) the correlation length $\xi$, and tune $\beta$ and $N$
so that the ratio $N / \xi$ is kept (approximately) fixed.

Since we are interested in studying gauge-fixing algorithms, i.e.\
minimizing the functional \reff{eq:spingl} for a given
lattice configuration $\{ U_{\mu}(\bx) \}$, a simpler
possibility \cite{H} is to work with lattices at $\beta = \infty$.
This corresponds to using the Gibbs weight $\prod_{\bx,\mu}\,\delta(\1 - 
U_{\mu}(\bx))$, namely the lattice configuration is completely 
ordered. In this case, the string tension $\kappa$ is zero for any
lattice side $N$, and therefore $N / \xi = N \sqrt{\kappa}$ is constant
and equal to zero for all the pairs $(\beta = \infty\mbox{,}\;N)$.
Note that with this particular choice the link variables
$U_{\mu}(\bx)$ are set to $\1$ and the configuration $\{g(\bx)\}$
is chosen randomly. Thus, $U^{(g)}_{\mu}(\bx)$ is a pure gauge
configuration, i.e.\ $U^{(g)}_{\mu}(\bx) = g(\bx)\, g^{-1}(\bx +
\bun_{\mu})$, which should be driven by the gauge fixing to the
the {\em vacuum} configuration $U^{(g)}_{\mu}(\bx) = \1$.
As we will see in Section \ref{Infinito}, the
advantage of using this particular case --- for which we know
the final gauge-fixed configuration --- is that we can study
analytically the efficiency of the gauge-fixing algorithms.

\vskip 3mm

In Section \ref{sec:gf} we briefly review the five gauge-fixing
algorithms considered in Ref.\ \cite{CM}. The case $\beta = \infty$
is studied analytically for general dimension $d$ in Sections
\ref{Infinito}, \ref{CSD} and \ref{Better} and the results
are checked numerically in two,
three and in four dimensions in Section \ref{InfinitoRes}.
The analytic approach (see Section \ref{Infinito})
is done by mapping --- in the limit of large number
of gauge-fixing sweeps $t$ --- the original problem \reff{eq:spingl} into
the minimization of the action of a three-vector massless-free-field model
[see eq.\ \reff{E2ter}]. In this way we obtain that the Cornell
method coincides with the overrelaxation method, in agreement with
the result found in Ref.\ \cite{CM}, and we show that the Fourier
acceleration method can eliminate critical slowing-down completely.
In Section \ref{CSD} we review the analysis of CSD done in Refs.\
\cite{N,Wo} for the thermalization of the Gaussian model
and we apply it to the four local gauge-fixing
algorithms considered here.
This analysis confirms the results obtained in Ref.\ \cite{CM} and
will give us a better understanding of the tuning for the three improved
local algorithms.
Moreover, in Section \ref{Better}, we
study generalizations of the overrelaxation algorithm
and of the stochastic overrelaxation algorithm and check that
we cannot have a dynamic critical exponent $z$ smaller than 1 with
these local algorithms.
Note that this result applies directly to the problem of
thermalizing a massless free field \cite{N,Wo,BN}.
Finally, the analytic approach is extended
to the so-called $\lambda$ gauges (see for example \cite{lambda}) in
Section \ref{Lambda} and the results are verified numerically
in the two-dimensional case.

Let us remark that the case $\beta = \infty$ gives information
valid also for the algorithms at finite $\beta$.
In fact, for each algorithm and ``constant physics'',
the evaluated relaxation times should be fitted by a 
function of the type
\be
\tau \, = \, c \, N^{z}
\label{eq:taufit}
\;\mbox{,}
\ee
where the dynamic critical exponent $z$ {\bf should not} depend on 
the constant physics, i.e.\ it should be the same\footnote{~Of course,
if the system undergoes a phase transition going from $\beta = 0$ to
$\beta = \infty$, then an algorithm can show a different behavior in 
different phases (see for example Ref.\ \cite{XY}).}
at finite $\beta$
and at $\beta = \infty$. On the contrary, the constant $c$ 
{\bf should} depend on the ratio $N / \xi$. In particular, for
each algorithm, we should find a faster convergence of the
gauge-fixing procedure --- and therefore a smaller constant $c$ ---
when the configuration $U_{\mu}(\bx)$
is less disordered, i.e.\ for larger values of $\beta$. This
implies
\be
c ( \beta = \infty )\; < \; c ( \,\mbox{finite}\, \beta)
\;\mbox{.}
\ee
In Section \ref{Concl} we check
this inequality in two dimensions using the data
obtained in Ref.\ \cite{CM} at finite $\beta$ and the data
produced for the present work. In the same
section we also give our final comments and conclusions.
 

\section{Gauge-Fixing Algorithms}
\label{sec:gf}

In this Section we review briefly the five gauge-fixing
algorithms we want to study, i.e.\ the so-called {\em Los Alamos}
method, the {\em overrelaxation} method, the {\em stochastic
overrelaxation} method, the so-called {\em Cornell} method and the
{\em Fourier acceleration} method. To this end, let us recall
that the updated gauge transformation at each step
of the algorithm can be written for the four local methods as
follows\footnote{~See Ref.\ \cite{CM} for more details.}
\ba
g^{(LosAl.)}(\bx) &\equiv&
{\widetilde h}^{\dagger}(\bx)
\label{eq:gLosAl}
\\[0.2cm]
g^{(cornell)}(\bx) &\equiv& \frac{
   \alpha \, {\cal N}(\bx) \, {\widetilde h}^{\dagger}(\bx)
       \, + \, \left[ 1 \, - \, \alpha\, {\cal N}(\bx) \,
        {\cal T}(\bx)  / 2\, \right] \, g(\bx)}{
     \sqrt{\, 1 + \alpha^2\, {\cal N}^{2}(\bx)  \left[
            \, 1 - {\cal T}^{2}(\bx) / 4 \, \right]
        }}
\\[0.2cm]
g^{(over)}(\bx) &\equiv& \frac{ \omega \,\, {\widetilde h}^{\dagger}(\bx)
       \,+\, \left(1 \,-\, \omega\right)\, g(\bx) 
          }{ \sqrt{ 1 + \omega
            \left( \omega - 1\right) \left[\,2 - {\cal T}(\bx)
             \,\right] }}
\label{eq:goverre}
\\[0.2cm]
g^{(stoc)}(\bx) &\equiv &
   \left\{ \begin{array}{ll}
     {\widetilde h}^{\dagger}(\bx) \, {\cal T}(\bx) \, - \, g(\bx) &
                   \quad \mbox{with probability $p$} \\
      \phantom{ } & \phantom{ } \\
          {\widetilde h}^{\dagger}(\bx) &
                   \quad \mbox{with probability $1 - p$}
        \end{array} \right.
\label{eq:gstocas}
\;\mbox{.}
\ea
Notice that, with the exception of the Los Alamos method,
all the above algorithms have a tuning parameter, i.e.\ 
$\alpha$ for the Cornell method, $\omega$ for
the overrelaxation algorithm and $p$ for the stochastic overrelaxation
algorithm. Also, in eqs.\ \reff{eq:gLosAl}--\reff{eq:gstocas}
we have used the definition
\be
h(\bx) \,\equiv\,
        \sum_{\mu = 1}^{d}
            \left[ \; U_{\mu}(\bx) \, g^{\dagger}(\bx + \bun_{\mu}) +
        U_{\mu}^{\dagger}(\bx - \bun_{\mu}) \,
	g^{\dagger}(\bx - \bun_{\mu}) \; \right]
\label{eq:hdefi}
\ee
and the fact that the matrix $h(\bx)$ is proportional
to an $SU(2)$ matrix, namely it can be written as
\be
h(\bx) \,\equiv\, {\cal N}(\bx) \; {\widetilde h}(\bx)
\label{def_ztilde}
\ee
with ${\widetilde h}(\bx) \in SU(2)$ and
\be
{\cal N}(\bx)\,=\,
\sqrt{\det h(\bx)} 
\, = \, \sqrt{\frac{\Tr}{2} \,h(\bx) \,h^{\dagger}(\bx) \,}
   \; \mbox{.}
\label{eq:defN}
\ee
We also define
\be
{\mathrm w}(\bx) \,\equiv\, g(\bx) \, h(\bx)
\label{eq:wdefi}
\ee
and
\be
{\cal T}(\bx) 
\, = \,\frac{1}{{\cal N}(\bx)} \, \mbox{Tr}\,{\mathrm w}(\bx)
\; \mbox{.}
\label{eq:calT}
\ee
Using the definition of $U^{(g)}_{\mu}(\bx)$ [see eq.\ \reff{eq:spingl}]
and eq.\ \reff{eq:hdefi} one can check that
\be
{\mathrm w}(\bx) \,=\,\sum_{\mu = 1}^{d}
            \left[ \, U^{(g)}_{\mu}(\bx) +
        {U^{(g)}_{\mu}}^{\dagger}(\bx - \bun_{\mu}) \,
        \right]     
\; \mbox{.}
\label{eq:wofU}
\ee

\vskip 0.3cm

For the Fourier acceleration algorithm one usually
writes \cite{CM}
\be
g^{(Fourier)}(\bx) \, \propto \,
\left\{ \1 \, - \, \alpha \left[ \, {\widehat F}^{-1}
         \, \frac{1}{p^{2}(\bk)} {\widehat F} \,
              \left(\nabla\cdot A^{(g)} \right) \right]\!(\bx) \right\}\,
      g(\bx)
\;\mbox{,}
\label{eq:gFouriersimpl}
\ee
where $\alpha$ is a tuning parameter, $\1$ is the identity matrix,
${\widehat F}$ indicates the Fourier transform,
${\widehat F}^{-1}$ is its inverse,
\be
\left(\nabla\cdot A^{(g)} \right)(\bx) \, \equiv \,
  \sum_{\mu = 1}^{d} \, A^{(g)}_{\mu}(\bx) -
                  A^{(g)}_{\mu}(\bx - \bun_{\mu})
\label{eq:diverA}
\ee
is the lattice divergence of the (gauge-transformed)
gauge field
\be
A^{(g)}_{\mu}(\bx) \,\equiv\, \frac{1}{2}
              \left[ \; U^{(g)}_{\mu}(\bx) -
                    {U^{(g)}_{\mu}}^{\dagger}(\bx)
              \; \right]
\label{eq:defA}
\ee
and
\be
p^{2}(\bk) \equiv 4 \, \sum_{\mu = 1}^{d} \,
\sin^{2}\left( \, \frac{p_{\mu}}{2}\, \right)
 \equiv  4 \, \sum_{\mu = 1}^{d} \,
\sin^{2}\left( \, \pi \,k_{\mu} \, \right)
\label{eq:p2def}
\ee
is the squared magnitude of the lattice momentum. The vector $\bk$
has components $k_{\mu}$ $(\mu = 1, \, \dots, d)$ given by
$k_{\mu}\,N\, =\, 0\mbox{,}\,1\mbox{,}\,\ldots \mbox{,}\,
N - 1\,$, where $N$ is the lattice side.
Sometimes eq.\ \reff{eq:gFouriersimpl} is also written
with $\, p^{2}_{max} \,/\, p^{2}(\bk)\, $ instead of
$\, 1 \,/\, p^{2}(\bk)\, $. However, $\, p^{2}_{max} = 4\,d\, $ is a
constant and can therefore be absorbed into the tuning
parameter $\alpha$.

Clearly, using eqs.\ \reff{eq:wofU},
\reff{eq:diverA} and \reff{eq:defA}, we have
\ba
\left(\nabla\cdot A^{(g)} \right)(\bx) &=&
\frac{1}{2} \,\sum_{\mu = 1}^{d} \,
U^{(g)}_{\mu}(\bx) - {U^{(g)}_{\mu}}^{\dagger}(\bx) -
U^{(g)}_{\mu}(\bx - \bun_{\mu}) + {U^{(g)}_{\mu}}^{\dagger}(\bx - \bun_{\mu})
\\
&=& \frac{1}{2} \, \left[\, {\mathrm w}(\bx) \,-\,
                          {\mathrm w}^{\dagger}(\bx) \, \right]
\;\;\mbox{.}
\ea
For a matrix $g \in SU(2)$ we use the parametrization
\be
g \, \equiv \, g_{0} \, \1 + i \bsig \cdot \bvg \, = \,
    \left( \begin{array}{rr}
          g_{0} + i g_{3} & g_{2} + i g_{1} \\
        - g_{2} + i g_{1} & g_{0} - i g_{3}
    \end{array} \right)
\;\;\mbox{,}
\label{eq:su2para}
\ee
where the components of $\bsig \equiv ( \sigma^{1} \mbox{,} \,
\sigma^{2} \mbox{,} \, \sigma^{3} ) $ are the three Pauli matrices.
Since ${\mathrm w}(\bx)$ is proportional to an $SU(2)$
matrix we can write ${\mathrm w}(\bx) = {\mathrm w}_{0}(\bx)
\1 + i \bsig\cdot \bvw(\bx)$ and 
\be
\left(\nabla\cdot A^{(g)} \right)(\bx) \,=\,
 i\,\bsig\cdot \bvw(\bx)
\label{eq:nablaAofw}
\;\mbox{.}
\ee
Defining now
\be
\bvu(\bx) \,\equiv\, \left\{ {\widehat F}^{-1}\left[
         \, \frac{1}{p^{2}(\bk)} \,
                {\widehat F} \bvw \right] \right\}\!(\bx)
\label{eq:bvu}
\;\mbox{,}
\ee
we have
\be
g^{(Fourier)}(\bx) \, \propto \,
\left\{ \1 \, - \,i\,\alpha\,\bsig\cdot \bvu(\bx)
              \right\}\, g(\bx)
\ee
and the Fourier acceleration update can be written as
\be
  g^{(Fourier)}(\bx) \equiv
\frac{\1 - \,i\,\alpha\,\bsig\cdot \bvu(\bx)}{\sqrt{
                1 + \alpha^2\,\left|\, \bvu(\bx)\,\right|^2 }} \, g(\bx)
\label{eq:gFourier}
\;\mbox{.}
\ee
In order to simplify the expression for $\bvu(\bx)$ in
eq.\ \reff{eq:bvu}, recall that (minus)
the lattice Laplacian $ \Delta $ [defined in eq.\
\reff{eq:laplaciano} below] is diagonal in momentum space, with
elements given by $p^2(\bk)$. Therefore we have
\be
{\widehat F}^{-1} \,\frac{1}{p^{2}(\bk)}\,{\widehat F}
 \;=\;\left(\,- \Delta\,\right)^{-1}
\label{eq:Dm1}
\ee
and
\be
\bvu(\bx) \;=\;\left[\,\left(\,- \Delta\,\right)^{-1} \,
 \bvw\,\right]\!(\bx)
\; \mbox{.}
\label{eq:bvubis}
\ee
Clearly, the inverse of the (minus) Laplacian is defined only in
the sub-space orthogonal to the constant mode $\bk = 0$.
From the previous equation it is evident that the
Fourier acceleration method is actually a Laplacian preconditioning
algorithm.


\section{The Case $\beta = \infty$}
\label{Infinito}

As said in the Introduction, we consider here the case in which all
the link variables $U_{\mu}(\bx)$ are set equal to the identity
matrix. Then, the minimizing functional \reff{eq:spingl} becomes
\ba
{\cal E}_{U}\left[\, g \, \right] & = & 1 \,-\,
\frac{1}{d\,V} \sum_{\mu = 1}^{d} \sum_{\bx} \,
         \frac{\Tr}{2} \left[ \;
             g(\bx) \; g^{\dagger}(\bx + \bun_{\mu})
                   \; \right]
\nonumber \\
& = &
\frac{1}{V}\sum_{\bx}\, \frac{\Tr}{2} \, \frac{1}{2\,d}\,
\sum_{\mu = 1}^{d} \left\{ \left[ \, g(\bx)\,-\,g(\bx + \bun_{\mu})
                             \,\right]\,
\left[ \, g(\bx)\,-\,g(\bx + \bun_{\mu})\,\right]^{\dagger}
\right\} \label{E1} \\
& = & \frac{1}{V}\sum_{\bx}\,\frac{1}{2\,d}\,
\sum_{\mu = 1}^{d}  \| \, \bg(\bx)\,-\,\bg(\bx + \bun_{\mu})\,\|^{2}
\label{E2}
\;\mbox{.}
\ea
Here we have used $\Tr\,g\,=\,\Tr\,g^{\dagger}$, $g\,g^{\dagger}\,
=\,\1$ and [using eq.\ \reff{eq:su2para}]
\be
\frac{1}{2}\, \Tr \; ( g \; g^{\dagger} ) \; = \;
   g_{0} \, g_{0} \, + \,
        \bvg \cdot \bvg \, \equiv 
        \, \bg \cdot \bg \, = \,\| \bg\|^2
\label{eq:scalarp}
\;\mbox{,}
\ee
where $\bg \equiv (g_{0} \mbox{,} \, \bvg )$ is a four-dimensional
unit vector.

Equation \reff{E2} is very similar to the action of a four-vector
massless free field $\bg(\bx)$. The only difference is that our 
field must satisfy the constraint $\| \bg(\bx) \|^{2} = 1$, namely
we are dealing with an $O(4)$ nonlinear $\sigma$-model. However,
after a few sweeps, we expect the (gauge-transformed) gauge configuration 
$U^{(g)}_{\mu}(\bx)\, = \,g(\bx) \, g^{\dagger}(\bx + \bun_{\mu})$
to be very close to the vacuum configuration, i.e.\ we expect the
matrices $g(\bx)$ to be very close to a constant matrix.
Since the minimizing functional (\protect\ref{eq:spingl2}) is
invariant under global gauge transformations, i.e.\ 
transformations in which $g(x)$ is
$x$-independent, we can set this constant matrix equal to the identity
matrix $\1$. Therefore, in the limit of large
number of gauge-fixing sweeps $t$ we can write\footnote{~Note
that with the parametrization $\,g(\bx) \, = \,
g_{0}(\bx) \, \1 + i \,\epsilon\,\bfv(\bx) \cdot \bsig \,
+\,{\cal O}(\epsilon^2)$
the unitarity condition for $g(\bx)$ implies that the
corrections to $g_0(\bx) \approx 1$ start at order
$\epsilon^2$.}
\be
g(\bx) \; = \;
  \1 +\,i\, \epsilon\, \bsig\cdot\bfv(\bx) \,+\,
{\cal O}(\epsilon^2)
\label{eq:gI}
\;\mbox{,}
\ee
with $\epsilon \ll 1$.
This approximation is justified because it is exactly in the limit
of large $t$ --- in which only the slowest mode survives --- that
we evaluate the relaxation time of each algorithm.

Now, by using eq.\ \reff{eq:gI}, we can rewrite the minimizing
functional \reff{E2} as
\be
{\cal E}\left[\, f\, \right]
\, = \, \frac{1}{V}\sum_{\bx}\,\frac{\epsilon^2}{2\,d}\,
\sum_{\mu = 1}^{d}  \left| \, \bfv(\bx)\,-\,\bfv(\bx + \bun_{\mu})
\,\right|^{2} \; +\,{\cal O}(\epsilon^3) 
\label{E2bis}
\ee
and, using the periodicity of the lattice,
\be
{\cal E}\left[\, f\, \right]
\, = \, \frac{1}{V}\frac{\epsilon^2}{2\,d}\sum_{\bx}\,\bfv(\bx)\cdot
\left(\,- \, \Delta\,\bfv\,\right)\!(\bx)\; +\,{\cal O}(\epsilon^3)
\;\mbox{,}
\label{E2ter}
\ee
where $- \Delta$ is (minus) the lattice Laplacian, defined by
\be
\left(\,-\,\Delta \,\bfv\,\right)\!(\bx) \equiv 
\sum_{\mu = 1}^{d}\left[ \;2\,\bfv(\bx)\,-
\,\bfv(\bx + \bun_{\mu}) -
\bfv(\bx - \bun_{\mu}) \; \right]
\label{eq:laplaciano}
\;\mbox{.}
\ee
Therefore, by consistently keeping only terms up to order $\epsilon$
in $g(\bx)$, we have that ${\cal E}\left[ \,f \,\right]$ ---
at the lowest order in $\epsilon$ --- is the action of the 
three-vector massless free field $\bfv(\bx)$. In particular, we can 
update this field, in order to minimize the action, without taking 
into account the problem of the unitarity of $g$. In fact, from eq.\ 
\reff{eq:gI} we have $g(\bx)\,g^{\dagger}(\bx) \,=\,
\1\,+\,{\cal O}(\epsilon^2)$ for {\em any} field $\bfv(\bx)$.

It is not difficult to translate the update of the $g(\bx)$ variables
for the five algorithms considered here into an update for the
$\bfv(\bx)$ field. To this end, we use the approximation in
eq.\ \reff{eq:gI} and obtain [see eqs.\ \reff{eq:hdefi} and
\reff{eq:wdefi}]
\ba
h(\bx) &=&
        \sum_{\mu = 1}^{d}
            \left[ \; g^{\dagger}(\bx + \bun_{\mu}) +
        g^{\dagger}(\bx - \bun_{\mu}) \; \right] \nonumber \\
       &=&\, 2\,d\,\1\,-\,i\,\epsilon\,\bsig\cdot\,
        \sum_{\mu = 1}^{d}
            \left[ \; \bfv(\bx + \bun_{\mu}) +
        \bfv(\bx - \bun_{\mu}) \; \right]\,+\,{\cal O}(\epsilon^2) 
\label{eq:happrox} \\[0.2cm]
{\mathrm w}(\bx) &=&  \,g(\bx)\, h(\bx) \nonumber \\[0.2cm]
       &=& \, 2\,d\,\1\,+\,i\,\epsilon\,\bsig\cdot\,
        \sum_{\mu = 1}^{d}
            \left[ \;2\,\bfv(\bx)\,-\, \bfv(\bx + \bun_{\mu})
                \,-\, \bfv(\bx - \bun_{\mu}) \; 
                  \right]\,+\,{\cal O}(\epsilon^2)
\ea
and
\be
\bvw(\bx)
\, = \, \,\epsilon\,\left(\,- \Delta\,\bfv\,\right)\!(\bx)
\,+\,{\cal O}(\epsilon^2)
\; \mbox{,}
\ee
where we used eq.\  \reff{eq:laplaciano}. From these
expressions it follows that
[see eqs.\ \reff{eq:defN}, \reff{eq:calT} and
\reff{eq:bvubis}]
\ba
{\cal N}(\bx)& = & \, 2\,d\,+\,{\cal O}(\epsilon^2)
\label{eq:Nd}
\; \mbox{,}
\\[0.2cm]
{\cal T}(\bx)& = & \, 2\,+\,{\cal O}(\epsilon^2)
\label{eq:Td}
\; \mbox{,}
\\[0.2cm]
\bvu(\bx) & =& \epsilon\,
\bfv(\bx)\,+\,{\cal O}(\epsilon^2)
\;\mbox{.}
\label{eq:bvu2}
\ea
Actually, to be precise [see comment after eq.\ \reff{eq:bvubis}], we 
should write eq.\ \reff{eq:bvu2} as
\be
\bvu(\bx) \,=\, \epsilon\,
\left[ \, \bfv(\bx) \,-\, \bfv_{0}\,\right]\,+\,{\cal O}(\epsilon^2)
\;\mbox{,}
\ee
where
\be
\bfv_{0} \,\equiv\, \frac{1}{V} \, \sum_{\bx}\, \bfv(\bx)
\;\mbox{.}
\ee
However, the zero mode of $g(\bx)$, and therefore of $\bfv(\bx)$, does
not contribute to the value of the minimizing functional \reff{E1}. 
Thus, in eq.\ \reff{eq:gI}, we can always consider a field $\bfv(\bx)$
with zero constant mode $\bfv_{0}$.

Now, by substituting eq.\ \reff{eq:gI} and eqs.\ 
\reff{eq:happrox}--\reff{eq:bvu2} into
eqs.\ \reff{eq:gLosAl}--\reff{eq:gstocas} and \reff{eq:gFourier} 
we obtain for the updated $\bfv(\bx)$ variables
(at the lowest order in $\epsilon$)
\ba
\bfv^{(LosAl.)}(\bx) &=& \frac{1}{2\,d}\,
        \sum_{\mu = 1}^{d} \left[ \; \bfv(\bx + \bun_{\mu}) +
        \bfv(\bx - \bun_{\mu}) \; \right]
\label{eq:LOSbfv}
\\[0.1cm]
\bfv^{(cornell)}(\bx) &=& 2\,d\,\alpha\, \bfv^{(LosAl.)}(\bx) \,+\,
                         \left(\,1\,-\,2\,d\,\alpha\,\right) \,\bfv(\bx) 
\label{eq:bfvCor}
\\[0.2cm]
\bfv^{(over)}(\bx) &=& \omega\,\bfv^{(LosAl.)}(\bx) \,+\,
                      \left(\,1\,-\,\omega\,\right)
                      \,\bfv(\bx)
\label{eq:bfvOve}
\\[0.2cm]
\bfv^{(stoc)}(\bx) &=&
   \left\{ \begin{array}{ll}
          2\,\bfv^{(LosAl.)}(\bx) \, - \, \bfv(\bx) &
                   \quad \mbox{with probability $p$} \\
      \phantom{ } & \phantom{ } \\ \bfv^{(LosAl.)}(\bx) &
                   \quad \mbox{with probability $1 - p$}
        \end{array} \right.
\label{eq:bfvstoc}
\\[0.2cm]
\bfv^{(Fourier)}(\bx) &=& \left(\,1\,-\,\alpha\,\right)\,\bfv(\bx)
\;\mbox{.}
\label{eq:FOUbfv}
\ea
The interpretation of these updates is clear if we consider the
{\em local} minimization of ${\cal E}\left[\, f\, \right]$,
i.e.\ we consider all $f(\bx)$'s fixed for $\bx \neq \by$, and
we make explicit the
dependence of the minimizing functional \reff{E2bis} on the value of
the field $\bfv$ at site $\by$, namely (to order $\epsilon^2$)
\be
{\cal E}\left[\, f(\by) \,\right] \,=\,
    \frac{\epsilon^2}{V}\, \bfv(\by) \cdot \left[\,
   \bfv(\by) \,-\, 2\, \bfv^{(LosAl.)}(\by)\,\right]\,+\,
\mbox{constant terms}
\;\mbox{,}
\label{eq:Eatsitey}
\ee
where we have used eq.\ \reff{eq:LOSbfv}.
Since $\bfv^{(LosAl.)}(\by)$ does not depend on $\bfv(\by)$ we
can complete the square and write
\be
{\cal E}\left[\, f(\by) \,\right] \,=\,
    \frac{\epsilon^2}{V}\, \left|\,
   \bfv(\by) \,-\,  \bfv^{(LosAl.)}(\by)\,\right|^2\,+\,
\mbox{constant terms}
\;\mbox{.}
\label{eq:Eatsiteybis}
\ee
Then it is clear that the Los Alamos update brings 
this local functional to its minimum, while the choice
$\bfv(\by) \, \to \, 2\,\bfv^{(LosAl.)}(\by) \, - \, \bfv(\by)$
does not change the value of ${\cal E}\left[ f \right]$.
From formulae \reff{eq:LOSbfv}--\reff{eq:FOUbfv}
it is also evident that:
\begin{itemize}
\item The Los Alamos method corresponds to the usual 
      {\em Gauss-Seidel} method \cite{Axe,Sa}.
\item The Cornell method coincides with the overrelaxation method
      if we use the relation [see eqs.\ \reff{eq:Nd}, \reff{eq:bfvCor}
      and \reff{eq:bfvOve}]
\be
\omega\,=\,2\,d\,\alpha\,=\,\alpha\,{\cal N}(\bx)\, + 
{\cal O}(\epsilon^2)
\label{eq:omegaalpha}
\;\mbox{.}
\ee
      This confirms the result found analytically and
      numerically in two dimensions at
      finite $\beta$ (see Sections 5 and 7.3 in Ref.\ \cite{CM}).
\item The overrelaxation method corresponds to the usual 
      {\em successive overrelaxation} method \cite{Axe,Sa}.
\item Since the vacuum configuration $\{ U_{\mu}(\bx) = \1 \}$ 
      corresponds to $\bfv(\bx) = 0$ for all $\bx$, it is clear 
      that the Fourier method can minimize ${\cal E}\left[\,
      f \,\right]$ in only one step if we set $\alpha = 1$. 
      Thus, in this case, critical slowing-down is completely 
      eliminated. Note that the tuning condition, i.e.\ $\alpha = 1$,
      does not depend on the dimension $d$ of the lattice.
\end{itemize}
Moreover, it is now possible to study analytically the critical 
slowing-down for the four local algorithms following the analyses in
Refs.\ \cite{N,Wo} (see next section). As we will see, the
results of this analytic approach confirm the dynamic critical
exponents found numerically in two dimensions at finite $\beta$
\cite{CM}. Also, these results are
particularly interesting with respect to 
the problem of tuning the improved local algorithms (see Section
\ref{InfinitoTun}).

\vskip 0.3cm

Let us notice that for $t$ going to infinity
all components of $\bfv(\bx)$ must go to zero for the
algorithm to converge.
Therefore, if in some basis we can write
\be
f_{t} \,=\, C f_{t-1}
\label{eq:fupdateC}
\;\mbox{,}
\ee
where $C$ is the updating matrix, then we should have
\be
\lim_{t \to \infty} \,
\| f_t \| \, = \,
\lim_{t \to \infty} \, \| C^t f_0 \| \,=\, 0
\;\mbox{,}
\label{eq:flimit}
\ee
for any reasonable definition of the norm $\| f \|$
and of the initial condition $f_0$.
From the definition of the norm of a matrix (see for example
Section 1.5 in \cite{Sa}) it follows that
\be
\| C^t f_0 \| \,\leq\, \| C^t \| \, \| f_0 \|
\label{eq:Climit}
\;\mbox{.}
\ee
Thus, the limit in eq.\ \reff{eq:flimit} is verified if
$\| C^t \|$ goes to zero when $t$ goes
to infinity, i.e.\ if the matrix $C^t$ goes to 0 in the
same limit. This happens if and only if (see theorem 1.4 in \cite{Sa})
the {\em spectral radius} $\rho(C)$ of the matrix $C$ is smaller than 1,
where
\be
\rho(C) \equiv \, \max_{\lambda \in \sigma(C)} \,\left| \lambda \right|
\ee
and $\sigma(C)$ is the set of all the eigenvalues of
the matrix $C$. One can also prove (see theorem 1.6 in \cite{Sa}) that
\be
\rho(C) \,=\, \lim_{t \to \infty} \| C^t \|^{1/t}
\label{eq:Candrho}
\ee
for any matrix norm $\| \,.\, \|$.
It follows that the algorithm converges if all the eigenvalues 
of the updating matrix $C$ are smaller than 1 (in absolute value),
i.e.\ if $\rho(C) < 1$. From now on, we consider only the case
$\rho(C) < 1$. Then, we can define the relaxation time $\tau > 0$
through the relation
\be
e^{- 1 / \tau} \,\equiv\, \rho(C)
\label{eq:tauandrho}
\;\mbox{,}
\ee
namely
\be
\tau \,=\,  \frac{- 1}{\log{\rho(C)}}
\label{eq:tauandrho2}
\;\mbox{.} \\[0.1cm]
\ee
Clearly if $\rho(C)$ is very close to 1, i.e.\ if
\be
\rho(C)
 \, \approx\, 1\, -\, \delta
\ee
with $0 < \delta \ll 1$, we find
\be
\tau \, \approx\, \frac{1}{\delta} \,\approx\, \frac{1}{1\, -\,
    \rho(C)} \;\mbox{.}
\label{eq:tau}
\ee
One can also consider the inequality
\be
1 \,-\, |\,\lambda\,| \,\leq\,
 |\,1\,-\,\lambda\,|
\;\mbox{,}
\ee
which becomes an equality if the eigenvalue $\lambda$ is real
and positive (since we are considering $| \,\lambda\,| < 1$).
This implies
\be
1\,-\, \rho(C) \,=\, 1\,-\,
\max_{\lambda \in \sigma(C)} \,\left| \lambda \right|
\,=\,
\min_{\lambda \in \sigma(C)} \, \left(
1\,-\,\left| \lambda \right| \right)
\,\leq\,
\min_{\lambda \in \sigma(C)} \,
\left| 1\,-\,
\lambda \right|
\;\mbox{.}
\ee
Thus, if $\rho(C)$ is very close to 1 we obtain
\be
\tau \, \approx\,\frac{1}{1\, -\,
    \rho(C)} \,\geq\,
\frac{1}{\min_{\lambda \in \sigma(C)} \,
\left| 1\,-\,
\lambda \right|}
\label{eq:tauanddelta}
\;\mbox{.}
\ee

Notice that eqs.\ \reff{eq:fupdateC}, \reff{eq:Climit},
\reff{eq:Candrho} and \reff{eq:tauandrho} imply that,
in the limit of large $t$,
\be
\| f \| \sim e^{-t /\tau}
\label{eq:deftau}
\;\mbox{.}
\ee
Thus, if we know the matrix $C$ and we can find its eigenvalues
$\lambda$, we can evaluate the relaxation time $\tau$ using
eq.\ \reff{eq:tauandrho2} [or the approximate expressions \reff{eq:tau},
\reff{eq:tauanddelta}]. On the contrary, for a numerical
determination of $\tau$ one should use eq.\ \reff{eq:deftau}
(for some definition of the norm $\| f \|$).


\section{Analysis of Critical Slowing-Down}
\label{CSD}

In this Section we review the analyses of critical slowing-down 
done in Refs.\ \cite{N,Wo} and we apply them to the four local
algorithms considered in this paper.\footnote{~See also
Ref.\ \protect\cite{HK} for a careful analysis of CSD for the
local hybrid Monte Carlo algorithm (which is equivalent to the
stochastic overrelaxation method) applied to the free-field case
using various updating schemes. Here we will consider only
the so-called even/odd update.} The only difference with
respect to those references is that, in our case, we minimize the 
free-field action \reff{E2bis} instead of thermalizing the 
configuration $\{ \bfv(\bx) \}$. Thus, their analyses can be applied 
directly to our case by setting the Gaussian noise $\eta(\bx)$
to zero.
We stress that the results presented in this section
are a straightforward application of the analyses reported
in Refs.\ \cite{N,Wo} and that most of these results, but not all
of them, can be found there and in other articles. However, we believe
that our presentation has some interesting insights, which can
help the reader understand how local algorithms deal with
the problem of critical slowing-down. Moreover, these
results will be extensively used in Section \ref{Better},
where we study generalizations of the overrelaxation
algorithm and of the stochastic overrelaxation algorithm.

\vskip 0.3cm

Let us start by considering the {\bf overrelaxation update}
[see eqs.\ \reff{eq:LOSbfv} and \reff{eq:bfvOve}]
\be
\bfv^{(over)}(\bx) \; = \; 
   - \left(\,\omega\,-\,1\,\right)\,\bfv(\bx)\,+\,
\frac{\omega}{2\,d}\,
        \sum_{\mu = 1}^{d}
            \left[ \; \bfv(\bx + \bun_{\mu}) +
        \bfv(\bx - \bun_{\mu}) \; \right]
\label{upfove2}
\;\mbox{.}
\ee
One can check that the condition $\left(\omega\,-\,1\right)^2
< 1$ is sufficient to prove that this update
never increases the value of the massless-free-field
action [see eq.\ \reff{eq:Eatsiteybis}]. Therefore we should
have\footnote{~As shown in Section 3.3
of Ref.\ \cite{CM}, this result also applies to
the update given in eq.\ \protect\reff{eq:goverre} when considering 
the minimizing functional \reff{eq:spingl}.}
$\omega \in (0\mbox{,}\, 2)$.
However, only when $\,\omega \in (1\mbox{,}\, 2)\,$ does
one obtain \cite{CM} that the overrelaxation
method performs better than the Los Alamos method, which corresponds
to the case $\omega = 1$. For the Cornell method, it follows
from eq.\ \reff{eq:omegaalpha} that one should have
$\alpha\,{\cal N}(\bx) \in (1\mbox{,}\, 2)$, which gives
$\alpha\,\in (1/ 2 d\mbox{,}\, 1/d)$ for the case $\beta = \infty$.

Clearly, using eq.\ \reff{upfove2}, we need to know the
value of the field $\bfv$ only at the site
$\bx$ and at the nearest-neighbor sites $\bx \pm \bun_{\mu}$
in order to update $\bfv(\bx)$. Note that a site is defined to be
even or odd according to whether the quantity
\be
|\, \bx\, |\, \equiv \, \sum_{\mu = 1}^{d}\,x_{\mu}
\ee
is even or odd.
Thus, if we consider lattices with an even number of sites in each
direction and a checkerboard ordering,
then we can first update all the even sites and 
subsequently all the odd ones, and so on.
In order to implement the checkerboard update we can 
rewrite the update \reff{upfove2} as
\ba
f^a_{t+1}(\bx) & = &
       - \left(\,\omega\,-\,1\,\right)\,f^a_{t}(\bx) \nonumber \\
                   & + & \frac{\omega}{4\,d}\,
        \sum_{\mu = 1}^{d} \left\{\, \left[\,
              1\,+\,(\,-1\,)^{|\, \bx \,|} \,\right]\,
            \left[ \; f^a_{t}(\bx + \bun_{\mu}) +
      f^a_{t}(\bx - \bun_{\mu}) \; \right] \right. \nonumber \\
& & \left. \quad \;\;\;\;\;\;\, + \, \left[\,
              1\,-\,(\,-1\,)^{|\, \bx \,|} \,\right]\,
            \left[ \; f^a_{t+1}(\bx + \bun_{\mu}) +
        f^a_{t+1}(\bx - \bun_{\mu}) \; \right]\,\right\}
\;\mbox{,}
\label{eq:Ebis}
\ea
where $f^a(\bx)$ are the three ``color'' components of
$\bfv(\bx)$ and $t$ denotes the number of
sweeps through the entire lattice.
Notice that when we update the odd sites, i.e.\ when
$\,(\,-1\,)^{|\, \bx \,|} = -1$, we use for the update
the value of the newly updated field $f^a_{t+1}$ at even sites.

The idea in Neuberger's article \cite{N} is to consider the Fourier
transform
\be
{\tilde f}^{a}(\bk) \,\equiv\, \sum_{\bx} \; f^{a}(\bx)\,
 \exp{\left( \,- \, 2\,\pi\,i\,\bk\,\cdot\,\bx\,\right)}
\ee
of eq.\ \reff{eq:Ebis}. To this end one can use the relation
\be
(\,-1\,)^{|\, \bx \,|} \, = \, \exp{\left( \,-\,2\,\pi\,i\,
         \bT\,\cdot\,\bx \,\right)}
\label{eq:defT2}
\;\mbox{,}
\ee
where the vector $\bT$ has components given by
\be
T_{\mu} \, = \, \frac{1}{2}\qquad\qquad\quad
               \,\mu\,=\,1\mbox{,}\, \ldots \, d
\label{eq:defT}
\;\;\mbox{.}
\ee
In this way one obtains
\ba
{\tilde f}^{a}_{t+1}(\bk) \!&=& \!
   - \left(\,\omega\,-\,1\,\right)\,{\tilde f}^{a}_{t}(\bk)
         \nonumber \\[0.3cm]
& & \qquad
\,+\, \omega\, c(\bk)\, \left[\,
{\tilde f}^{a}_{t}(\bk)\,+\,
{\tilde f}^{a}_{t+1}(\bk)\,-\,
{\tilde f}^{a}_{t}(\bk \, +\,\bT)\,+\,
{\tilde f}^{a}_{t+1}(\bk\, +\,\bT)\,\right]
\label{eq:ftildeb}
\;\mbox{,}
\ea
where we defined
\be
c(\bk)\equiv
\frac{1}{2\,d} \,
            \sum_{\mu = 1}^{d} \,\cos{(2\,\pi\,k_{\mu})}
\label{eq:cdefini}
\;\mbox{.}
\ee
One can write eq.\ \reff{eq:ftildeb} also for ${\tilde f}^{a}_{t+1}
(\bk\, +\,\bT)$, namely
\ba
{\tilde f}^{a}_{t+1}(\bk\, +\,\bT) \!&=& \!
   - \left(\,\omega\,-\,1\,\right)\,{\tilde f}^{a}_{t}(\bk\, +\,\bT)
   \nonumber \\[0.3cm]
& & \qquad
\,-\,\omega\,c(\bk)\,
\left[\,
{\tilde f}^{a}_{t}(\bk\, +\,\bT)\,+\,
{\tilde f}^{a}_{t+1}(\bk\, +\,\bT)\,-\,
{\tilde f}^{a}_{t}(\bk)\,+\,
{\tilde f}^{a}_{t+1}(\bk)\,\right]
\label{eq:ftildeb2}
\;\mbox{.}
\ea
Then, equations \reff{eq:ftildeb} and \reff{eq:ftildeb2} can be written
as a system of two equations
\be
A(\bk\mbox{,}\,\omega)\, \left( \begin{array}{ll} 
{\tilde f}^{a}_{t+1}(\bk) \\[0.2cm]
{\tilde f}^{a}_{t+1}(\bk\, +\,\bT) \end{array} \right)
\; = \; B(\bk\mbox{,}\,\omega) \, \left( \begin{array}{ll} 
   {\tilde f}^{a}_{t}(\bk) \\[0.2cm]
{\tilde f}^{b}_{a}(\bk\, +\,\bT) \end{array} \right)
\label{eq:systemAB}
\;\mbox{,}
\ee
where the $2 \times 2$ matrices $A(\bk\mbox{,}\,\omega)$ and 
$B(\bk\mbox{,}\,\omega)$ are given by
\ba
A(\bk\mbox{,}\,\omega) & \equiv&  \1 \, +\, \omega\, c(\bk)\,
\left( \begin{array}{cc}
          - 1 & - 1 \\
          \phantom{-} 1 & \phantom{-} 1
    \end{array} \right)
\label{eq:A} \\
B(\bk\mbox{,}\,\omega)&  \equiv&  - \left(\,\omega\,-\,1\,\right)\, \1 \,
         +\,\omega\, c(\bk)\,
\left( \begin{array}{cc}
            1 & - 1 \\
            1 & - 1
    \end{array} \right)
\;\mbox{.}
\label{eq:B}
\ea
Let us now define
\be
C(\bk\mbox{,}\,\omega) \equiv A^{- 1}(\bk\mbox{,}\,\omega)\, 
B(\bk\mbox{,}\,\omega)
\label{eq:defC}
\;\mbox{;}
\ee
then, the update of the field ${\tilde f}^{a}$, namely
one even update followed by an odd update, can be written 
(in momentum space) as [see eq.\ \reff{eq:systemAB}]
\be
\left( \begin{array}{ll} {\tilde f}^{a}_{t+1}(\bk) \\[0.2cm]
{\tilde f}^{a}_{t+1}(\bk\, +\,\bT) \end{array} \right)
\; = \; C(\bk\mbox{,}\,\omega) \, \left( \begin{array}{ll} 
{\tilde f}^{a}_{t}(\bk) \\[0.2cm]
{\tilde f}^{a}_{t}(\bk\, +\,\bT) \end{array} \right)
\label{eq:ftp1}
\;\mbox{.}
\ee
Notice that the determinant of $A(\bk\mbox{,}\omega)$ is equal
to $1$ for any $\omega$, i.e.\ this matrix can always be
inverted:
\be
A^{-1}(\bk\mbox{,}\,\omega) \,=\,  \1 \, +\,\omega\, c(\bk)\,
\left( \begin{array}{cc}
          \phantom{-} 1 & \phantom{-} 1 \\
          - 1 & - 1
    \end{array} \right)
\label{eq:Ainv}
\;\mbox{.}
\ee
This gives
\be
C(\bk\mbox{,}\,\omega) \,=\,  \left[\,2 \,\omega^2\, c^2(\bk)\,-\,
  \left(\,\omega\,-\,1\,\right)\, \right]\, \1 \,
         +\, \omega\, c(\bk)\,
\left( \begin{array}{cc}
            2\,-\,\omega & -\,\omega\,\left[ 1 \,+\, 2\,c(\bk) \right] \\
            \omega\,\left[ 1 \, -\, 2\,c(\bk) \right] & 
          - \left( 2 \,-\, \omega \right)
    \end{array} \right)
\;\mbox{.}
\label{eq:C}
\ee
It is easy to check that the eigenvalues of the matrix 
$C(\bk\mbox{,}\, \omega)$ are given by
\ba
\lambda_{\pm}(\bk\mbox{,}\,\omega) & = &
\left[\,
2\,\omega^2\,c^{2}(\bk)\,-\,
(\,\omega\,-\,1\,)\,\right]\,
\pm\,\sqrt{\,
\left[\,
2\,\omega^2\,c^{2}(\bk)\,-\,
(\,\omega\,-\,1\,)\,\right]^2\,
-\,(\,\omega\,-\,1\,)^2\,} \label{eq:lambdapmold} \;\;\;\;\;\;\;\;\;\; 
\\[0.2cm]
   & = & \left[\,
2\,\omega^2\,c^{2}(\bk)\,-\,
(\,\omega\,-\,1\,)\,\right]\,
\pm\,2\,\omega\,
  \sqrt{\,\omega^2\,c^{4}(\bk)\,-\, (\,\omega\,-\,1\,)\,c^2(\bk)\,}
\label{eq:lambdapm}
\;\mbox{.}
\ea
Then, if we can prove that
$| \lambda_{\pm}(\bk\mbox{,}\,\omega) | < 1$, we can
use eq.\ \reff{eq:tauandrho2} and write
\be
\tau = \frac{-1}{\log{( \max_{\bk \neq 0} |
\lambda_{\pm}(\bk\mbox{,}\,\omega) | )}}
\label{eq:deftaugen}
\;\mbox{,}
\ee
where we don't consider the constant (or zero) mode
because it does not contribute to the action
\reff{E2bis}. 

We can obtain the eigenvalues $\lambda_{\pm}(\bk\mbox{,}\,\omega)$
also working in a slightly different way, namely following
now Ref.\ \cite{Wo}. The main difference with
respect to the approach used above (based on Ref.\ \cite{N})
is that, instead of considering 
the Fourier transform of eq.\ \reff{eq:Ebis}, one applies to
eq.\ \reff{upfove2} the Fourier-like transform
\be
f^{a, \pm}(\bk) \,\equiv\, \sum_{\bx} \; f^a(\bx)\,
     \left[\, 1 \, \pm \, e^{-\,2\,\pi\,i\,
         \bT\,\cdot\,\bx\, } \, \right] \,
 \exp{\left( \,- \, 2\,\pi\,i\,\bk\,\cdot\,\bx\,\right)}
\label{eq:fourierlike}
\;\mbox{,}
\ee
with $\bT$ defined in eqs.\ \reff{eq:defT2} and \reff{eq:defT}. 
By using the result $\exp{(\pm\,2\,\pi\,i\,T_{\mu})} = - 1\,$
and the periodicity of the lattice
one can verify that
\ba
\sum_{\bx} \; f^a(\bx + e_{\mu})\,
     \left[\, 1 \, \pm \, e^{-\,2\,\pi\,i\,
         \bT\,\cdot\,\bx\, } \, \right] \,
 \exp{\left( \,- \, 2\,\pi\,i\,\bk\,\cdot\,\bx\,\right)}
&=& e^{+\,2\,\pi\,i\,k_{\mu}}\,
         f^{a, \mp}(\bk)
\;\mbox{,}
\label{eq:fourierlike+} \\[0.2cm]
\sum_{\bx} \; f^a(\bx - e_{\mu})\,
     \left[\, 1 \, \pm \, e^{-\,2\,\pi\,i\,
         \bT\,\cdot\,\bx\, } \, \right] \,
 \exp{\left( \,- \, 2\,\pi\,i\,\bk\,\cdot\,\bx\,\right)}
&=& e^{-\,2\,\pi\,i\,k_{\mu}}\,
         f^{a, \mp}(\bk)
\label{eq:fourierlike-}
\;\mbox{.}
\ea
Thus, using eq.\ \reff{upfove2} and eqs.\
\reff{eq:fourierlike}--\reff{eq:fourierlike-}
and by updating first the $\,f^{a, +}(\bk)\,$ components and then
the $\,f^{a, -}(\bk)\,$ components, we obtain
\be
\left( \begin{array}{ll} f^{a, +}_{t+1}(\bk) \\[0.2cm]
f^{a, -}_{t+1}(\bk) \end{array} \right)
\; = \; M(\bk\mbox{,}\,\omega) \,
\left( \begin{array}{ll} f^{a, +}_{t}(\bk) \\[0.2cm]
f^{a, -}_{t}(\bk) \end{array} \right)
\;\mbox{,}
\ee
where
\be
M(\bk\mbox{,}\,\omega)  \equiv  - \left(\,\omega\,-\,1\,\right)\, \1 \,+\,
2\, \omega\, c(\bk)\,
\left( \begin{array}{cc}
            0          & 1                            \\[0.1cm]
            - \left(\,\omega\,-\,1\,\right) & 2\,\omega\,c(\bk)
    \end{array} \right)
\;\mbox{.}
\label{eq:M}
\ee
It is straightforward to check that the matrix $M(\bk\mbox{,}\,\omega)$ also 
has eigenvalues $\lambda_{\pm}(\bk\mbox{,}\,\omega)$ [given in eqs.\
\reff{eq:lambdapmold} and \reff{eq:lambdapm}],
namely $M(\bk\mbox{,}\,\omega)$ is
the matrix $C(\bk\mbox{,}\,\omega)$ [see eq.\ \reff{eq:C}]
written in a different basis.
Indeed, with
\be
R \,\equiv\,\left( \begin{array}{rr}
            1 & 1      \\[0.1cm]
            1 & -1
    \end{array} \right)
\label{eq:D}
\;\mbox{,}
\ee
we have
\be
M(\bk\mbox{,}\,\omega)\, =\, R\, C(\bk\mbox{,}\,\omega)
\,R^{-1}
\label{eq:DCD}
\;\mbox{.}
\ee

\vskip 0.3cm

Note that we started the analysis of the overrelaxation
update from one equation, namely eq.\ \reff{upfove2}, and we
are now considering a system of two equations. Thus, we should
avoid double counting for the momenta \cite{Wo} and set, for 
example, $k_{d}\,N\,=\, 0\mbox{,}\,1\mbox{,}\,\ldots \mbox{,}\,
N/2 - 1\,$ and $k_i\,N\,=\,0\mbox{,}\,1 \mbox{,}\, 
\ldots \mbox{,}\, N - 1\,$ for $i = 1\mbox{,}\, \ldots 
\mbox{,}\,d-1$. Therefore, in the limit of infinite lattice
side $N$, the quantity $c(\bk)$ defined in eq.\ \reff{eq:cdefini}
takes values in the interval $(-1/2\mbox{,}\, 1/2]$ and
the lattice momentum $p^2(\bk)$ [see eq.\ \reff{eq:p2def}]
takes values in $[0\mbox{,}\, 4\,d)$. Also,
since we know that CSD is due to the long-wavelength modes,
which are usually the ones with slowest relaxation, we should
consider $c(\bk)$ in the limit of small momenta. Obviously,
if $k_{\mu} \,=\, 0$ for all $\mu$ then $c(\bk) = 1/2$ and,
using the fact that $\omega \in (1\mbox{,}\,2)\,$, one can
check that $\lambda_+(0\mbox{,}\,\omega) = 1$
and $\lambda_-(0\mbox{,}\,\omega) = (1\,-\,\omega)^2 < 1$, i.e.\
the constant (or zero) mode does not converge to zero. However,
as said above, this mode does not contribute to the action
\reff{E2bis} and it is therefore not relevant when studying CSD.
On the contrary, for the smallest non-zero momentum --- corresponding
to $k_i = 0$ for $i = 1\mbox{,}\, \ldots
\mbox{,}\,d-1$ and $k_{d} = 1/N$ --- we have
\be
c_{s m}(N) \,=\,
\frac{1}{2\,d}\left[\,d\,-\,1\,+\,\cos{\left(\,\frac{2\,\pi}{N}
\,\right)}\,\right]
\label{eq:slargeN}
\ee
and, in the limit of large lattice side $N$,
\be
c_{s m}(N) \,\approx\, \frac{1}{2} \left[\,1\,-\,\frac{2\, \pi^2}{d\, N^2}
\,\right] \,\equiv\, \frac{1}{2} \left[\,1\,-\,\zeta(N)
\,\right]
\;\mbox{,}
\label{eq:slargeN2}
\ee
namely we get the largest value of $c(\bk)$ smaller than $1/2$.
This case is important because, as we will see below,
the largest eigenvalue of the matrix
$C(\bk\mbox{,}\,\omega)$ is very close to 1 exactly
for the smallest non-zero momentum in the limit of
large $N$. Thus, the quantity $\zeta(N)$ will play a central
role in the study of CSD for the four local algorithms
considered here.

Note that eqs.\ \reff{eq:slargeN} and \reff{eq:slargeN2} are valid also
in the case of asymmetric lattices if we set $N = \max_{\mu} N_{\mu}$.
It is also interesting to check that $c(\bk)$
switches sign when $\bk_{\mu}$ goes to $\bk_{\mu} + T_{\mu}$
for all $\mu$ [see eq.\ \reff{eq:cdefini}]. Thus, we have
$|\,c(\bk)\,| \approx 1/2$ not only for small
momenta $p^2(\bk) \approx 0$ but also for the largest momenta
$p^2(\bk) \approx 4 \, d$, corresponding to small-wavelength modes.
This is obvious if we observe that
\be
c(\bk)\equiv
\frac{1}{2} \left[ \frac{1}{d}\,
            \sum_{\mu = 1}^{d} \,\cos{(2\,\pi\,k_{\mu})} \right]
\,=\,
\frac{1}{2} \left[\,1\,-\,\frac{2}{d}
\sum_{\mu = 1}^{d} \,  
\sin^2{(\pi\,k_{\mu})}\,\right]
\,=\,
\frac{1}{2} \left[\,1\,-\,\frac{p^2(\bk)}{2 d}
\,\right]
\label{eq:sdefini}
\;\mbox{,}
\ee
where we used eq.\ \reff{eq:p2def}.
Indeed, if we consider the largest momentum $p^2(\bk)$,
namely if we set $k_i = 1/2$ for $i = 1\mbox{,}\, \ldots
\mbox{,}\,d-1$ and $k_{d} = 1/2 - 1/N$, then we obtain
\be
c_{l m}(N) \,=\,
\frac{1}{2\,d} \, \left[\,-d\,+\,1\,+\,\cos{\left(\,\pi\,-\,\frac{2\,\pi}{N}
\,\right)}\,\right] \,=\,
\frac{1}{2\,d}\left[\,-d\,+\,1\,-\,\cos{\left(\,\frac{2\,\pi}{N}
\,\right)}\,\right] \,=\, - c_{s m}(N)
\label{eq:slargeNbis}
\;\mbox{.}
\ee
Therefore, in the limit of large lattice side $N$, the largest value
of $|\,c(\bk)\,|$ smaller than $1/2$ is given by
\be
|\,c(\bk(N))\,| \,=\,\frac{1}{2}\,\left|\,1\,-\,\zeta(N)\,\right|
\label{eq:clargeN}
\;\mbox{,}
\ee
corresponding both to the smallest non-zero momentum and to the
largest momentum.
Since the eigenvalues $\lambda_{\pm}(\bk\mbox{,}\,\omega)$
are functions only of $c^2(\bk)$ --- or equivalently of $|\,c(\bk)\,|$
--- and since they do not depend on the sign of the quantity $c(\bk)$
[see eqs.\ \reff{eq:lambdapmold} and \reff{eq:lambdapm}],
the previous result implies that these large momenta contribute
to CSD too. This unexpected effect is due to the even/odd
update which couples the low- and high-frequency modes \cite{HK}.


\subsection{Los Alamos Method}
\label{LosAl}

As said before, the Los Alamos update is
obtained from eq.\ \reff{upfove2} by
setting $\omega = 1$. Then, the eigenvalues $\lambda_{\pm}(\bk
\mbox{,}\,\omega)$ of the matrix $C(\bk\mbox{,}\,\omega)$ become
[see eqs.\ \reff{eq:cdefini} and \reff{eq:lambdapmold}]
\ba
\lambda_{-}(\bk\mbox{,}\,1) &\!\! = \!\!& 0 \\
\lambda(\bk) \equiv \lambda_{+}(\bk\mbox{,}\,1)
 &\! = \!& 4\,c^{2}(\bk) \,=\,
 \left[ \frac{1}{d}\,
            \sum_{\mu = 1}^{d} \,\cos{(2\,\pi\,k_{\mu})} \right]^2
\; \mbox{,}
\label{eq:eigen}
\ea
and, in the limit of infinite lattice side $N$,
we can consider $k_{d}$ taking values in the interval
$[ 0\mbox{,}\, 1/2 ) $ and $k_i$ in the interval
$[ 0\mbox{,}\, 1 ) $ for $i = 1\mbox{,}\,\ldots \mbox{,}\,d-1$
[see comment after eq.\ \reff{eq:DCD}]. It is obvious that,
if $k_{\mu} \,=\, 0$ for all $\mu$, then $\lambda(\bk)$ is equal
to $1$, while in any other case it is strictly smaller than $1$.
Thus, all the Fourier modes relax, except for the constant
(or zero) mode which, however, does not contribute to the action
\reff{E2bis}. Moreover, $\lambda(\bk)$ is always nonnegative.
Therefore we can write
\be
0 \,\leq\,|\, \lambda(\bk)\,| \,=\, \lambda(\bk) \, \leq 1
\; \mbox{.}
\ee
It follows that [see eq.\ \reff{eq:deftaugen}]
the relaxation time of the Los Alamos method is given by
\be
\tau = \frac{-1}{\log{( \max_{\bk \neq 0} \lambda(\bk))}}
\ee
and if $\lambda(\bk)$ is very close to 1 --- i.e.\ if
$\left| c(\bk) \right| \approx 1/2$ ---
we have [see eq.\ \reff{eq:tau}]
\be
\tau \,\approx\,
\frac{1}{1\,-\,\max_{\bk \neq 0}\,\lambda(\bk)}
\;\mbox{.}
\ee
From the previous section we know that this is the case when one
considers the smallest non-zero momentum --- or the largest momentum
--- in the limit of large
lattice side $N$. Then, using eq.\ \reff{eq:clargeN} we obtain
\be
\max_{\bk\,\neq\,0} \,
\lambda(\bk)\,=\,  \left[\,1\,-\,\zeta(N)
\,\right]^2 \,\approx\, 1\,-\,2\,\zeta(N)
\;\mbox{,}
\ee
from which follows
\be
\tau_{Los Alamos}
  \,\approx\, \frac{1}{2\,\zeta(N)} \,=\,
 \frac{d}{4\,\pi^{2}}\,N^{2}
\label{eq:tauLOS}
\;\mbox{.}
\ee
So, as expected, the dynamic critical exponent $z$ is equal 
to $2$.


\subsection{Overrelaxation Method\protect\footnote{~The results 
presented in this section apply also to the Cornell method [see 
eqs.\ \protect\reff{eq:bfvCor}, \protect\reff{eq:bfvOve} and 
\protect\reff{eq:omegaalpha}].}}
\label{Overr}

As seen above, the update
for the overrelaxation algorithm is given by eq.\
\reff{upfove2} with the parameter $\omega$ taking values in the
interval $( 1\mbox{,}\, 2 )$  and its relaxation time is related to
the eigenvalues $\lambda_{\pm}(\bk\mbox{,}\,\omega)$ of the matrix
$C(\bk\mbox{,}\, \omega)\,$ [see eqs.\ \reff{eq:lambdapmold}
and \reff{eq:lambdapm}] through the
relation [see eq.\ \reff{eq:deftaugen}]
\be
\tau = \frac{-1}{\log{( \max_{\bk \neq 0} | \lambda_{\pm}(\bk) | )}}
\;\mbox{,}
\ee
provided that $\max_{\bk \neq 0} | \lambda_{\pm}(\bk) | < 1$.
As said at the end of Section \ref{Infinito}, we must verify
that this condition is satisfied to ensure the convergence
of the algorithm. To this end, following Ref.\ \cite{N}, we define
\be
r(\bk\mbox{,}\,\omega)\,\equiv\,1\,-\,
\frac{2\,\omega^2\,c^{2}(\bk)}{\omega\,-\,1}
\label{eq:defr}
\ee
and, using the fact that $\omega \in ( 1\mbox{,}\, 2 )$,
we can write eq.\ \reff{eq:lambdapmold} as
\be
\lambda_{\pm}(\bk\mbox{,}\,\omega)\,=\,(\,\omega\,-\,1\,)\,
\left[\,-\,r(\bk\mbox{,}\,\omega)\,\pm
\,\sqrt{\,r^{2}(\bk\mbox{,}\,\omega)\,-\,1\,}\,\right]
\label{eq:lambdaoverre}
\;\mbox{.}
\ee

Note that $r(\bk\mbox{,}\,\omega) \in (-\infty\mbox{,}\,1]$. There are
therefore two possibilities, described below.
\begin{itemize}
\item[1)] $r(\bk\mbox{,}\,\omega) \in [-1\mbox{,}\,1]$
          for all values of $\bk$, namely we have to
          impose the condition $r(\bk\mbox{,}\,\omega) \geq - 1$
          for all $\bk$. In this
          case the eigenvalues $\lambda_{\pm}(\bk\mbox{,}\,\omega)$
          are complex conjugates of each other and we have
\be
|\,\lambda_{\pm}(\bk\mbox{,}\,\omega)\,| \,=\,
|\,\omega\,-\,1\,|
\;\mbox{,}
\ee
          namely $| \lambda_{\pm\mbox{,}\,\omega}(\bk) |$ is independent of
          $\bk$ (see Refs.\ \cite{A,N,Wo}). Also note that 
\be
|\,\omega \,-\,1\,| \,=\, \omega \,-\,1\,< \,1
\label{eq:lpm}
\;\mbox{.}
\ee
\item[2)] $r({\bar \bk}\mbox{,}\,\omega) < -1$ for some values of $\bk$,
          denoted as ${\bar \bk}$. In this case the corresponding
          eigenvalues $\lambda_{\pm}({\bar \bk}\mbox{,}\,\omega)$
          are real and we can verify that $0 < \lambda_{-}({\bar \bk}\mbox{,}\,
          \omega) < \lambda_{+}({\bar \bk}\mbox{,}\,\omega)$.
          From eq.\ \reff{eq:defr} it is clear that we can obtain
          $r(\bk\mbox{,}\,\omega) < -1$ only for the largest values of
          $c^2(\bk)$, i.e.\ $c^2(\bk) \approx 1/4$, corresponding to
          momenta $p^2(\bk) \approx 0$ or
          to momenta $p^2(\bk) \approx 4\,d$ (see comment at the end of
          Section \ref{CSD}). Considering now eq.\ \reff{eq:defr}
          with $r({\bar \bk}\mbox{,}\,\omega) < -1$ we get
\be
0\,<\,\omega \,-\,1\,<\, \omega^2\,c^{2}({\bar \bk})
\label{eq:ineqso}
\ee
          and
\be
\omega^2\,c^{2}({\bar \bk}) \,<\,
2\,\omega^2\,c^{2}({\bar \bk})\,-\,
(\,\omega \,-\,1\,)
\label{eq:ineqso2}
\;\mbox{.}
\ee
          Also, since
          $c(\bk) \in (-1/2\mbox{,}\, 1/2]$ for all values of
          $\bk$, we have
\be
c^2(\bk) \leq \frac{1}{4}
\ee
          and one can check that
\be
0\,<\,
2\,\omega^2\,c^{2}({\bar \bk}) \,-\,
(\,\omega \,-\,1\,)\,\leq\,
\frac{1}{2} (\,\omega \,-\,1\,)^2 \,+\, \frac{1}{2}
\ee
          and
\be
 \omega^2\,c^{2}({\bar \bk})\,-\,
(\,\omega \,-\,1\,)
 \,\leq\,
\frac{(\,2\,-\,\omega\,)^2}{4}
\;\mbox{.}
\ee
          These inequalities imply [see eq.\ \reff{eq:lambdapm}]
\be
\lambda_{+}({\bar \bk}\mbox{,}\,\omega) \leq 
\frac{1}{2} (\,\omega \,-\,1\,)^2 \,+\, \frac{1}{2} \,+\,
\omega\,\frac{2\,-\,\omega}{2}
\leq 1
\;\mbox{;}
\ee
          moreover, $\lambda_{+}({\bar \bk}\mbox{,}\,\omega)$ is equal to
          $1$ only if $c^2({\bar \bk}) = 1 / 4$,
          i.e.\ if ${\bar \bk} = 0$. Finally, from eqs.\
          \reff{eq:lambdapmold}, \reff{eq:ineqso} and \reff{eq:ineqso2}
          one can prove that
          $\lambda_{+}({\bar \bk}\mbox{,}\,\omega) > \omega\,-\,1$.
\end{itemize}

To sum up we can say that, with $\omega \in (1\mbox{,}\,2)\,$, there are
two different situations.
If the inequality $r(\bk\mbox{,}\,\omega) \geq - 1$ is satisfied for
all values of $\bk$, then the eigenvalues
$\lambda_{\pm}(\bk\mbox{,}\,\omega)$
are complex conjugates of each other with absolute value independent of
$\bk$ and given by $\,\omega-1\,$. In this case we have that
the largest eigenvalue is close to 1 --- which is the
interesting case when one studies CSD --- only when $\omega$ is close
to 2. If, on the contrary,
we have that $r({\bar \bk}\mbox{,}\,\omega) < -1$ for some values
${\bar \bk}$, then the largest eigenvalue of the matrix
$C(\bk\mbox{,}\, \omega)\,$ is real and given by
$\max_{{\bar \bk} \neq 0} \lambda_{+}({\bar \bk}\mbox{,}\,\omega)$.
(Note that $\max_{{\bar \bk} \neq 0}$ is taken only over the
values ${\bar \bk}$.)
In this case it is easy to check that,
in the limit of large $N$,
\be
\max_{{\bar \bk} \,\neq\,0}\,\lambda_{+}({\bar \bk}\mbox{,}\,\omega)\,
   \approx\,1\,-\,\frac{2\,\omega\,\zeta(N)}{2\,-\,\omega} \,=\,
    1\,-\,\frac{4\,\omega\,\pi^2}{(2\,-\,\omega)\,d\,N^2}
\label{eq:lambdamax}
\;\mbox{,}
\ee
where we used eqs.\ \reff{eq:lambdapmold} and \reff{eq:clargeN}.

\vskip 0.3cm

In order to study CSD for the overrelaxation algorithm
one usually writes
\be
\omega\,=\, \frac{2}{1\,+\,\Omega}
\label{eq:omegaC}
\ee
with $\Omega \in (0\mbox{,}\,1)$.
By using eq.\ \reff{eq:omegaC}
we can then rewrite $r(\bk\mbox{,}\,\omega)$ as
\be
r(\bk\mbox{,}\,\Omega)\,\equiv\,1\,-\,
\frac{8\,c^2(\bk)}{1\,-\,\Omega^{2}}
\label{eq:deftilder}
\;\mbox{.}
\ee
We now discuss separately the two cases considered above.

Case $1)$, i.e.\ if $r(\bk\mbox{,}\,\omega) \geq -1$ for all values of
$\bk$, corresponds to
\be
\Omega^2\,\leq\,1\,-\,4\,c^2(\bk)
\;\mbox{.}
\label{eq:C2ineq}
\ee
Since this condition should be satisfied for all values of
$\bk$ we must have
\be
\Omega^2\,\leq\,1\,-\,4\,\max_{\bk \neq 0}\,c^2(\bk)
\;\mbox{,}
\label{eq:omega2andc}
\ee
where again we don't consider the zero mode $\bk = 0$.
When $N$ goes to infinity we find [using eq.\
\reff{eq:clargeN}]
\be
\Omega^2\,\leq\,2\,\zeta(N) \,=\,
\frac{4\,\pi^2}{d\,N^2}
\;\mbox{,}
\label{eq:Omegadiz}
\ee
namely $\Omega$ goes to zero.
In this case we have
\be
|\,\lambda_{\pm}(\bk\mbox{,}\,\omega)\,|\;=\;\omega\,-\,1\;=\;
\frac{1\,-\,
\Omega}{1\,+\,\Omega}
\;\mbox{.}
\ee
In the limit of small $\Omega$ we obtain
\be
|\,\lambda_{\pm}(\bk\mbox{,}\,\omega)\,|\,\approx\, 1\,-\,2
\,\Omega
\ee
and the relaxation time becomes
[see eq.\ \reff{eq:tau}]
\be
\tau \,\approx\,
\frac{1}{1\,-\,\max_{\bk \neq 0}\,|\,\lambda_{\pm}(\bk\mbox{,}\,\omega)\,|}
  \,\approx\, \frac{1}{2\,\Omega}
\;\mbox{.}
\label{tauconOmega}
\ee
So, if we want to minimize the relaxation time $\tau$ we
have to maximize the value of $\Omega$ allowed
by the inequality \reff{eq:Omegadiz}, i.e\ we have to set
\be
\Omega \,=\, \frac{2\,\pi}{\sqrt{d}\,N}
\;\mbox{.}
\label{eq:Omegadis}
\ee
In this way we get
\be
\tau_{over}
  \,\approx\, \frac{\sqrt{d}}{4\,\pi}\,N
\label{eq:tauoverfin}
\ee
and, as expected, we have $z = 1$.
Note that the tuning for $\Omega$ given by eq.\ 
\reff{eq:Omegadis} implies
\be
\omega \,=\, \frac{2}{1\,+\,\Omega} \, = \,
      2\,\left(1\,+\,\frac{2\,\pi}{\sqrt{d}\,N}\right)^{-1}
\label{eq:omegatun}
\;\mbox{.}
\ee

Case $2)$ corresponds to the existence of values ${\bar \bk}$
satisfying the inequality
\be
\Omega^2\,>\,1\,-\,4\,c^2({\bar \bk})
\label{eq:case2}
\;\mbox{.}
\ee
In this case, as we saw above, we have
\be
\max_{{\bar \bk} \,\neq\,0}\,\lambda_{+}({\bar \bk}\mbox{,}\,\omega)\,
   \approx\, 1\,-\,\frac{4\,\omega\,\pi^2}{(2\,-\,\omega)\,d\,N^2}
\;\mbox{.}
\ee
This implies
\be
\tau \,\approx\,
\frac{1}{1\,-\,\max_{{\bar \bk}  \,\neq\,0}\,
     \lambda_{+}({\bar \bk}\mbox{,}\, \omega)}
 \,\approx\,
\frac{(2\,-\,\omega)\,d\,N^2}{4\,\omega\,\pi^2}
\label{eq:taunoo}
\ee
and in order to minimize the relaxation time $\tau$ we have
to minimize $(2\,-\,\omega) / \omega = \Omega$. This can be
done by choosing $\Omega \sim 1/N^{m}$, yielding
$\tau$ proportional to $N^{2 - m}$ and
$z = 2 - m$. In particular, it might seem possible
to set $m = 2$ so that $\tau$ becomes constant in $N$ and CSD is
completely eliminated. However, we note that the tuning
of $\Omega$ must be
done while still satisfying eq.\ \reff{eq:case2}, since it
defines what we are calling ``case $2)$''. This implies the
condition
\be
\frac{1}{N^{2 m}} \,\sim\,\Omega^2 \,>\,1\,-\,4\,c^2({\bar \bk})
\,\sim\, \frac{1}{N^2}
\;\mbox{,}
\ee
namely $m \leq 1$ and $z \geq 1$. Thus, since we want to minimize
$\Omega$, we have to set $m = 1$ and impose that the inequality
\reff{eq:case2} become the equality
\be
\Omega^2\,=\,1\,-\,4\,\max_{\bk \neq 0}\,c^2(\bk)
\label{eq:omega2andcbis}
\;\mbox{.}
\ee

We can conclude by saying that --- in both cases
considered above ---
the {\bf best} tuning for the overrelaxation
algorithm is obtained from the condition \reff{eq:omega2andcbis}.
Then, the largest eigenvalues 
$\lambda_{\pm}(\bk\mbox{,}\,\omega)$ of the matrix
$C(\bk\mbox{,}\, \omega)\,$ with $\bk \neq 0$ are real
and both equal to $\,\omega-1\,$ [corresponding to
$r(\bk\mbox{,}\,\Omega) = -1$], while for all the
other non-zero momenta these eigenvalues
are complex conjugates of each other [corresponding to
$r(\bk\mbox{,}\,\Omega) > -1$].

\vskip 0.3cm

If we do {\bf not} tune the value of the
parameter $\Omega$ (or equivalently of the parameter $\omega$),
we can consider two limiting cases. When $\Omega$ is so small
that the condition
\be
\Omega^2\,<\,2\,\zeta(N)
\ee
is satisfied for all the lattice sides $N$ considered, then
$r(\bk \mbox{,}\,\omega) > - 1$ for all momenta
and we obtain [see eq.\ \reff{tauconOmega}]
\be
\tau \,\approx\,
  \frac{1}{2\,\Omega}
\;\mbox{.}  
\ee
Thus, the relaxation time $\tau$ is constant in $N$ and
we get $z = 0$. However, $\tau$ is very large, and
even though the number of gauge-fixing
sweeps $n_{gf}$ is in this case independent of the lattice side $N$,
we need a very large $n_{gf}$ in order to complete
the gauge fixing even for relatively small lattice volumes.
On the contrary if $\Omega$ is large and, for all lattice sides
$N$ considered, we can find non-zero momenta $p^2({\bar \bk})$
such that the condition
\be
\Omega^2\,>\,1\,-\,4\,c^2({\bar \bk})
\ee
is satisfied, namely $r({\bar \bk} \mbox{,}\,\omega) < - 1$,
then the relaxation time $\tau$ is given by
[see eq.\ \reff{eq:taunoo}]
\be
\tau \,\approx\,
\frac{(2\,-\,\omega)\,d\,N^2}{4\,\omega\,\pi^2}
\label{eq:tauovenotun}
\ee
and $z = 2$. Nevertheless, in this case, the 
overrelaxation algorithm works better than the Los Alamos method:
in fact, using eq.\ \reff{eq:tauLOS}, we obtain
\be
\tau \,\approx\,
\frac{2 - \omega}{\omega} \, \tau_{Los Alamos}
\,=\, \Omega \, \tau_{Los Alamos}
\ee
and $\tau$ is always smaller
[for $\omega \in (1\mbox{,}\,2)$, i.e.\ $\Omega \in (0\mbox{,}\,1)$]
than the Los Alamos relaxation time.
Clearly, when $N$ goes to infinity with $\Omega$
fixed, we always obtain eq.\ \reff{eq:tauovenotun} and find $z=2$.

As a final remark, we note that this analysis implies that the
dynamic critical exponent $z$ of the overrelaxation algorithm
depends only on the relation
between $\Omega^2$ and $2\,\zeta(N)$, namely
if for all values of $N$ considered we set $\Omega^2$ much
smaller than, equal to, or much larger than $2\,\zeta(N)$
we have that the dynamic critical exponent $z$ is equal to
$0\mbox{,}\, 1$ or $2$ respectively.


\subsection{Stochastic Overrelaxation Method}
\label{Stoch}

In this section we want to analyze the critical slowing-down of the
stochastic overrelaxation method. As explained in Ref.\ 
\cite{CM}, this algorithm is similar in spirit to the idea behind 
the so-called {\em hybrid} version of overrelaxed algorithms (HOR), 
which are used to speed up Monte Carlo simulations of spin models, 
lattice gauge theory, etc.\ \cite{A,Wo,BW}. In these algorithms, $m$ 
micro-canonical (or energy conserving) update sweeps are done 
followed by one standard local ergodic update (such as Metropolis or 
heat-bath) sweep over the lattice. Actually, for the Gaussian 
model, it has been proven \cite{Wo} that the best result is obtained 
when the micro-canonical steps are chosen at random, namely when 
$m$ is the average number of micro-canonical sweeps between two 
subsequent ergodic updates. This is essentially what is done in the 
stochastic overrelaxation method [see eq.\ \reff{eq:gstocas}],
with $m / (m + 1)$ equal in 
average to $p$ or, equivalently, $m$ equal on average to $p / 
(1 - p)$.

\vskip 0.3cm

In order to follow the analysis in Ref.\ \cite{Wo}, we suppose
that the stochastic overrelaxation method is implemented as an HOR
algorithm: $m$ sweeps using the ``micro-canonical''
update [see eq.\ \reff{upfove2} 
with $\omega = 2$]
\be
\bfv(\bx) \;=\; - \, \bfv(\bx) \,+\,
\frac{1}{d}\,\sum_{\mu = 1}^{d}
\left[ \; \bfv(\bx + \bun_{\mu}) +
        \bfv(\bx - \bun_{\mu}) \; \right]
\;\mbox{,}
\label{eq:up02}
\ee
which does not change the value of the minimizing functional, and 
one sweep using the Los Alamos update [see eq.\ \reff{upfove2}
with $\omega = 1$]
\be
\bfv(\bx) \;=\; 
\frac{1}{2\,d}\, \sum_{\mu = 1}^{d}
\left[ \; \bfv(\bx + \bun_{\mu}) +
        \bfv(\bx - \bun_{\mu}) \; \right]
\;\mbox{,}
\label{eq:up01}
\ee
which brings the minimizing functional to its local absolute
minimum. Let us notice that, for $\omega = 2$, the eigenvalues
$\lambda_{\pm}(\bk\mbox{,}\,\omega)$ in eq.\ \reff{eq:lambdapm}
become
\be
\lambda_{\pm}(\bk\mbox{,}\,2)\,=\,
\left[\,- 1\,+\,8\,c^{2}(\bk)\,\right]\,\pm\,i\,
4\,\sqrt{\,c^{2}(\bk)\, -\,4\, c^{4}(\bk)}
\,=\,\exp{[\,\pm\,i\,\theta(\bk)\,]}
\;\mbox{,}
\label{eq:lambdapm2}
\ee
where we define $\theta(\bk)$ such that
\be
\cos{\left[ \, \theta(\bk) \,/\, 2 \, \right]}\,
\equiv\,2\,c(\bk)
\label{eq:deftheta}
\;\mbox{.}
\ee
Clearly, in this case, we have
$| \lambda_{\pm}(\bk\mbox{,}\,2) | = 1$ for all
values of $\bk$ and, as expected, none of the Fourier modes relaxes.

If we consider as {\em one sweep} of the lattice the combination
of $m$ sweeps using the update \reff{eq:up02} and one sweep using the update
\reff{eq:up01}, then the matrix that defines this combined update (in momentum
space) is given by
\be
{\widetilde C}(\bk\mbox{,}\, m)
\,=\, C(\bk\mbox{,}\, 1)\; \left[\,C(\bk\mbox{,}\, 2)\,\right]^{m}
\label{eq:defCStoc}
\;\mbox{,}
\ee
where $C(\bk\mbox{,}\, \omega)$ is defined in eq.\ \reff{eq:C},
or by
\be
{\widetilde M}(\bk\mbox{,}\, m)
\,=\, M(\bk\mbox{,}\, 1)\; \left[\,M(\bk\mbox{,}\, 2)\,\right]^{m}
\label{eq:defMStoc}
\;\mbox{,}
\ee
with $M(\bk\mbox{,}\, \omega)$ defined in eq.\ \reff{eq:M}. Since
it is easier to work with $M(\bk\mbox{,}\, \omega)$ than with 
$C(\bk\mbox{,}\, \omega)$ we will use eq.\ \reff{eq:defMStoc}.
However, one can check that ${\widetilde C}(\bk\mbox{,}\, m)$ and 
${\widetilde M}(\bk\mbox{,}\, m)$ have the same
eigenvalues\footnote{~This is immediate if we consider that
the matrix $R$ defined in eq.\ \protect\reff{eq:D} is
independent of $\bk$ and $\omega$. Therefore the relation
\protect\reff{eq:DCD} between the matrices $M(\bk\mbox{,}\, \omega)$
and $C(\bk\mbox{,}\, \omega)$ is also valid for the
matrices ${\widetilde M}(\bk\mbox{,}\, m)$ and
${\widetilde C}(\bk\mbox{,}\, m)$.}.
Then, following Ref.\ \cite{Wo}, we can write
\ba
M(\bk\mbox{,}\,2) &=& \left( \begin{array}{cc}
                  -1 & 4\, c(\bk)       \\[0.1cm]
           -4\, c(\bk) & -1\, + \,16\, c^2(\bk)
                             \end{array}
                      \right) \\
                  &=& V(\bk)\, 
                      \left( \begin{array}{cc} 
\exp{[ \, - \,i\, \theta(\bk) \, ]} & 0 \\[0.1cm]
                                  0 & \exp{[ \, i\,\theta(\bk) \, ]}
                             \end{array} 
                      \right)\, V^{- 1}(\bk)
\;\mbox{,}
\ea
where $\theta(\bk)$ is defined in eq.\ \reff{eq:deftheta} and the 
matrix $V(\bk)$ is given by
\be
V(\bk) \equiv
\left( \begin{array}{cc}
       \exp{[ \,i\,\theta(\bk) / 2\, ]} & \exp{[\, -\,i\,
              \theta(\bk) / 2\, ]} \\[0.1cm]
       1 & 1
\end{array} \right)
\;\mbox{.}
\ee
This implies that
\ba
\left[\,M(\bk\mbox{,}\, 2)\,\right]^{m} &=& V(\bk)\, 
                      \left( \begin{array}{cc}
\exp{[ \, - \,i\, \theta(\bk) \, ]} & 0 \\[0.1cm]
                                  0 & \exp{[ \, i\,\theta(\bk) \, ]}
                             \end{array}
                      \right)^m \, V^{- 1}(\bk) \\
&=& V(\bk)\,
                      \left( \begin{array}{cc}
\exp{[ \, - \,i\, m\, \theta(\bk) \, ]} & 0 \\[0.1cm]
                                  0 & \exp{[ \, i\,m\,\theta(\bk) \, ]}
                             \end{array}
                      \right) \, V^{- 1}(\bk)
\label{eq:MStoctom}
\;\mbox{.}
\ea
One can also write the matrix $M(\bk\mbox{,}\,1)$ in the 
dyadic form
\be
M(\bk\mbox{,}\,1) \,=\,
    2\,c(\bk)\,
    \left( \begin{array}{cc}
                 0 & 1 \\[0.1cm]
                 0 & 2\,c(\bk) \end{array}
                  \right) \, = \, 2\,c(\bk)\,
    \left( \begin{array}{c} 1 \\[0.1cm]
         2\, c(\bk) \end{array} \right)\,\cdot\,
\left(\,0\mbox{,}\; 1\,\right)
\label{eq:Mo2}
\ee
and, using eqs.\ \reff{eq:defMStoc}, \reff{eq:MStoctom} and
\reff{eq:Mo2} we obtain
\be
{\widetilde M}(\bk\mbox{,}\, m)
\,=\,\frac{2\,c(\bk)}{\sin{\left[\, \theta(\bk)\, /\, 
2\, \right]}}
\,\left( \begin{array}{c} 1 \\[0.1cm]
          2\,c(\bk) \end{array} \right)\,\cdot\,
\left(\,-\,\sin{[\, m\,\theta(\bk) \, ]}\mbox{,}\;
      \sin{\left[\,\left( m + \frac{1}{2} \right)\,
    \theta(\bk)\,\right]}\,\right)
\;\mbox{.}
\label{eq:Mtildeofm}
\ee
The eigenvalues of this matrix are equal to zero and to
\be
\cos{\left[ \,\frac{\theta(\bk)}{2} \, \right]}\,
   \cos{\left[\,\left( m + \frac{1}{2} \right)\,\theta(\bk)\,\right]}
\;\mbox{.}
\ee

However, $m$ is not fixed but varies between zero and infinity
with probability $p^{m}\,( 1 \,-\, p )$.
This gives an average value
\be
\langle \,m\,\rangle \, \equiv \, \sum_{m=0}^{\infty}\,
m\,p^{m}\,( 1 \,-\, p )\,=\, \frac{\sum_{m=0}^{\infty}\,
m\,p^{m}}{\sum_{m=0}^{\infty}\,p^{m}}
\,=\,p\, \frac{d}{dp}\,\log{ \sum_{m=0}^{\infty}\,p^{m} }
\,=\, \frac{p}{1\,-\,p}
\label{eq:mmedio}
\;\mbox{,}
\ee
as said above. Thus, instead of looking for the eigenvalues of the
matrix ${\widetilde M}(\bk\mbox{,}
\, m)$ we should consider the matrix
\be
{\cal M}(\bk\mbox{,}\, p) \, \equiv \,
( 1\,-\,p ) \, \sum_{m = 0}^{\infty}\, p^{m}\,
{\widetilde M}(\bk\mbox{,}\, m)
\;\mbox{.}
\label{eq:defineMdip}
\ee
After writing eq.\ \reff{eq:Mtildeofm} as
\ba
{\widetilde M}(\bk\mbox{,}\, m)
& =& \frac{2\,c(\bk)}{\sin{\left[\, \theta(\bk)\, /\,
2\, \right]}} \nonumber \\[2mm]
& & \qquad \times \,
\left( \begin{array}{c} 1 \\[0.1cm]
          2\,c(\bk) \end{array} \right)\,\cdot\,
\left[ \,\mbox{${\mathrm Im}$} \, 
          \left(\,-\,\exp{[\,i\, m\,\theta(\bk) \, ]}\mbox{,}\;
      \exp{\left[\,i\,\left( m + \frac{1}{2} \right)\,
    \theta(\bk)\,\right]}\,\right) \right]
\;\mbox{,} \qquad\;\;\;\;\;
\ea
it is straightforward to
check that the matrix ${\cal M}(\bk\mbox{,}\, p) $
can be written in the dyadic form
\ba
{\cal M}(\bk\mbox{,}\, p) & = & 
\frac{( 1\,-\,p ) \,\cos{\left[\, \theta(\bk)\, /\, 2\, \right]} }{
   \left[\,1\,+\,p^{2}\,-\,2\,p\,\cos{ \theta(\bk)} \,\right]\,
      \sin{\left[\, \theta(\bk)\, /\, 2\, \right]} } \nonumber \\[2mm]
& & \qquad \times \,
   \left( \begin{array}{c} 1 \\[0.1cm]
   \cos{\left[\, \theta(\bk)\, /\, 2\, \right]} \end{array} \right)\,
   \cdot\,
        \left(\,-\,p\,\sin{ \theta(\bk) } \mbox{,}\; ( 1\,+\,p ) \,
               \sin{\left[\, \theta(\bk)\, /\, 2\, \right]}
  \,\right)
\label{eq:newmatrix}
\ea
and has eigenvalues zero and
\be
\lambda(\bk\mbox{,}\,p) \,=\,
  \frac{( 1\,-\,p )^{2}\,\cos^{2}{\left[\, \theta(\bk)\, /\, 2\, 
   \right]} }{ ( 1\,-\,p )^{2}\,+\,4\,p\,\sin^{2}{\left[\, \theta(\bk)
       \, /\, 2\, \right]} }
\;\mbox{.}
\label{eq:neweigen}
\ee
Clearly this eigenvalue is nonnegative for any $p \in ( 0\mbox{,}
\, 1 )$ and for $p = 0$ (i.e.\ $\langle \,m\,\rangle = 0$)
we obtain the non-zero
eigenvalue $\lambda(\bk\mbox{,}\,0) = \cos^{2}[\, \theta(\bk) / 2 \,] =
4\, c^2(\bk)$ of the Los Alamos method [see eq.\ \reff{eq:eigen}].

Note that, since the matrix ${\cal M}(\bk\mbox{,}\, p)$
describes (on average) $\langle m \rangle \,+
\,1$ sweeps of the lattice, eq.\ \reff{eq:tauandrho} is not correct
for the stochastic overrelaxation algorithm. In fact, in this case,
the relaxation time $\tau$ is related to the eigenvalue 
$\lambda(\bk\mbox{,}\,p)$ by the expression \cite{Wo}
\be
\max_{\bk \neq 0}\,|\,\lambda(\bk\mbox{,}\,p)\,| \,=\,
 \max_{\bk \neq 0}\,\lambda(\bk\mbox{,}\,p) \,\equiv\,
e^{- \,( \langle m \rangle \,+\,1 ) / \tau}
\;\mbox{,}
\ee
namely
\be
\frac{\tau}{\langle m \rangle\,+\,1} \,=\,
\frac{-1}{\log{( \max_{\bk \neq 0}
\lambda(\bk\mbox{,}\,p) )}}
\;\mbox{.}
\ee

\vskip 0.3cm

\begin{figure}[t]
\vspace*{-3cm} \hspace*{-0cm}
\begin{center}
\epsfxsize = 6.0in
\leavevmode\epsffile{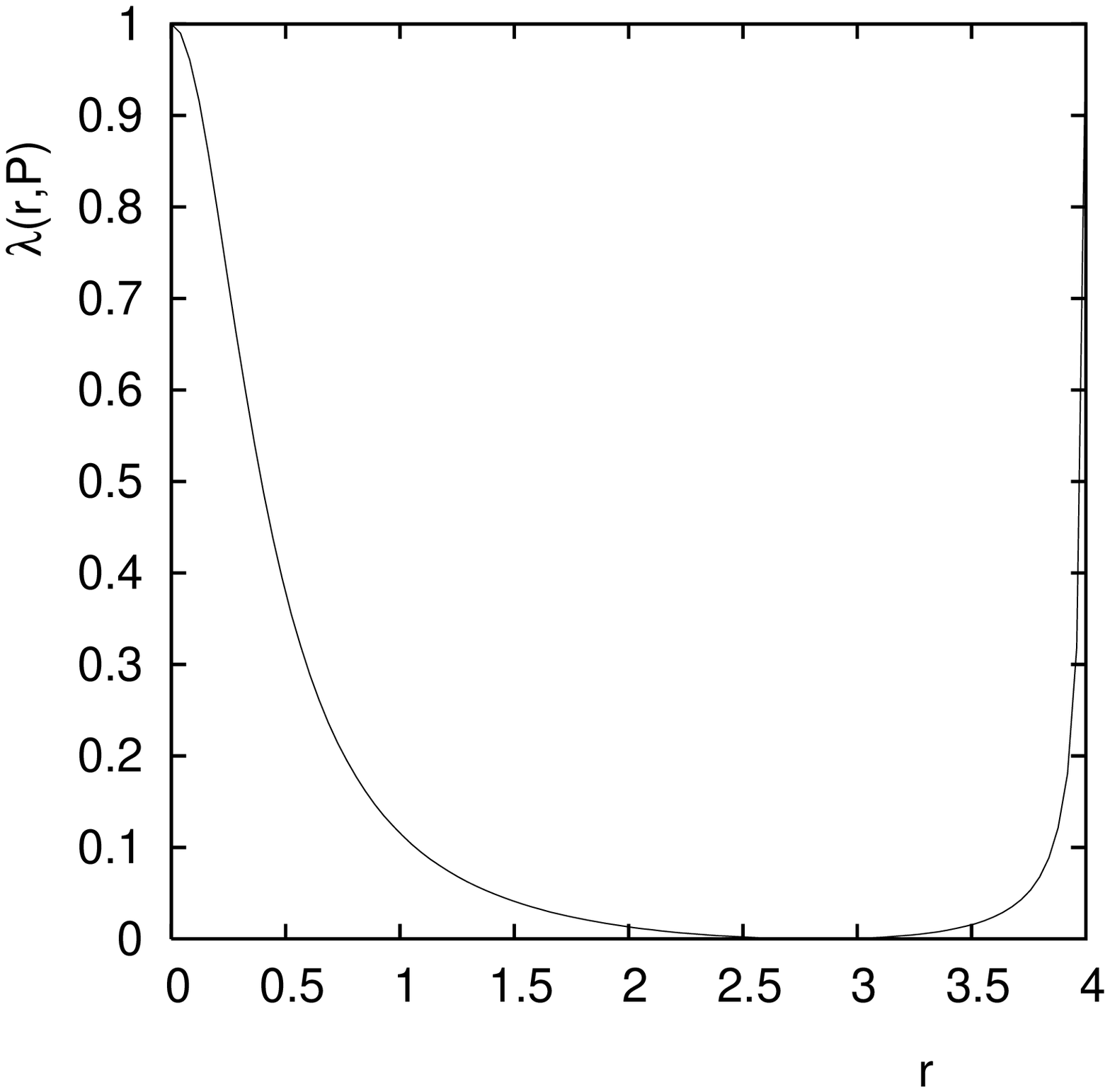} \\
\end{center}
\caption{~Plot of the eigenvalue $\lambda(r\mbox{,}\,P)$
     [see eq.\ \protect\reff{eq:lambdaP}] as a function 
     of $r = |\,p(\bk)\,|$
     for the case $d = 4$ and $P = 0.2$.
}
\label{fig:lambda}
\end{figure}

In order to study CSD for the stochastic overrelaxation algorithm,
we introduce $P \in ( 0\mbox{,}\, 1 )$ and write
\be
p \,=\,\frac{1\,-\,P}{1\,+\,P}
\label{eq:defPdep}
\;\mbox{.}
\ee
Then, the eigenvalue in eq.\ \reff{eq:neweigen} becomes
\be
\lambda(\bk\mbox{,}\,P) \,=\, \frac{P^2\,
         \cos^{2}{\left[\, \theta(\bk)\, /\, 2\, \right]}}{
   1\, +\, \left(\, P^2\,-\,1\,\right)\, 
   \cos^{2}{\left[\, \theta(\bk)\, /\, 2\, \right]}}
\;\mbox{.}
\label{eq:newlambda}
\ee
Also, using eqs.\ 
\reff{eq:sdefini} and \reff{eq:deftheta}, one obtains
\be
\cos{\left[\, \theta(\bk)\, /\, 2\, \right]}
\,=\, 2\,c(\bk)
\,=\,1\,-\,\frac{p^2(\bk)}{2\,d}
\,\equiv\,1\,-\,\frac{r^2}{2\,d}
\;\mbox{,}
\label{eq:variousrel}
\ee
where $r$ is the magnitude $|\,p(\bk)\,|$ of the lattice momentum.
Note that, since $c(\bk)$ takes values in the interval
$(-1/2\mbox{,}\,1/2]$, we have $\cos{[\,\theta(\bk)\, /\, 2\, ]}
\in (-1\mbox{,}\,1]$, $\theta(\bk) \in [0\mbox{,}\,2\,\pi)$
and $r \in [0\mbox{,}\,2 \sqrt{d})$. 
Thus, we can rewrite this eigenvalue as
\be
\lambda(r\mbox{,}\,P) \,=\,
\frac{P^2 \, \left( 1 \,-\, \frac{r^2}{2\,d} \right)^2}{
      1 \, + \, \left( P^2 \,-\, 1 \right) \,
                \left( 1 \,-\, \frac{r^2}{2\,d} \right)^2}
\label{eq:lambdaP}
\;\mbox{.}
\ee
It is easy to check that
\begin{itemize}
\item $\lambda(r \mbox{,}\,P) = 1$ for $r = 0$,
\item the derivative of $\lambda(r\mbox{,}\,P)$ with respect to $r$ is zero
      for $r = 0$ and $r = \sqrt{2\,d}$,
\item this derivative is negative (respectively positive) 
      for $r < \sqrt{2\,d}$ (respectively for $r > \sqrt{2\,d}$).
\end{itemize}
In Fig.\ \ref{fig:lambda} we plot $\lambda(r\mbox{,}\,P)$ as a function
of $r$ for the case $d = 4$ and $P = 0.2$. We note that the 
eigenvalue $\lambda(r\mbox{,}\,P)$ does not show the oscillatory
behavior that can be observed when one considers a probability
distribution that is uniform\footnote{~Note that with this
distribution we automatically have $\langle \,m\,\rangle$
equal to $\,{\overline m}\,$.} for $m$
in the interval $[1\mbox{,}\, 2 \,{\overline m}\, - 1]$
(see Figs.\ 1 and 2 in Ref.\ \cite{Wo}). Also note
that this eigenvalue is close to $1$ for small momenta
$r \approx 0$ and
for very large momenta, i.e.\ with $r \approx 2\,\sqrt{d}$, and that
in both cases we have $c^2(\bk) \approx 1/4$ and
$\cos^2{[\,\theta(\bk)\, /\, 2\, ]} \approx 1\,$.
As explained at the end of Section \ref{CSD},
this is a natural consequence of the even/odd updating scheme.

It follows that $\max_{\bk 
\neq 0} \, \lambda(\bk\mbox{,}\,P) < 1$ is obtained for
$c^2(\bk) \approx 1/4$.
Then, using eqs.\ \reff{eq:clargeN} and \reff{eq:deftheta}
we have, in the limit of large lattice side $N$,
\be
\cos^{2}{\left[\, \theta(\bk)\, /\, 2\, \right]}
\,=\, 4\,c^2(\bk)
\,\approx\,1\,-\, 2\,\zeta(N)
\label{eq:sto0}
\ee
and from eq.\ \reff{eq:newlambda} we find
\be
\lambda(\bk\mbox{,}\,P) \,\approx\,1\,-\, \frac{2\,\zeta(N)}{
                            P^2}
\label{eq:lambdato1}
\;\mbox{.}
\ee
Thus, $\lambda(\bk \mbox{,}\,P)$ is very close to 1 and we obtain
\be
\frac{\tau}{\langle m \rangle\,+\,1} \,\approx\,
\frac{1}{1\,-\,\max_{\bk \neq 0}\,\lambda(\bk\mbox{,}\,P)}
\;\mbox{.}
\label{eq:tmp1}
\ee
Using again eq.\ \reff{eq:newlambda} and eq.\ \reff{eq:sto0}
we get
\be
1\,-\,\max_{\bk \neq 0}\,\lambda(\bk\mbox{,}\,P) \,\approx\,
\frac{2\,\zeta(N)}{2\,\zeta(N) \,\left(\,1\,-\,P^2\,\right)\,+\, P^2}
\;\mbox{,}
\ee
so that\footnote{~Since at this point we don't know the relation
between $P$ and $\zeta(N)$ we have to keep all terms in this equation.}
\be
\frac{\tau}{\langle m \rangle\,+\,1} \,\approx\,
\left(\,1\,-\,P^2\,\right)\,+\,\frac{P^2}{2\,\zeta(N)}
\;\mbox{.}
\ee
Also, from eqs.\ \reff{eq:mmedio} and \reff{eq:defPdep}, we have
\be
\langle m \rangle \, = \, \frac{p}{1 - p} \,
   =\, \frac{1 - P}{2 \, P}
\ee
and
\be
\langle m \rangle \,+\,1\, = \, 
    \frac{1 + P}{2 \, P}   
\;\mbox{.}
\label{eq:mpiu1}
\ee
These equations give
\be
\tau\,\approx\,\left[\,\left(\,1\,-\,P^2\,\right)\,+\,
             \frac{P^2}{2\,\zeta(N)}
  \,\right]\, \frac{1 + P}{2 \, P}
\;\mbox{.}
\label{eq:taudiPandz}
\ee
We can now fix $P$ by minimizing the value of $\tau$. In this
way we obtain
\be
P^2 \,=\, \frac{2\,\zeta(N)}{\left[\,1\,-\,2\,\zeta(N)\,\right]
\,\left(\,1\,+\,2\,P\,\right)}\,\approx\,
\frac{2\,\zeta(N)}{\left(\,1\,+\,2\,P\,\right)}
\;\mbox{.}
\ee
Therefore, in the limit of large lattice side $N$, we
get that $P$ goes to zero (and $p$ goes to $1$) as
\be
P\,\approx\,\sqrt{2\,\zeta(N)}
 \,=\,\frac{2\,\pi}{\sqrt{d}\, N}
\label{eq:Ptun}
\ee
and
\be
\tau_{stoc} \,\approx\, \left[\,1\,+\,
             \frac{P^2}{2\,\zeta(N)}
  \,\right]\, \frac{1}{2 \, P} \,\approx\,\frac{1}{P}
\,\approx\,\frac{\sqrt{d}\,N}{2\,\pi}
\label{eq:taustocfin}
\;\mbox{.}
\ee
Thus, as expected, we have $z = 1$.
Let us notice that the tuning for $P$ given in eq.\ \reff{eq:Ptun}
coincides with the tuning for $\Omega$ obtained in eq.\
\reff{eq:Omegadis} for the overrelaxation algorithm.
Also we have
\be
p\,=\, \frac{1 \,-\,P}{1 \,+\,P} \,\approx \,
       1\,-\,2\,P \,=\,
       1\,-\,\frac{4\,\pi}{\sqrt{d}\, N}
\ee
and
\be
1\,+\,p\,=\,\frac{2}{1 \,+\,P}\,\approx\,
\frac{2}{1 \,+\,\Omega}\,=\,\omega
\;\mbox{,}
\ee
namely we find the relation
$\,p\,\approx\,\omega - 1\,$. This is in agreement with the result obtained
analytically and numerically in two dimensions at finite
$\beta$ (see Sections 5 and 7.3 in Ref.\ \cite{CM}).

\vskip 0.3cm

Finally, if we do {\bf not} tune the value of the
parameter $P$ (or equivalently of the parameter $p$) then
we have again two limiting cases. In fact, if $P$ is very
small and such that
\be
P^2 \,<\,2\,\zeta(N)
\ee
then from eq.\ \reff{eq:taudiPandz} we obtain
\be
\tau\,\approx\,\frac{1}{2\,P}
\;\mbox{,}
\ee
namely the relaxation time is constant in $N$ and $z =0$.
On the contrary, if $P$ is large and
\be
P^2 \,>\,2\,\zeta(N)
\;\mbox{,}
\ee
then we find
\be
\tau \, \approx\,
\frac{1\,+\,P}{2\,P}\,\frac{P^2}{2\,\zeta(N)}
\,=\,
\frac{(1\,+\,P)\, P}{2} \, \frac{d\,N^2}{4\,\pi^2}
\label{eq:taustocnotun}
\ee
and [using eq.\ \reff{eq:tauLOS}]
\be
\tau \,\approx\,
\frac{(1\,+\,P)\, P}{2} \, \tau_{Los Alamos}
\;\mbox{.}
\label{eq:taunop}
\ee
This gives a dynamic critical exponent $z$ equal to $2$. However,
in this case, the improved local
algorithm without tuning works better than the Los Alamos
method: the relaxation time
$\tau$ given in eq.\ \reff{eq:taunop} is always smaller [for $P \in 
(0\mbox{,}\,1)$] than the Los Alamos relaxation time.
As in the overrelaxation case, when $N$ goes to infinity
(with $P$ fixed) we always get \reff{eq:taustocnotun} and find $z=2$.
Also, this analysis implies that the
dynamic critical exponent $z$ depends only on the relation
between $P^2$ and $2\,\zeta(N)$. In fact,
if we set $P^2$ much smaller than, equal to,
or much larger than $2\,\zeta(N)$
we have that the dynamic critical exponent $z$ is equal to
$0\mbox{,}\, 1$ or $2$ respectively.


\section{Generalized Local Algorithms}
\label{Better}

In Sections \ref{sec:gf}, \ref{Infinito} and \ref{CSD} we
have studied CSD for two main types of local algorithms:
the overrelaxation algorithm --- which coincides with the Los
Alamos algorithm for $\omega = 1$ and with the Cornell
algorithm with the choice [see eq.\ \reff{eq:omegaalpha}]
$\omega \approx \,\alpha\,{\cal N}(\by)$ --- and the stochastic
overrelaxation algorithm. In both cases we have seen that
with a careful tuning (of the parameters $\omega$ and $p$
respectively) we obtain a dynamic critical exponent $z=1$.
In this section we want to see if it possible to generalize
these algorithms in order to get $z < 1$. To the best of
our knowledge, the results presented in Sections
\ref{sec:4x4} and \ref{generalstochoverr} below are new.


\subsection{Generalized Overrelaxation Algorithm}
\label{generaloverr}

Let us consider the following generalization of the standard
overrelaxation update [see eq.\ \reff{eq:goverre}]
\be
g^{(linear)}(\bx) \, \equiv \, \frac{a(\omega) \, 
       {\widetilde h}^{\dagger}(\bx)\, + \, b(\omega)\,g(\bx)}{ \sqrt{\,
      \left[\, a(\omega)\,+\, b(\omega)\,\right]^2 \,-\,
           a(\omega)\,b(\omega) \left[ \, 2 - {\cal T}(\bx) \, \right]
                 }}
\label{eq:glini}
\;\mbox{.}
\ee
One can check that $g^{(linear)}(\bx) \in SU(2)$ and that the
overrelaxation update corresponds to the choices
$a(\omega) = \omega$ and $b(\omega) =\,1\,-\,\omega$.
In the limit of large
number of gauge-fixing sweeps $t$ [see Section \ref{Infinito}] we
can use eq.\ \reff{eq:Td} and obtain
\be
g^{(linear)}(\bx) \, \approx \,
    \frac{a(\omega)}{\left| \,a(\omega)\,+\, b(\omega)\,\right|}\,
              {\widetilde h}^{\dagger}(\bx) \,+
   \, \frac{b(\omega)}{\left| \,a(\omega)\,+\, b(\omega)\,\right|}\,
       g(\bx)
\;\mbox{,}
\ee
where terms of order $\epsilon^{2}$ have been neglected. This
udate can be written in the simpler form\footnote{~We have considered
this type of update also in Section 5 of Ref.\ \protect\cite{CM}.}
\be
g^{(linear)}(\bx) \, \approx \, a(\omega)\,{\widetilde h}^{\dagger}(\bx)
    \,+\, \,b(\omega)\,g(\bx)
\ee
if we assume the condition $a(\omega)\,+\,b(\omega)\,=\,1$
or, equivalently, by a redefinition of the coefficients
$a(\omega)$ and $b(\omega)$. Finally,
using eq.\ \reff{eq:gI}, we find the update for the $\bfv(\bx)$ field
\be
\bfv^{(linear)}(\bx) \, = \, a(\omega) \,\bfv^{(LosAl.)}(\bx)\,+\,
     b(\omega)\,\bfv(\bx)
\label{eq:bfvgen}
\;\mbox{.}
\ee
If we consider the dependence of the massless-free-field action
on the value of the field $\bfv$ at a given site $\by$ [see eqs.\ 
\reff{eq:Eatsitey} and \reff{eq:Eatsiteybis}],
it is clear that this is the most general 
local linear update (with site-independent coefficients) of the 
field $\bfv(\by)$. One can also check that the condition
$ b^2(\omega) < 1 $ is sufficient to prove that the update
\reff{eq:bfvgen} never increases the value of the massless-free-field
action [see eq.\ \reff{eq:Eatsiteybis}].
Since the definition of $\omega$ is in principle arbitrary, we
can at this point set $a(\omega) = \omega$ and obtain that
the standard overrelaxation algorithm is the most general
local linear update with side-independent coefficients.
Therefore, following the analysis
presented in the previous section we have at best
that $z = 1$.

In order to understand why, in this case, one cannot get a
dynamic critical exponent $z$ smaller than $1$,
we can follow Ref.\ \cite{BN} and consider the
inequality [see eq.\ \reff{eq:tauanddelta}]
\be
\tau \,\geq\,
\frac{1}{
\min_{\bk \neq 0} \, \left|\,1\,-\,\lambda_{\pm}(\bk\mbox{,}\,\omega)\,
\right|}
\;\mbox{.}
\ee
When $r^2(\bk\mbox{,}\,\omega) \leq 1$, namely when these eigenvalues
are complex, one finds [see eqs.\ \reff{eq:defr}
and \reff{eq:lambdaoverre}]
\be
|\,1\,-\,\lambda_{\pm}(\bk\mbox{,}\,\omega)\,| \,=\,
\omega 
\,\sqrt{\,1\,-\,4\,c^2(\bk)\,}
\label{eq:modoulodi1ml}
\;\mbox{.}
\ee
Since the tuning condition, i.e.\ $r^2(\bk\mbox{,}\,\omega) \leq 1$,
is equivalent to the relation
[see eq.\ \reff{eq:C2ineq}]
\be
\Omega\,\leq\,\sqrt{\,1\,-\,4\,c^2(\bk)\,}
\label{eq:omegaminimum}
\ee
[where the inequality becomes an equality for the largest value of
$c^2(\bk) < 1/4$], we have [using eq.\ \reff{eq:omegaC}]
\be
\min_{\bk \neq 0} \, |\,1\,-\,\lambda_{\pm}(
    \bk\mbox{,}\,\omega)\,| \,=\, \frac{2\,\Omega}{1\,+\,\Omega}
\;\mbox{.}
\ee
In the limit of large $N$ this implies [see eq.\ \reff{eq:clargeN}]
\be
\Omega\,\ltapprox\,\sqrt{ 2\,\zeta(N)}
\ee
and
\be
\min_{\bk \neq 0} \, |\,1\,-\,\lambda_{\pm}(\bk\mbox{,}\,\omega)\,| \,
\approx\, 2 \,\sqrt{ 2\,\zeta(N)}
\,=\,\frac{4\,\pi}{\sqrt{d}\,N}
\label{eq:distan}
\;\mbox{,}
\ee
so that [in agreement with eq.\ \reff{eq:tauoverfin}]
\be
\tau \,\approx\, 
\frac{\sqrt{d}}{4\,\pi}\,N
\;\mbox{,}
\ee
which gives $z = 1$. 

Equations \reff{eq:modoulodi1ml} and
\reff{eq:distan} are the starting point of
Ref.\ \cite{BN} (see their Fig.\ 1). The idea in
that article is that one should look for an update characterized
by a matrix $C(\bk)$ whose eigenvalues $\lambda(\bk)$ satisfy,
in the limit of large $N$, the relation
\be
\min_{\bk \neq 0} \, |\,1\,-\,\lambda(\bk)\,| \,
\propto\, \left( \frac{4\,\pi^2}{d\,N^2} \right)^{1/m} 
\ee
with $m > 2$. In fact, this would imply
\be
\tau \,\sim\,N^{2/m}
\;\mbox{,}
\ee
namely $z < 1$. From eq.\ \reff{eq:lambdaoverre} it is clear
that we have $m = 2$, and therefore $z = 1$, because
the eigenvalues $\lambda_{\pm}(\bk\mbox{,}\,\omega)$
are solutions of a quadratic
equation, i.e.\ because $C(\bk\mbox{,}\, \omega)$ is a
$2 \times 2$ matrix. What is proven in Ref.\ \cite{BN} is that, 
unfortunately, we cannot get $m > 2$ because any general
update matrix $C$ can be shown to be block-diagonal with
blocks of size not larger than $2 \times 2$.
In the next section we will consider explicitly local algorithms
characterized by updating matrices of size $4 \times 4$ and we
will check that we cannot obtain $z$ smaller than 1 in that case.


\subsection{A $4 \times 4$ Updating Matrix}
\label{sec:4x4}

Following what was done in
Section \ref{CSD}, we consider here the update
\reff{upfove2} and the Fourier-like transformation
defined in eq.\ \reff{eq:fourierlike} and
generalize that procedure
in a way that produces an updating matrix that is $4 \times 4$
instead of $2 \times 2$. 
To this end, let us recall that if the vector $\bT$
has components $T_{\mu} = 1/2$ for all $\mu$
[see eq.\ \reff{eq:defT}] then we have
\ba
1\,+\,\exp{\left( \,-\,2\,\pi\,i\, \bT\,\cdot\,\bx \,\right)}
   & = & \left\{ \begin{array}{l}
                  2\quad \mbox{if}\;\; |\, \bx \,|\;\; \mbox{is even} \\
                  0\quad \mbox{if}\;\; |\, \bx \,|\;\; \mbox{is odd}
                \end{array} \right. \\
1\,-\,\exp{\left( \,-\,2\,\pi\,i\, \bT\,\cdot\,\bx \,\right)}
   & = & \left\{ \begin{array}{l}
                  0\quad \mbox{if}\;\; |\, \bx \,|\;\; \mbox{is even} \\
                  2\quad \mbox{if}\;\; |\, \bx \,|\;\; \mbox{is odd}
                \end{array} \right.
\ea
These two linear combinations automatically select
the even and odd sub-lattices and can be used to
construct the two-component field $f^{b, \pm}(\bk)$
[see eq.\ \reff{eq:fourierlike}].
Let us notice that one can also rewrite the above relations as
\be
\exp{\left( \,-\,2\,\pi\,i\, \bT_1\,\cdot\,\bx \,\right)}
  \,\pm\,
\exp{\left( \,-\,2\,\pi\,i\, \bT_2\,\cdot\,\bx \,\right)}
\ee
with $T_{1, \mu} = 1$ and $T_{2, \mu} = 1/2$
for all $\mu$.
We can easily generalize this result and
divide the lattice in, for example, four sub-lattices.\footnote{~In
this section, in order to simplify the notation,
we suppose that the lattice side $N$ is a multiple of $4$.}
In fact with $T_{1, \mu} = 1\mbox{,}\;
T_{2, \mu} = 1/2\mbox{,}\; T_{3, \mu} = 3/4$ and
$T_{4, \mu} = 1/4$
we have that the linear combinations
\ba
L_{ee}(\bx) & = & 
  e^{\left( \,-\,2\,\pi\,i\, \bT_1\,\cdot\,\bx \,\right)}
  \,+\,
  e^{\left( \,-\,2\,\pi\,i\, \bT_2\,\cdot\,\bx \,\right)}
  \,+\,\;\;
  e^{\left( \,-\,2\,\pi\,i\, \bT_3\,\cdot\,\bx \,\right)}
  \,+\,\;\;
  e^{\left( \,-\,2\,\pi\,i\, \bT_4\,\cdot\,\bx \,\right)} 
\label{eq:Lee} \\
L_{eo}(\bx) & = & 
  e^{\left( \,-\,2\,\pi\,i\, \bT_1\,\cdot\,\bx \,\right)}
  \,+\,
  e^{\left( \,-\,2\,\pi\,i\, \bT_2\,\cdot\,\bx \,\right)}
  \,-\,\;\;
  e^{\left( \,-\,2\,\pi\,i\, \bT_3\,\cdot\,\bx \,\right)}
  \,-\,\;\;
  e^{\left( \,-\,2\,\pi\,i\, \bT_4\,\cdot\,\bx \,\right)} 
\label{eq:Leo} \\
L_{oe}(\bx) & = & 
  e^{\left( \,-\,2\,\pi\,i\, \bT_1\,\cdot\,\bx \,\right)}
  \,-\,
  e^{\left( \,-\,2\,\pi\,i\, \bT_2\,\cdot\,\bx \,\right)}
  \,+\,i\,
  e^{\left( \,-\,2\,\pi\,i\, \bT_3\,\cdot\,\bx \,\right)}
  \,-\,i\,
  e^{\left( \,-\,2\,\pi\,i\, \bT_4\,\cdot\,\bx \,\right)} 
\label{eq:Loe} \\
L_{oo}(\bx) & = & 
  e^{\left( \,-\,2\,\pi\,i\, \bT_1\,\cdot\,\bx \,\right)}
  \,-\,
  e^{\left( \,-\,2\,\pi\,i\, \bT_2\,\cdot\,\bx \,\right)}
  \,-\,i\,
  e^{\left( \,-\,2\,\pi\,i\, \bT_3\,\cdot\,\bx \,\right)}
  \,+\,i\,
  e^{\left( \,-\,2\,\pi\,i\, \bT_4\,\cdot\,\bx \,\right)}
\;\mbox{,}
\label{eq:Loo}
\ea
called respectively
even-even, even-odd, odd-even and odd-odd, 
are always zero but for the following cases
\ba
L_{ee}(\bx) & = & 4 \qquad \mbox{if} \;\;\;\;
                      |\, \bx \,|\,\bmod\,4\,=\,0 \\
L_{eo}(\bx) & = & 4 \qquad \mbox{if} \;\;\;\; 
                      |\, \bx \,|\,\bmod\,4\,=\,2 \\
L_{oe}(\bx) & = & 4 \qquad \mbox{if} \;\;\;\; 
                      |\, \bx \,|\,\bmod\,4\,=\,1 \\
L_{oo}(\bx) & = & 4 \qquad \mbox{if} \;\;\;\; 
                      |\, \bx \,|\,\bmod\,4\,=\,3 \;\mbox{.}
\ea
Thus, in this way we can automatically divide the
lattice in four sub-lattices and define a four-component
field $\,(f^{a}_{ee}(\bk)\mbox{,}\, f^{a}_{eo}(\bk)\mbox{,}\,
f^{a}_{oe}(\bk)\mbox{,}\, f^{a}_{oo}(\bk))\,$ by \\[0.1cm]
\be
f^{a}_{ee}(\bk) \,\equiv\, \sum_{\bx} \; f^a(\bx)\, L_{ee}(\bx) \,
 \exp{\left( \,- \, 2\,\pi\,i\,\bk\,\cdot\,\bx\,\right)}
\label{eq:fourierlike4ee}
\;\mbox{,}
\ee
and analogously for the other three components.
Also, in analogy with
eqs.\ \reff{eq:fourierlike+} and \reff{eq:fourierlike-},
one can check that
\ba
\sum_{\bx} \; f^a(\bx + e_{\mu})\, L_{ee}(\bx) \,
 \exp{\left( \,- \, 2\,\pi\,i\,\bk\,\cdot\,\bx\,\right)}
&=& e^{+\,2\,\pi\,i\,k_{\mu}}\, f^{a}_{oo}(\bk)
\label{eq:fourierlike+ee} \\
\sum_{\bx} \; f^a(\bx - e_{\mu})\, L_{ee}(\bx) \,
 \exp{\left( \,- \, 2\,\pi\,i\,\bk\,\cdot\,\bx\,\right)}
&=& e^{-\,2\,\pi\,i\,k_{\mu}}\, f^{a}_{oe}(\bk)
\label{eq:fourierlike-ee}
\ea
and similarly using the other linear combinations defined 
in eqs.\ \reff{eq:Leo}--\reff{eq:Loo}. Then, for any
updating sequence of the four components
$(f^{a}_{ee}(\bk)\mbox{,}\, f^{a}_{eo}(\bk)\mbox{,}\,
f^{a}_{oe}(\bk)\mbox{,}\, f^{a}_{oo}(\bk))$ we obtain
an updating matrix which is $4 \times 4$. For example,
if we update these components in the order 
even-even, even-odd, odd-even and odd-odd one can verify that
the updating matrix is given by\footnote{~In order to
avoid over-counting for the momenta we should set, for
example, $k_{d}\,N\,=\, 0\mbox{,}\,1\mbox{,}\,\ldots \mbox{,}\,
N/4 - 1\,$ and $k_i\,N\,=\,0\mbox{,}\,1 \mbox{,}\,
\ldots \mbox{,}\, N - 1\,$ for $i = 1\mbox{,}\, \ldots
\mbox{,}\,d-1$.}
\be
 (1\,-\,\omega)\, \1 \,+\, \omega\,
 \left( \begin{array}{cc}
              0 & E^{-}(\bk) \\[0.1cm]
              (1\,-\,\omega)\,E^{+}(\bk) & \omega \, E^{+}(\bk)\,E^{-}(\bk)
                 \end{array} \right) 
\;\mbox{,}
\label{eq:4x4matrix}
\ee
where $\1$ is the $4 \times 4$ identity matrix and
$E^{\pm}(\bk)$ are $2 \times 2$ matrices defined by
\be
E^{\pm}(\bk)\,\equiv\,\left( \begin{array}{cc}
        e^{\pm}(\bk) & e^{\mp}(\bk) \\[0.1cm]
        e^{\mp}(\bk) & e^{\pm}(\bk)\end{array}
                  \right)
\ee
with
\be
e^{\pm}(\bk) \,\equiv\, \frac{1}{2 \,d}\,\sum_{\mu = 1}^{d}\,
  e^{\pm 2\,\pi\,i\,k_{\mu}}\,\equiv\, e_r(\bk) \,\pm\,i\, e_i(\bk)
\;\mbox{.}
\label{eq:erei}
\ee
Clearly the matrix in eq.\ \reff{eq:4x4matrix} has a structure
very similar to that of the $2 \times 2$ matrix given
in eq.\ \reff{eq:M}. In fact, in this case,
we are still using a checkerboard ordering (first all even sites
and then all odd sites), but with the even and odd sub-lattices
divided in turn into two sub-lattices. The eigenvalues of this
$4 \times 4$ matrix are, as expected from Ref.\ \cite{BN},
solutions of two different second-order equations and are given by
\ba
\lambda_{\pm}^r(\bk\mbox{,}\,\omega) &=&
(1\,-\,\omega) \,+\,2\,\omega^2\,e_r^2(\bk)\,\pm\,
   2\,\omega\,\sqrt{(1\,-\,\omega)^2 \,e_r^2(\bk)+\,
               \omega^2\,e_r^4(\bk)}
\label{eq:lambdarpm}\\[0.2cm]
\lambda_{\pm}^i(\bk\mbox{,}\,\omega) &=&
(1\,-\,\omega) \,+\,2\,\omega^2\,e_i^2(\bk)\,\pm\,
   2\,\omega\,\sqrt{(1\,-\,\omega)^2 \,e_i^2(\bk)+\,
               \omega^2\,e_i^4(\bk)}
\label{eq:lambdaipm}
\;\mbox{.}
\ea
Since $e_r(\bk)$ is equal to the quantity $c(\bk)$ defined in eq.\
\reff{eq:cdefini}, we have that the eigenvalues
$\lambda_{\pm}^r(\bk\mbox{,}\,\omega)$ coincide with the
two eigenvalues of the overrelaxation method obtained
in Section \ref{CSD}. Also, for $\omega = 1$, we have
$\lambda_{\pm}^{r, i}(\bk\mbox{,}\,\omega) = 0\mbox{,}\,
4\,e^2_{r, i}(\bk)\,$ and $\,4\,e^2_{r}(\bk) =
4\,c^2(\bk)$ is the non-zero eigenvalue of the Los Alamos method
found in Section \ref{LosAl}. We note that
\be
e_i(\bk)\,=\,
\frac{1}{2\,d} \,
            \sum_{\mu = 1}^{d}\,
              \sin{(\,2\,\pi\,k_{\mu}\,)}
\label{eq:eidefinit}
\ee
assumes its largest value (equal to $1/2$) when $k_{\mu} = 1/4$
(for all directions $\mu$). This implies [see eqs.\
\reff{eq:lambdarpm} and \reff{eq:lambdaipm}] that CSD is now
due not only to the long-wavelength modes --- i.e.\ small momenta
$p^2(\bk) \approx 0$, for which $4\,c^2(\bk) \approx 1$ --- but
also to the modes with $k_{\mu} \approx 1/4$,
corresponding to momenta $p^2(\bk) \approx 2\,d$ and
for which $4\,e^2_i(\bk) \approx 1$.
This result can be explained by observing that
the division of the lattice in four sub-lattices couples the
modes with $k_{\mu} \approx 0$ to the modes with
$k_{\mu} \approx 1/4$. In fact, $e_i(\bk)$ becomes
$e_r(\bk) = c(\bk)$ when $k_{\mu}$ goes to $k_{\mu} + 1/4$
(for all directions $\mu$).

\vskip 0.3cm

Of course, if one considers a different updating sequence
for the four components of the field
$f^{a}(\bk)$, then the
updating matrix will be different from that reported in
eq.\ \reff{eq:4x4matrix}. For example,
if we update these components in the order
even-even, odd-even, even-odd and odd-odd one can verify that
the updating matrix is given by
\small
\be
{\cal M}_4 \,=\,
 \left( \begin{array}{cccc}
  (1\,-\,\omega)  &  0  &  \omega\, e^-(\bk)  &  \omega\, e^+(\bk) \\[0.2cm]
  (1\,-\,\omega)\,\omega^2\,{e^+}^2(\bk)  & f(\bk\mbox{,}\, \omega)  &
  \omega\,e^+(\bk)\,f(\bk\mbox{,}\, \omega)  &
  g(\bk\mbox{,}\, \omega) \\[0.2cm]
  (1\,-\,\omega)\,\omega\,e^+(\bk)  &  \omega \, e^-(\bk)  & 
  f(\bk\mbox{,}\, \omega)  &  \omega^2\,{e^+}^2(\bk)  \\[0.2cm]
  (1\,-\,\omega)\,g(\bk\mbox{,}\, \omega)  &
  \omega\,e^+(\bk)\,f(\bk\mbox{,}\, \omega)  &
  h(\bk\mbox{,}\, \omega) &
  f(\bk\mbox{,}\, \omega)\,+\,
  \omega\,e^+(\bk)\,g(\bk\mbox{,}\, \omega) 
  \end{array} \right)
\;\mbox{,}
\label{eq:4x4matrixnew}
\ee
\normalsize
where
\ba
f(\bk\mbox{,}\, \omega) & \equiv &
(1\,-\,\omega)\,+\,\omega^2\,e^+(\bk)\,e^-(\bk) \\[0.2cm]
g(\bk\mbox{,}\, \omega) & \equiv &
\omega\,e^-(\bk)\,+\,\omega^3\,{e^+}^3(\bk) \\[0.2cm]
h(\bk\mbox{,}\, \omega) & \equiv &
\omega^2\,\left[
{e^-}^2(\bk) \,+\, {e^+}^2(\bk)\, f(\bk\mbox{,}\, \omega)
\right]
\;\mbox{.}
\ea
In this case we were not able to prove that the characteristic
equation of the above $4 \times 4$ matrix can be
factorized in two different second-order equations and we
could not find any simple expression for the
four eigenvalues. [On the other hand, from
\reff{eq:4x4matrixnew} it is obvious that one of
the eigenvalues of the matrix ${\cal M}_4$ is zero when
$\omega = 1$.] However, after some manipulations
one can verify that the characteristic
equation of ${\cal M}_4$ can be factorized into two
``almost'' second-order equations given by
\be
\left(1 - \omega - \lambda\right)^2 \,-\,
2\,\omega^2\,\lambda\,e^+(\bk)\,e^-(\bk)\,
\pm \, \sqrt{\lambda}\,\omega^2\,
\left[{e^-}^2(\bk)\,+\,\lambda\,{e^+}^2(\bk)\right]
\,=\,0
\;\mbox{.}
\ee
By using eq.\ \reff{eq:erei} and the relation $e_r(\bk)
= c(\bk)$ one can re-write the above equations as
\ba
\left(1 - \omega - \lambda\right)^2 \,-\,
4\,\omega^2\,\lambda\,c^2(\bk) \! &+& \!
\sqrt{\lambda}\,\omega^2\,\left[ 
c^2(\bk)\left(1\,+\,\sqrt{\lambda}\right)^2 \right.
\nonumber
\\[0.2cm]
& & \left. \!\!\!\!\!\!
-\, e_i^2(\bk)\left(1\,+\,\sqrt{\lambda}\right)^2\,-\,
2\,i\,c(\bk)\,e_i(\bk)\left(1\,-\,\lambda\right)
\right]\,=\,0 \qquad \quad
\label{eq:equationofM4plus}
\ea
and
\ba
\left(1 - \omega - \lambda\right)^2 \,-\,
4\,\omega^2\,\lambda\,c^2(\bk) \! &-& \! 
\sqrt{\lambda}\,\omega^2\,\left[
c^2(\bk)\left(1\,-\,\sqrt{\lambda}\right)^2 \right.
\nonumber
\\[0.2cm]
& & \left. \!\!\!\!\!\!
-\, e_i^2(\bk)\left(1\,-\,\sqrt{\lambda}\right)^2\,-\,
2\,i\,c(\bk)\,e_i(\bk)\left(1\,-\,\lambda\right)
\right]\,=\,0  
\;\mbox{.} \qquad \quad
\label{eq:equationofM4minus}
\ea
Let us notice that, written in this way, these two
equations are very similar to the characteristic equation 
of the matrix $M(\bk\mbox{,}\,\omega)$
given in eq.\ \reff{eq:M} [or equivalently of the
matrix $C(\bk\mbox{,}\,\omega)$ given in eq.\ \reff{eq:C}]:
\be
\left(1\,-\,\omega\,-\,\lambda\right)^2\,-\,4\,
\omega^2\,\lambda\, c^2(\bk)\,=\,0
\label{eq:characeq}
\;\mbox{.}
\ee

Before considering the equations \reff{eq:equationofM4plus}
and \reff{eq:equationofM4minus}
it is interesting to study in more detail eq.\ \reff{eq:characeq}
and see how we can estimate the dynamic critical exponent $z$.
To this end we can re-write the previous equation as
\be
\left[\left(1\,-\,\lambda\right)\,-\,\omega\,\sqrt{1\,-\,4\,c^2(\bk)}
       \right]^2\,+\,2\,\omega\,\left(1\,-\,\lambda\right)\,
\left\{\sqrt{1\,-\,4\,c^2(\bk)} \,-\,\left[1\,-\,2\,\omega\,
               c^2(\bk)\right]\right\}\,=\,0
\ee
and for the largest value of $c^2(\bk) < 1/4$, in the limit of
large lattice side $N$, we obtain [using eq.\ \reff{eq:clargeN}]
\ba
\!\!\!\!\!\!
  & & \left[\left(1\,-\,\lambda\right)\,-\,\omega\,\sqrt{2\,\zeta(N)} \,+\,
   {\cal O}(\zeta^{3/2}(N))
       \right]^2\, \qquad \nonumber  \\[0.2cm]
& & \qquad \qquad \qquad \qquad +\,2\,\omega\,\left(1\,-\,\lambda\right)\,
\left\{\sqrt{2\,\zeta(N)} \,-\,\left(1\,-\,\frac{\omega}{2}
\right)\,+\, {\cal O}(\zeta(N))\right\}\,=\,0 
\;\mbox{.} \qquad \qquad
\ea
Then by setting $\lambda = 1\,-\,\delta$ we get
\be
\delta^2\,+\,2\,\omega^2\,\zeta(N)\,
+\,2\,\omega\,\delta\,\left(\frac{\omega}{2} \,-\,1\right)
\,+\,
{\cal O}(\delta\,\zeta(N)\mbox{,}\, \zeta^{2}(N)) \,=\,0
\label{eq:eqfordelta}
\;\mbox{.}
\ee
Thus, if we do not tune the parameter $\omega$ we have
that $\delta$ is the solution of the equation
\be
\omega\,\zeta(N)\,
+\,\delta\,\left(\frac{\omega}{2} \,-\,1\right)
\,+\,
{\cal O}(\delta^2\mbox{,}\,\delta\,\zeta(N)
\mbox{,}\, \zeta^{2}(N)) \,=\,0
\;\mbox{,}
\ee
namely $\delta \propto \zeta(N) \sim 1/N^2$. This implies
[see end of Section \ref{Infinito}]
\be
\tau \, \gtapprox
\frac{1}{\left| \,\delta \, \right|}
\,\sim\,N^2
\ee
and $z = 2$. On the contrary if we tune
$\omega = 2/ (1\,+\,\Omega)$ by setting
\be
\Omega =  {\overline \Omega}\; \zeta^{m}(N)
\;\mbox{,}
\label{eq:Omegatuning}
\ee
with $m > 0$, then from eq.\ \reff{eq:eqfordelta} we obtain
\be
\delta^2\,-\,4\,\delta\,{\overline \Omega}\, \zeta^{m}(N)\,+\,
8\,\zeta(N)\,+\,
{\cal O}(\delta\,\zeta(N)\mbox{,}\,
\delta\,\zeta^{2 m}(N)\mbox{,}\,
\zeta^{1+m}(N))\mbox{,}\, \zeta^{2}(N)) \,=\,0
\label{eq:eqfordelta2}
\;\mbox{.}
\ee
Note that for $m < 1/2$ the previous equation
simplifies to 
\be
\delta^2\,-\,4\,\delta\,{\overline \Omega}\,
\zeta^{m}(N) \,+\,
{\cal O}(\delta\,\zeta^{2 m}(N)\mbox{,}\,
\zeta(N)) \,=\,0
\;\mbox{,}
\ee
with solutions $\delta = 0$ and
$\delta = 4\,{\overline \Omega}\, \zeta^{m}(N)$.
Thus, in this case one of the eigenvalues is
equal to 1 and the algorithm does not converge.
On the other hand, for $m = 1/2$ we have
\be
\delta^2\,-\,4\,\delta\,{\overline \Omega}\, \sqrt{\zeta(N)}\,+\,
8\,\zeta(N)\,+\,
{\cal O}(\delta\,\zeta(N)\mbox{,}\,
\zeta^{3/2}(N)) 
\,=\,0  
\ee
and the solutions $\delta$ are clearly
proportional to $\sqrt{\zeta(N)}$, giving
$z = 1$. One can also check that the tuning
condition ${\overline \Omega} = \sqrt{2}$
[see eq.\ \reff{eq:Omegadiz}] reduces the previous equation
to a perfect square, namely
\be
\left[\delta\,-\,2\,\sqrt{2\,\zeta(N)}\right]^2
\,+\,
{\cal O}(\delta\,\zeta(N)\mbox{,}\,
\zeta^{3/2}(N)) 
\,=\,0
\;\mbox{.}
\ee

Clearly, this analysis can also be applied to the two
equations \reff{eq:equationofM4plus}
and \reff{eq:equationofM4minus}. In this way we will check
that the dynamic critical exponent $z$ cannot be
smaller than 1 when using the updating matrix ${\cal M}_4$.
To this end, we have to consider the largest value of
$c^2(\bk) < 1/4$ --- corresponding to 
$k_i = 0$ for $i = 1\mbox{,}\, \ldots
\mbox{,}\,d-1$ and $k_{d} = 1/N$ --- in the limit of large
lattice side $N$, i.e.\ we use eq.\ \reff{eq:clargeN}.
For the same $\bk$, in the limit of large $N$, we
also have the relation
\be
e_i(\bk) \,=\, \frac{1}{2\,d}\,\sin{\left(\frac{
2\,\pi}{N}\right)} \approx \frac{\pi}{d\,N} \,+\,
{\cal O}(N^{-3})
=\,\sqrt{\frac{\zeta(N)}{2\,d}}\,+\,
{\cal O}(\zeta^{3/2}(N))
\;\mbox{.}   
\label{eq:eismall}
\ee
Then, for $N = \infty$ and by considering the equation
\reff{eq:equationofM4plus} we get
\be
\left(1 - \lambda\right)^2 \,+\,
2\,\omega\,\left(1 - \lambda\right)\,
\left(\frac{\omega}{2}\,-\,1\right)\,+\,
\sqrt{\lambda}\,\frac{\omega^2}{4}\,\left(
1\,+\,\sqrt{\lambda}\right)^2  \,=\, 0
\;\mbox{.}   
\ee
Obviously, with $\omega \in (1\mbox{,}\, 2)$ there is no
solution $\lambda \approx 1$ that can satisfy this equation.
In other words, if we set $\lambda = 1\,-\,\delta$, then
$\delta$ stays finite and we don't get a divergent
relaxation time $\tau$. Thus, critical
slowing-down should be related to the solutions
of the equation \reff{eq:equationofM4minus}.
For this equation, using eqs.\ \reff{eq:clargeN} and
\reff{eq:eismall} and setting again $\lambda = 1\,-\,
\delta$, we find
\ba
\!\!\!\!\!\!  & & 
\left(1\,-\,\frac{\omega^2}{16}\right)\,\delta^2\,+\,
2\,\omega^2\,\zeta(N)\,
+\,2\,\omega\,\delta\,\left(\frac{\omega}{2} \,-\,1\right)
\qquad \nonumber \\[0.2cm]
& & \qquad \qquad \qquad \qquad
\,+\,i\,\omega^2\,\delta\,\sqrt{\frac{\zeta(N)}{2\,d}}
\,+\,
{\cal O}(\delta^3\mbox{,}\, \delta^2 \sqrt{\zeta(N)}
\mbox{,}\, \delta\,\zeta(N)\mbox{,}\, 
\zeta^{2}(N))\,=\,0 
\label{eq:eqfordeltanew}
\;\mbox{.}  \qquad \qquad
\ea
This is very similar to eq.\ \reff{eq:eqfordelta} and
we can check that the two new terms $\,- \omega^2\,\delta^2 / 16\,$
and $\,i\,\omega^2\,\delta\,\sqrt{\zeta(N) / (2\,d)}\,$ do
not spoil the analysis already done for that equation.
In fact, if we do not tune the parameter
$\omega$ we have
\be
\omega\,\zeta(N)\,
+\,\delta\,\left(\frac{\omega}{2} \,-\,1\right)
\,+\,
{\cal O}(\delta^2\mbox{,}\, \delta \sqrt{\zeta(N)}
\mbox{,}\, \zeta^{2}(N)) \,=\,0
\;\mbox{,}
\ee
namely $\delta \propto \zeta(N) \sim 1/N^2$ and $z = 2$. 
On the contrary, if we use the tuning given in eq.\
\reff{eq:Omegatuning} and consider $0 < m < 1/2$ we have
\be
\frac{3}{4}\,\delta^2\,-\,4\,\delta\,{\overline \Omega}\,
\zeta^{m}(N) \,+\,
{\cal O}(\delta^3\mbox{,}\,
\delta^2\,\zeta^{m}(N)\mbox{,}\,
\delta\,\zeta^{2 m}(N)\mbox{,}\,
\zeta(N)) \,=\,0
\ee
with solutions $\delta = 0$ and
$\delta = 16\,{\overline \Omega}\, \zeta^{m}(N) / 3$.
Finally, with the tuning condition \reff{eq:Omegatuning} and
$m = 1/2$ we have
\be
\frac{3}{4}\,\delta^2\,+\,8\,\zeta(N)\,-\,
4\,\delta\,\sqrt{\zeta(N)}
\,\left({\overline \Omega}\,-\,\frac{i}{2\,d}\right)
 \,+\,
{\cal O}(\delta^3\mbox{,}\,
\delta^2\,\sqrt{\zeta(N)}\mbox{,}\,
\delta\,\zeta(N)\mbox{,}\,
\zeta^{3/2}(N)) \,=\,0
\ee
with the solution $\delta \sim \sqrt{\zeta(N)}$  and
$z = 1$. 

Note that one arrives at the same results by considering
$k_i = 1/4$ for $i = 1\mbox{,}\, \ldots
\mbox{,}\,d-1$ and $k_{d} =  1/4 - 1/N$, 
corresponding to momenta $p^2(\bk) \approx 2\,d$,
for which $4\,e^2_i(\bk) \approx 1$
[see comment after eq.\ \reff{eq:eidefinit}]. In fact, in this
case, we have
\be
c(\bk) \,=\,
\sqrt{\frac{\zeta(N)}{2\,d}}\,+\,
{\cal O}(\zeta^{3/2}(N))
\ee
and
\be
e_i(\bk) \,=\,\frac{1}{2}\,\left[\,1\,-\,\zeta(N)\,\right]
\;\mbox{.}
\ee
In particular, for this value of $\bk$ one can verify that
the equation \reff{eq:equationofM4minus} has
solutions $\lambda = 1 - \delta$ with $\delta$
finite. Thus, the corresponding relaxation times $\tau$
do not diverge and CSD is related to the solutions of the equation
\reff{eq:equationofM4plus}, yielding $z$ not smaller than 1.

We checked numerically these results in two, three and four
dimensions at $\beta = \infty$ for several lattice sides,
obtaining indeed $z = 1$ and a computational cost
equivalent to that of the standard overrelaxation
method. [This check was done by updating the components
of the $f^a(\bk)$ field
in the order even-even, odd-even, even-odd and odd-odd.]

\vskip 0.3cm

To sum up we can say that with the two matrices of
size $4 \times 4$ considered in this section, we cannot get
a dynamic critical exponent $z$ smaller than 1. In particular,
we found that the characteristic equation of the first
$4 \times 4$ matrix [see eq.\ \reff{eq:4x4matrix}]
can be factorized in two different second-order
equations --- as predicted in Ref.\ \cite{BN} --- giving
eigenvalues identical to those found when considering a matrix
of size $2 \times 2$ [see Section \ref{CSD}].
For the second $4 \times 4$ matrix [see eq.\ \reff{eq:4x4matrixnew}]
the same factorization leads to two ``almost''
second-order equations and one can check that, compared to
the $2 \times 2$ case, the extra term
(proportional to $\sqrt{\lambda}$) does not really
modify the structure of the equations. Thus, the
dynamic critical exponent $z$ must be the same in the
two cases, i.e.\ not smaller than 1.


\subsection{Generalized Stochastic Overrelaxation Algorithm}
\label{generalstochoverr}

One can try to generalize the stochastic overrelaxation
algorithm by considering a probability distribution
different from $p^{m}\,( 1 \,-\, p )$. For example, let
$f(p\mbox{,}\, m)$ be the probability of having
$m$ micro-canonical sweeps [see eq.\ \reff{eq:up02}]
followed by one standard Los Alamos update.
Then one has
\be
\langle \,m\,\rangle \, \equiv \, \sum_{m=0}^{\infty}\,
m\,f(p\mbox{,}\, m)
\,=\,m(p)
\;\mbox{,}
\ee
where we suppose that $\sum_{m=0}^{\infty}\,f(p\mbox{,}\, m) = 1$.
(Here $p$ is a tuning parameter or a set of tuning parameters.) 
Note that, if we set $f(p\mbox{,}\, m) = \delta_{m 0}$
where $\delta_{m 0}$ is the Kronecker delta, we obtain
the Los Alamos algorithm, i.e.\ we don't do any micro-canonical
update and $m(p) = 0$.

For this generalized stochastic overrelaxation algorithm
we should consider the matrix
\be
{\cal M}(\bk\mbox{,}\, p) \, \equiv \,
\sum_{m = 0}^{\infty}\, f(p\mbox{,}\, m)\,
{\widetilde M}(\bk\mbox{,}\, m)
\ee
instead of the matrix defined in eq.\ \reff{eq:defineMdip}.
It is easy to check that this matrix is given by
\be
{\cal M}(\bk\mbox{,}\, p)\, = \, 
\frac{\cos{\left[ \theta(\bk)\, /\,
2 \right]}}{\sin{\left[ \theta(\bk)\, /\,
2 \right]}}
  \left( \begin{array}{cc}   
 - \Sigma_1(\theta(\bk)\mbox{,}\, p) & 
   \Sigma_2(\theta(\bk)\mbox{,}\, p) \\[0.1cm]
 - \cos{\left[ \theta(\bk)\, /\, 2 \right]}
  \, \Sigma_1(\theta(\bk)\mbox{,}\, p) &
   \cos{\left[ \theta(\bk)\, /\, 2 \right]}
  \, \Sigma_2(\theta(\bk)\mbox{,}\, p)
                             \end{array} \right)
\label{eq:matrixforgsoa}
\;\mbox{,}
\ee
where
\ba
\Sigma_1(\theta(\bk)\mbox{,}\, p) & \equiv &
    \sum_{m = 0}^{\infty}\, f(p\mbox{,}\, m)\,
      \sin{\left[\,m\,\theta(\bk)\,\right]}
\label{eq:defsigma1} \\[0.2cm]
\Sigma_2(\theta(\bk)\mbox{,}\, p) & \equiv &
    \sum_{m = 0}^{\infty}\, f(p\mbox{,}\, m)\, 
      \sin{\left[\,\left(m\,+\,\frac{1}{2}\right) \,
                 \theta(\bk)\,\right]}
\;\mbox{.}
\ea
The matrix ${\cal M}(\bk\mbox{,}\, p)$ has eigenvalues $0$ and
\ba
\lambda(\theta(\bk)\mbox{,}\, p) &=&
\frac{\cos{\left[ \theta(\bk)\, /\,
2 \right]}}{\sin{\left[ \theta(\bk)\, /\,
2 \right]}} \, \left\{
   \cos{\left[ \theta(\bk)\, /\, 2 \right]}
  \, \Sigma_2(\theta(\bk)\mbox{,}\, p) \,-\,
     \Sigma_1(\theta(\bk)\mbox{,}\, p) \, \right\} \\[0.3cm]
 & = &
-\,\sin{\left[ \theta(\bk)\, /\, 2 \right]}\,
      \Sigma_1(\theta(\bk)\mbox{,}\, p) \,+\,
   \cos^2{\left[ \theta(\bk)\, /\, 2 \right]}\,
      \Sigma_3(\theta(\bk)\mbox{,}\, p)
\;\mbox{,}
\ea
where we define
\be
\Sigma_3(\theta(\bk)\mbox{,}\, p)  \equiv 
    \sum_{m = 0}^{\infty}\, f(p\mbox{,}\, m)\,
      \cos{\left[\,m\,\theta(\bk)\,\right]}
\label{eq:defsigma3}
\;\mbox{.}
\ee
For $f(p\mbox{,}\, m) = \delta_{m 0}$ we have
$\Sigma_1(\theta(\bk)\mbox{,}\, p) = 0$,
$\Sigma_3(\theta(\bk)\mbox{,}\, p) = 1$ and
[using eq.\ \reff{eq:deftheta}]
\be
\lambda(\theta(\bk)\mbox{,}\, p) = \cos^2{\left[
\theta(\bk)\, /\, 2 \right]} = 4\, c^2(\bk)
\;\mbox{,}
\ee
in agreement with the result obtained for the Los Alamos algorithm
in Section \ref{LosAl}.

\vskip 0.3cm

In order to study the CSD of this algorithm we should consider
the largest eigenvalue of the matrix ${\cal M}(\bk\mbox{,}\, p)$.
It is obvious that, if $\theta(\bk) = 0$ [corresponding
to $c(\bk) = 1/2$], we have $\Sigma_3(0\mbox{,}\, p) = 1$,
$\Sigma_1(0\mbox{,}\, p) = 0$ and $\lambda(0\mbox{,}\, p)
= 1$. At the same time, for a small angle $\theta(\bk) = \theta_s$,
we can expand the above expressions
in powers of $\theta_s$ and obtain\footnote{~A
similar analysis can be done for the case $\theta(\bk) = 
2\,\pi - \theta_s$, with $\theta_s$ small,
corresponding to $c(\bk) \approx - 1/2$.}
\ba
\Sigma_1(\theta_s\mbox{,}\, p) & = & \,
              \theta_s \,\langle \,m\,\rangle
  \,+\, {\cal O}(\theta_s^3)
\label{eq:s1small} \\ [0.2cm]
\Sigma_3(\theta_s\mbox{,}\, p) & = &
         1 \,-\,\frac{\theta_s^2}{2} \,
                    \langle \,m^2\,\rangle 
  \,+\, {\cal O}(\theta_s^4)
\label{eq:s3small} \\[0.2cm]
\lambda(\theta_s\mbox{,}\, p) & = &
   1\,-\,\frac{\theta_s^2}{2} \left(\,
        \langle \,m\,\rangle \,+\,
         \langle \,m^2\,\rangle \,+\, \frac{1}{2}\,
          \right)\,+\, {\cal O}(\theta_s^4) \;\mbox{,}
\ea
where $\langle \,m\,\rangle$ and $\langle \,m^2\,\rangle$ 
are, in general, functions of $p$.  This implies
\be
\frac{\tau}{\langle m \rangle\,+\,1} \,\approx\,
\frac{1}{1 \,-\,\lambda(\theta_s\mbox{,}\, p)} \,\approx\,
\frac{1}{\theta_s^2} \, \frac{2}{\langle \,m\,\rangle \,+\,
         \langle \,m^2\,\rangle \,+\, \frac{1}{2}}
\;\mbox{.}
\ee
From eqs.\ \reff{eq:clargeN} and \reff{eq:deftheta} and the
relation $\theta(\bk) = \theta_s$ we have
that
\be
\theta_s^2 \,\approx \, 8\,\zeta(N) \,=\, \frac{16 \pi^2}{d N^2}
\;\mbox{.}
\ee
Thus, if $\langle \,m\,\rangle$ and $\langle \,m^2\,\rangle$
stay finite we have that $\tau \propto N^2$ and $z = 2$.
In order to reduce CSD we have to tune $p$ so that
both $\langle \,m\,\rangle$ and $\langle \,m^2\,\rangle$
go to infinity as powers of $1/\theta_s$. Note that if
by tuning $p$ we have that $\langle \,m\,\rangle$ goes to
infinity, then from the inequality $\langle \,m\,\rangle \, \leq\,
\langle \,m^2\,\rangle$, which is a consequence of the positiveness
of the variance $\sigma^2 = \langle \,m^2\,\rangle \,-\,
\langle \,m\,\rangle^2\,$, we get 
that $\langle \,m^2\,\rangle$ goes to infinity too.
The same inequality allows, in principle, $\langle \,m^2\,\rangle$
to go to infinity while $\langle \,m\,\rangle$ stays finite,
but this cannot happen if $\sigma^2$ is finite. Moreover, if
$\sigma^2 < + \infty$ and $\langle \,m\,\rangle \sim
\theta_s^{-n}$ we should have that $\langle \,m^2\,\rangle \sim
\theta_s^{-2 n}$ and we obtain
\be
\tau \,\approx\,\frac{2}{\theta_s^2} \,
  \frac{1}{1 + \frac{\langle \,m^2\,\rangle}{\langle \,m\,\rangle}}
\,\approx\,\frac{2}{\theta_s^{2-n}}
\;\mbox{.}
\ee
From eqs.\ \reff{eq:s1small} and \reff{eq:s3small}
and from the fact that $\Sigma_1(\theta(\bk)\mbox{,}\, p)$
and $\Sigma_3(\theta(\bk)\mbox{,}\, p)$ are finite,\footnote{~Actually,
from eqs.\ \protect\reff{eq:defsigma1} and \protect\reff{eq:defsigma3},
from the relation $f(p\mbox{,}\, m) \geq 0$
and the normalization condition for the
probability distribution $f(p\mbox{,}\, m)$ we have that
$| \Sigma_1(\theta(\bk)\mbox{,}\, p) |$ and
$| \Sigma_3(\theta(\bk)\mbox{,}\, p) |$ are smaller
than or equal to $1$.}
we obtain that --- at least to order $\theta_s^2$ ---
the only tuning we can have is given by
\be
\langle m \rangle\, \sim\, \theta_s^{-1}
\qquad \qquad
\langle m^2 \rangle\, \sim\, \theta_s^{-2}
\;\mbox{,}
\label{eq:tuningform}
\ee
which implies [see eq.\ \reff{eq:taustocfin}]
\be
\tau \,\approx\,\frac{2}{\theta_s}
\,\approx\,\frac{\sqrt{d} N}{2 \pi}
\;\mbox{.}
\ee
This yields, as expected, $z = 1$. Thus, with an appropriate choice
of the distribution $f(p\mbox{,}\, m)$ and of the tuning,
one can only hope to reduce the factor that multiplies
$1 / \theta_s \approx \sqrt{d} N / \pi$ in the above equation,
but there is no way of having $z < 1$.

We can verify the tuning relations
\be
\langle m^2 \rangle \,\sim\, \langle m \rangle^2 \,\sim\,N^2
\label{eq:m2andm}
\ee
for the probability distribution
$f(p\mbox{,}\, m) = p^m (1\,-\,p)$, which has a tuning condition
$P \sim 1/N$ (see Section \ref{Stoch}), and for the
probability distribution $f(p\mbox{,}\, m)$ constant in the interval
$[1\mbox{,}\,2 \,{\overline m}\,-\,1]$, which has a
tuning condition ${\overline m} \sim N$ (see Ref.\ \cite{Wo}).
In the first case we know that
\be
\langle m \rangle\, =\, \frac{1 \,-\,P}{2\,P}
\ee
and one can check that
\be
\langle m^2 \rangle\,=\,
\frac{p\,(1\,+\,p)}{(1\,-\,p)^2} \,=\,
 \frac{1 \,-\,P}{2\,P^2}
\;\mbox{.}
\ee
Thus, when $P$ goes to zero as $1/N$, relations \reff{eq:m2andm}
are satisfied.
In the second case we have that the average value of
$m$ is ${\overline m}$ and one can verify
that
\be
\langle m^2 \rangle\, =\, \frac{{\overline m}}{3}\,
\left(4\,{\overline m}\,-\,1\right)
\;\mbox{.}
\ee
Thus, when ${\overline m}$ goes to infinity as $N$ we obtain again
relations \reff{eq:m2andm}.

Let us notice that the analysis presented in Ref.\ \cite{BN}
clearly applies also to the stochastic overrelaxation algorithm
for any probability distribution $f(p\mbox{,}\, m)$, as long as the
updating matrix ${\cal M}(\bk\mbox{,}\, p)$ is $2 \times 2$.
So, the result that $z$ cannot be smaller than $1$ for the
generalized stochastic overrelaxation algorithm given
by the matrix \reff{eq:matrixforgsoa} is not unexpected. However, we
believe that the previous analysis,
and especially the relations in eq.\ \reff{eq:tuningform},
clarify how critical slowing-down is reduced by this
algorithm.


\section{Numerical Results}
\label{InfinitoRes}

In order to check the analytical predictions presented in the previous
sections we have done numerical tests in two, three and four
dimensions. In each case we considered eight different lattice
sides $N$, namely $N = 16\mbox{,}\; 32\mbox{,}\;48 \mbox{,}\,\ldots
\mbox{,}\;128$ in two dimensions, $N = 8\mbox{,}\; 16\mbox{,}\;24
\mbox{,}\,\ldots\mbox{,}\;64$ in three dimensions and $N = 4\mbox{,}\;
8\mbox{,}\;12 \mbox{,}\,\ldots\mbox{,}\;32$ in four dimensions.
Also, for all the algorithms we have done
tests using both the lexicographic and the even/odd update.
For the Fourier acceleration method only lattice sides that are powers
of $2$ were considered and we used either the whole lattice
or even/odd sublattices
to implement the Laplacian preconditioning.

Simulations were done on the PC cluster installed in July 2001
at the IFSC-USP in connection with a grant from FAPESP (``Projeto
Jovem Pesquisador''). The system has 16 nodes and a server with 866
MHz Pentium III CPU and 256/512 MB RAM memory (working at 133 MHz) and
is operating with Linux Debian. The machines are connected with a
100 Mbps full-duplex network.
The total computer time used for the tests (including the runs described
in Sections \ref{sec:4x4}, \ref{sec:notun}, \ref{sec:fourierver} and
\ref{Lambda}) was equivalent to about $100$ days on one node.

\vskip 0.3cm

In Ref.\ \cite{CM} we checked the convergence of the gauge-fixing
algorithms by considering six different quantities. We found that,
for each given algorithm, these quantities relax to zero with the 
same speed, i.e.\ the same relaxation time. Here we consider only two 
of these six quantities, namely\footnote{~In Ref.\ \cite{CM}
these two quantities were called, respectively, $e_2$ and
$e_6$.}
\be
(\nabla A)^2 \, \equiv \,  \frac{1}{V} \sum_{\bx} \,
\sum_{c\, = 1}^{3} \, \Big[
    \left( \nabla \cdot A^{c} \right) (\bx) \Big]^{2}
\label{e2}
\;\mbox{,}
\ee
which is commonly used in numerical simulations, and
\be
\Sigma_Q \, \equiv \, \frac{1}{d} \,
  \sum_{\mu = 1}^{d} \, \frac{1}{3\,N}
\sum_{c\, = 1}^{3} \, \sum_{x_{\mu} = 1}^{N} \,
    \left[ \,  Q_{\mu}^{c}(x_{\mu}) - {\overline Q}_{\mu}^{c}  \,
      \right]^{2} \, \left[ {\overline Q}_{\mu}^{c} \right]^{- 2}
\;\mbox{,}
\label{eq:e6}
\ee
which provides a very sensitive test of the goodness of the 
gauge fixing \cite{CM}. Let us recall that we defined the lattice
gauge field as [see eq.\ \reff{eq:defA}]
\be
A_{\mu}(\bx) \,\equiv\, \frac{1}{2} \left[ \; U_{\mu}(\bx) -
                    U_{\mu}^{\dagger}(\bx) \; \right]
\ee
and that [see eq.\ \reff{eq:diverA}]
\be
\left(\nabla\cdot A^{c} \right)(\bx) \equiv
  \sum_{\mu = 1}^{d} \, \left[ A_{\mu}^{c} (\bx) -
                  A_{\mu}^{c} (\bx - \bun_{\mu}) \right]
\ee
is the lattice divergence of\\[0.1cm]
\be
A_{\mu}^{c} (\bx) \, \equiv \, \frac{1}{2\,i} \, 
  \mbox{Tr} \left[ A_{\mu}(\bx) \, \sigma^{c} \right]
    \; \mbox{,} \\[0.1cm]
\ee
where $\sigma^{c}$ is a Pauli matrix and $c = 1\mbox{,}\, 2
\mbox{,}\, 3$. We also define
\be
{\overline Q}_{\mu}^{c} \,
\equiv \, \frac{1}{N} \, \sum_{x_{\mu} = 1}^{N} \,
            Q_{\mu}^{c}(x_{\mu})
\; \mbox{,}
\ee
where the quantities
\be
Q_{\mu}^{c}(x_{\mu}) \, \equiv \, \sum_{\nu \neq \mu} \,
     \sum_{x_{\nu}} \, A_{\mu}^{c}(\bx)  \qquad \qquad
  \phantom{forall} \; \; \; \mu = 1\mbox{,}\ldots\,\mbox{,} d
\label{eq:charges}
\ee
are constant, i.e.\ independent of $x_{\mu}$, if the
Landau-gauge-fixing condition is satisfied \cite{CM}. Let us notice that, 
at $\beta = \infty$ and at a minimum of ${\cal E}_{U}\left[ g 
\right]$, one has ${\overline Q}_{\mu}^{c} = 0$. Therefore at $\beta = 
\infty$ the quantity $\Sigma_Q$ should be defined as
\be
\Sigma_Q \, \equiv \, \frac{1}{d} \,
  \sum_{\mu = 1}^{d} \, \frac{1}{3\,N}
\sum_{c\, = 1}^{3} \, \sum_{x_{\mu} = 1}^{N} \,
    \left[ \,  Q_{\mu}^{c}(x_{\mu}) \,
      \right]^{2}
\; \mbox{.}
\label{eq:e6inf}
\ee
We used $(\nabla A)^2 \,\leq \,10^{-15}$ as stopping
condition for the gauge-fixing algorithms. The quantity
$\Sigma_Q$ and the minimizing functional ${\cal E}_{U}\left[ g
\right]$ have been evaluated only for the final
gauge-fixed configuration.

Let us notice that, using eq.\ \reff{eq:nablaAofw}, we can re-write
$(\nabla A)^2$ as
\be
(\nabla A)^2 \, =\,\frac{1}{V} \sum_{\bx} \,
\sum_{c\, = 1}^{3} \, \left[ 
    \,{\mathrm w}^c(\bx)\,\right]^2
\;\mbox{.}
\ee
Also, from \reff{def_ztilde} and \reff{eq:wdefi} we
can write
\be
{\mathrm w}(\bx) \,=\, {\cal N}(\bx) \; {\widetilde {\mathrm w}}(\bx)
\label{eq:defwtilde}
\;\mbox{,}
\ee
where ${\widetilde {\mathrm w}}(\bx)$ is an $SU(2)$ matrix,
so that
\be
(\nabla A)^2 \, =\,\frac{1}{V} \sum_{\bx} \, {\cal N}^2(\bx)\,
\sum_{c\, = 1}^{3} \, \left[
    \,{\widetilde {\mathrm w}}^c(\bx)\,\right]^2
\;\mbox{.}
\ee
At the same time we can re-write the minimizing functional
[see eqs.\ \reff{eq:minfun} and \reff{eq:wofU}] as
\be
{\cal E}_{U}\left[ \, g\, \right] \, = \,
 1 \,-\, \frac{1}{2\,d\,V} \sum_{\mu = 1}^{d} \sum_{\bx} \,
         \frac{\Tr}{2} \left[ \;
               U_{\mu}^{\left( g \right)}(\bx) \,+\,
               {U_{\mu}^{\left( g \right)}}^{\dagger}(\bx - \bun_{\mu}) 
                        \; \right] \,=\,
 1 \,-\, \frac{1}{2\,d\,V}  \sum_{\bx} \,
         \frac{\Tr}{2} \,{\mathrm w}(\bx)
\ee
and using eq.\ \reff{eq:defwtilde} above we have
\be
{\cal E}_{U}\left[ \, g\, \right] \, = \,
  1 \,-\, \frac{1}{2\,d\,V}  \sum_{\bx} \,{\cal N}(\bx)\,
  \sqrt{\,1\,-\,\sum_{c\, = 1}^{3} \, \left[
    \,{\widetilde {\mathrm w}}^c(\bx)\,\right]^2}
\;\mbox{.}
\ee
Then, for $\beta = \infty$ and
in the limit of large number of gauge-fixing sweeps $t$,
namely by using eq.\ \reff{eq:Nd}, we obtain
\be
(\nabla A)^2 \, =\,\frac{4\,d^2}{V} \sum_{\bx} \,
\sum_{c\, = 1}^{3} \, \left[
    \,{\widetilde {\mathrm w}}^c(\bx)\,\right]^2
\,+\, {\cal O}(\epsilon^2)
\ee
and
\be
{\cal E}_{U}\left[ \, g\, \right] \, = \,
\frac{1}{2\,V} \sum_{\bx} \,
\sum_{c\, = 1}^{3} \, \left[
    \,{\widetilde {\mathrm w}}^c(\bx)\,\right]^2
\,+\, {\cal O}(\epsilon^2)
\;\mbox{.}
\ee
Thus, with the gauge fixing, $(\nabla A)^2$ goes to zero and
we should find $(\nabla A)^2 \gg   
{\cal E}_{U}\left[ \, g\, \right]$ in the
final gauge-fixed configuration.

\vskip 0.3cm

For the quantity $(\nabla A)^2(t)$, in the limit of 
large number of gauge-fixing sweeps $t$, we introduce a
relaxation time $\tau$ through 
the relation [see eq.\ \reff{eq:deftau}]
\be
(\nabla A)^2(t) \,\approx\,b\,\exp{( - \,t\,/\,\tau )}
\label{deftau}
\;\mbox{.}
\ee
The evaluation of $\,\tau\,$ is done using a chi-squared fit of
the function $\,\log (\nabla A)^2(t)\,$.
In order to get rid of the initial fluctuations,
this fit has been done five times using the data
corresponding to $t > n_{gf} (1 - 1 / t_{fac})$, where $n_{gf}$ is
the total number of sweeps necessary to fix the gauge and
$t_{fac} = 2\mbox{,}\, 4\mbox{,}\, 8\mbox{,}\, 16$ and $32$.
In most cases, $\tau$ increases for increasing $t_{fac}$, reaching
a plateau. We have chosen as the final value for $\tau$ the second
point where the $\tau$ values become stable within errors.

For the algorithms depending on a parameter, and therefore requiring
{\em tuning}, we used a procedure in three steps in order to find the optimal 
choice of the parameter, namely the value that minimizes the 
relaxation time $\tau$ at a fixed lattice side $N$. This procedure is
similar to the one described in Ref.\ \cite{CM}. We considered,
respectively, $5$ configurations in the first step, $10$ in the second
and $20$ in the third and final step.

Our final data for the  relaxation time $\tau$, the number of
gauge-fixing sweeps $n_{gf}$ and the time $t_{gf}$ (measured
in seconds) necessary to complete the gauge fixing are reported
in Tables \ref{Table.LosAlamos}--\ref{Table.Fourier} for the two-,
three- and the four-dimensional cases. When necessary we also report
the optimal choice for the tuning parameter. For all these
quantities we don't show the statistical error since it is
usually very small, namely less than $1 \%$. Recall that at
$\beta = \infty$ the link variables $U_{\mu}(\bx)$ are set equal to
the identity matrix and the initial $\{ g(\bx) \}$ configuration is
chosen randomly. Also, as said in the Introduction, in the limit of
large number of gauge-fixing sweeps $t$ the configuration
$U^{(g)}_{\mu}(\bx) = g(\bx)\, g^{-1}(\bx + \bun_{\mu})$ is driven by
the gauge fixing to the {\em vacuum} configuration
$U^{(g)}_{\mu}(\bx) = \1$, losing memory of the initial
$\{ g(\bx) \}$ configuration. This explains why there are 
usually very small fluctuations in the quantities that one is
interested in.


\subsection{Critical Exponents and Computational Cost of the
Algorithms}
\label{critexp}

From the data shown in Tables \ref{Table.LosAlamos}--\ref{Table.Fourier}
one can evaluate the dynamic critical exponents $z$ for the five
algorithms. In all cases --- and in two, three and four dimensions ---
the results are in agreement with our findings in Ref.\ \cite{CM}, namely
$z\approx 2$ for the Los Alamos method, $z \approx 1$ for the three
improved local algorithms and $z \approx 0$ for the Fourier acceleration
method. One can also check that the total number of gauge-fixing
sweeps $n_{gf}$ grows approximately as $N^z$, with the same values
of $z$ given above. The time $t_{gf}$ (measured in seconds) necessary
to complete the gauge fixing grows approximately as $N^{z+d}$,
as expected. The only exception is the Los Alamos method, for
which we find $n_{gf} \sim N^{1.5}$ and $t_{gf} \sim N^{1.5 + d}$,
suggesting that in the initial gauge-fixing sweeps this method
is more effective than the other local algorithms. This is not surprising,
since it is well known \cite{A} that the optimal strategy for the
overrelaxation algorithm is precisely to vary the parameter $\omega$ from an
initial value 1 (corresponding to the Los Alamos method) to a larger
asymptotic value $\omega_{opt}$. The computational cost of the Fourier
acceleration method will be discussed in more detail in Section
\ref{sec:fourierver} below.

We have also looked at the values of $\Sigma_Q\mbox{,}\;
(\nabla A)^2$ and ${\cal E}_{U}\left[ \, g\, \right]$ in the final
gauge-fixed configurations. We observed that:
\begin{itemize}
\item For all gauge-fixing algorithms and all dimensions
      considered one finds
      $\Sigma_Q > (\nabla A)^2 > {\cal E}_{U}\left[ \, g\, \right]$.
\item These inequalities become stronger as the lattice side $N$ increases.
\item As found in Ref.\ \cite{CM}, the Fourier acceleration
      method is very efficient in relaxing $\Sigma_Q$ and in this
      case one finds $\Sigma_Q \gtapprox (\nabla A)^2$.
\item The ratio $\Sigma_Q / (\nabla A)^2$ is usually smaller (or much
      smaller) for the even/odd update than for the lexicographic update.
\item If one uses the quantity $\Sigma_Q$ to check the convergence
      of the gauge fixing, then the stochastic overrelaxation
      algorithm is better than the other local algorithms when
      considering the lexicographic update, in agreement with our
      findings in Ref.\ \cite{CM}. On the contrary, if one considers
      the even/odd update, then the quality of the gauge fixing for
      the Cornell method becomes almost as
      good as for the stochastic overrelaxation update.
\item For the three improved local algorithms there is in general a
      gain in computational cost when using the even/odd update compared to the
      lexicographic update. For the Los Alamos method and the Fourier
      acceleration method the situation is reversed. (See next
      Section for a discussion of the Fourier
      acceleration method.)
\end{itemize}

We can conclude by saying that among the local algorithms the
best appears to be the Cornell method with even/odd update. In fact,
from the point of view of computational cost, this method is
equivalent to the overrelaxation algorithm and almost twice as fast
as the stochastic overrelaxation algorithm. At the same time, the quality
of the gauge fixing, especially when considering the relaxation of
the quantity $\Sigma_Q$, is better than what is obtained with the
overrelaxation algorithm and almost equivalent to the performance of
the stochastic overrelaxation algorithm. Let us recall that, in the
limit of large number of gauge-fixing sweeps $t$, the overrelaxation
method and the Cornell method coincide [see eqs.\ \reff{eq:bfvCor},
\reff{eq:bfvOve} and \reff{eq:omegaalpha}] at leading order in
$\epsilon$. Therefore, the different performances of the two
methods could be related to a difference in the behavior for small $t$.
This possibility is discussed in \cite[Section 7.1]{CM}
and we plan to investigate it further.


\subsection{Tuning of the Algorithms}
\label{InfinitoTun}

In this section we check numerically the analytic predictions for the
tuning of the three improved local algorithms and of the
Fourier acceleration method obtained in Sections \ref{Infinito}
and \ref{CSD}.

\vskip 0.3cm

\noindent {\bf Overrelaxation method: \hskip 1mm} In this case we have
the tuning condition [see eqs.\ \reff{eq:omegaC} and \reff{eq:Omegadis}]
\be
\omega_{opt}\, = \, \frac{2}{1 \, + \, C_{opt} / N }
\ee
with
\be
C_{opt} \,\approx\, \frac{2\,\pi}{d^{1 /2}}
\label{eq:cvalueeo}
\;\mbox{.}
\ee
In order to find the constant $C_{opt}$ one can write
\be
C_{opt} \,=\, N \;
\frac{2\,-\,\omega_{opt}}{\omega_{opt}} 
\label{eq:Copt}
\;\mbox{,}
\ee
which can be use to fit the numerical data. In this way,
considering the optimal choice of
$\omega$ for the three largest lattice sides $N$, we find for the
lexicographic update $C_{opt} = 5.0 \pm 0.2$ in two dimensions,
$4.26 \pm 0.08$ in three dimensions and
$3.65 \pm 0.05$ in four dimensions. 
The same fitting procedure gives $C_{opt} = 4.01 \pm 0.03$ in two dimensions,
$3.63 \pm 0.04$ in three dimensions and
$3.13 \pm 0.02$ in four dimensions for the even/odd update.
Notice that from eq.\ \reff{eq:cvalueeo} above, which is valid for the 
even/odd update, we have the analytic predictions
$C_{opt} \,\approx\,4.44$
for $d = 2$, $C_{opt} \,\approx\, 3.63$ for $d = 3$ and $C_{opt} \,\approx\,3.14$
for $d = 4$, in good agreement with our numerical results.

\vskip 0.3cm

\noindent {\bf Cornell method: \hskip 1mm} In Ref.\ \cite{CM} we have
found the relation
\be
\omega_{opt}\,=\,
\alpha_{opt}\,\langle\,{\cal N}\,\rangle
\,=\,\alpha_{opt}\,
2d \,(1\,-\,\langle\,{\cal E}_{min}\,\rangle)
\label{eq:omegalfa}
\;\mbox{,}
\ee
where $\langle\, {\cal E}_{min}\,\rangle$ is the average value of the 
minimizing functional at the minimum. At $\beta = \infty$ one
has $\langle\, {\cal E}_{min}\,\rangle = 0$ and the previous relation
becomes $\omega_{opt}\,=\, 2\,d\,\alpha_{opt}$, in agreement with
the analysis presented in Section \ref{Infinito}
[see eqs.\ \reff{eq:bfvCor}, \reff{eq:bfvOve} and 
\reff{eq:omegaalpha}]. From Tables \ref{Table.Cornell} and \ref{Table.Overrel}
one can check that this relation is very well verified by our data
in the two-, three- and four-dimensional cases.

\vskip 0.3cm

\noindent {\bf Stochastic overrelaxation method: \hskip 1mm} In Section
\ref{Stoch} we have seen that the tuning condition for the
stochastic overrelaxation algorithm is given by [see eqs.\
\reff{eq:defPdep} and \reff{eq:Ptun}]
\be
p \,=\,\frac{1\,-\,P}{1\,+\,P}
\ee
with
\be
P\,\approx\,\frac{2\,\pi}{\sqrt{d}\, N}
\;\mbox{.}
\ee
Moreover, by comparison with the overrelaxation algorithm, one
can write $p \approx \omega - 1$. It is immediate to check that this
relation is indeed verified by our data (see Tables
\ref{Table.Overrel} and \ref{Table.Stochas}).

\vskip 0.3cm

\noindent {\bf Fourier acceleration: \hskip 1mm} We did not discuss
the tuning of this algorithm in Ref.\ \cite{CM}.
The theoretical analysis in Section \ref{Infinito} gives
the simple result 
\be
\alpha_{opt} \, = \, 1
\label{eq:alpha1}
\ee
for any dimension $d$.  In Ref.\ \cite{CM} we obtained --- in two
dimensions, at finite $\beta$ and in the limit of large lattice
sides $N$ --- the value $\alpha_{opt} \approx 1.28$, in qualitative
agreement with \reff{eq:alpha1}. The agreement is better, as expected,
at $\beta = \infty$. In fact, using the lexicographic update we find
that $\alpha_{opt} = 1$ for any lattice side $N$ and dimension $d$.
On the contrary, using the even/odd update we have $\alpha_{opt}
\approx 1.1$, with a slow decrease of $\alpha_{opt}$ as $N$ increases.
Also note that $n_{gf}$ in the lexicographic case is about
two times smaller than for the even/odd update.
Thus, for the Fourier acceleration method, the even/odd update
does not help the convergence of the algorithm. This result
can be understood if one checks the size of the
Fourier components of the lattice divergence $\nabla\cdot A$. In
particular, one can check that with the even/odd update
the slowest relaxing mode corresponds to the shortest wavelength.
It is this mode that makes the Fourier acceleration method
perform worse in the even/odd case. This is related to the fact
that the even/odd update couples the low- and high-frequency modes
\cite{HK}.


\subsection{Overrelaxation and Stochastic Overrelaxation
without Tuning}
\label{sec:notun}

In order to check the analytic predictions
presented in Sections \ref{Overr} and \ref{Stoch},
we studied numerically the performance of the overrelaxation and of
the stochastic overrelaxation algorithms also in the case
without tuning. To this end, we did tests in
the two dimensional case (at $\beta = \infty$ with lattice
sides $N = 16\mbox{,}\; 32\mbox{,}\;48 \mbox{,}\,\ldots
\mbox{,}\;128$). For both algorithms we considered the two
limiting cases studied analytically, namely:
\begin{itemize}
\item for the overrelaxation algorithm we used $\omega = 1.98$
      (corresponding to the small value $\Omega \approx 0.01$) and
      $\omega = 1.3$ (corresponding to $\Omega \approx 0.54$);
\item for the stochastic overrelaxation we set $p = 0.96$ (namely
      $P \approx 0.02$) and $p = 0.3$ (corresponding to
      $P \approx 0.54$).
\end{itemize}

As a result we got that, for small values of $\Omega$
(respectively $P$),  $\tau$ and $n_{gf}$ are indeed constant
in $N$. In particular, for the overrelaxation algorithm, we
found $\tau \approx 50$ and $n_{gf}$ of the order of
$1500-1600$ using the lexicographic update and
$\tau \approx 25$ and $n_{gf} \approx 800$ using the
even/odd update. For the stochastic overrelaxation we have 
$\tau \approx 30$ and $n_{gf}$ of the order of
$1100$ using the lexicographic update and
$\tau \approx 25$ and $n_{gf} \approx 900$ using
the even/odd update.

For large values of $\Omega$ (respectively $P$), the relaxation
time $\tau$ is well fitted (for both algorithms and with both
types of update) by $\tau \approx 0.0136 \,N^z$ with $z \approx 2$. 
We have also checked that in this case these two algorithms
(without tuning) are better than the Los Alamos method,
showing a relaxation time about two times smaller.


\subsection{Alternative Implementations of the Fourier
Acceleration Method}
\label{sec:fourierver}

In Ref.\ \cite{MGFFT} we introduced a new implementation
of the Fourier acceleration method, in which the inversion
of the Laplacian is done using a multigrid algorithm,
avoiding the use of the fast Fourier transform. This makes
the method more flexible, i.e.\ it can efficiently work with
any lattice side $N$ and not only with $N$ equals to a power of $2$.
Moreover, the new implementation is well suited for vector
and parallel machines. In particular, we checked that the
computational cost shows a linear speedup with the number of
processors on an APE100 machine for the four-dimensional $SU(2)$
case. In that article, we have also implemented a version of the
method using conjugate gradient instead of multigrid, leading to
an algorithm that is efficient at intermediate lattice volumes.

In this Section we want to compare the performance of the Fourier
acceleration method based on a fast Fourier transform (FFT) to
the performance of the alternative implementations based on
multigrid (MG) or conjugate gradient (CG) algorithms.
To this end, let us recall that, even though the overhead
for the MG or the CG routine is likely to be larger
than the one for FFT, one can hope to reduce it by
exploiting the fact that multigrid and conjugate gradient
(as opposed to FFT) are iterative methods. In particular,
by changing the
stopping criterion for the inversion, the accuracy of the solution
can be suitably varied, while for FFT the accuracy is fixed by the
precision used in the numerical code. This is important, since
the tuning of the parameter $\alpha$ is usually done
only up to an accuracy of a few percent. Thus, the inversion of the
Laplacian most likely will not require the high
accuracy employed in the FFT case, making possible a substantial
reduction of the computational cost. We checked
in Ref.\ \cite{MGFFT} that this is indeed the case and we found
that, with an accuracy of about $10^{-5}$ for the inversion,
one obtains an algorithm equivalent to the original one
(based on FFT). For the present paper we checked again this
result and found a small bug in the code previously used.
After correcting it we got that the
accuracy necessary for the inversion is about $10^{-3}$,
yielding a substantial gain with respect to the old result
and in agreement with the intuitive argument above.
In order to compare different
implementations of the Fourier acceleration method
we used here the stopping criterion
\be
\frac{r_t}{r_0} \,\leq \,10^{-3}
\;\mbox{,}
\ee
where $r_t$ is the magnitude of the residual after $t$
iterations. Recall that we want to solve the equation
\be
-\Delta\;\phi^c(\bx) \,=\, \left(\nabla\cdot A\right)^c(\bx)
\;\mbox{,}
\ee
where $\phi^c(\bx)$ is the desired solution.
Then, the residual is defined by
\be
r^c(\bx) \,\equiv \,\left(\nabla\cdot A\right)^c(\bx) \,+\Delta\;
\phi^c(\bx)
\;\mbox{.}
\label{eq:residual}
\ee
In particular, we have tested six different implementations
of the Fourier acceleration method described
below, in addition to the original version (denoted by FFT-FA), based
on FFT (working in single precision).
\begin{enumerate}
\item{MG-FA:} The inversion of the Laplacian is done using MG in
              single precision; as in Ref.\ \cite{MGFFT}, we
              used a W-cycle with 2 Gauss-Seidel
              sweeps before coarsening and 2 after coarsening.
\item{MGCG-FA: } Same as the previous algorithm, but with a CG iteration
                 applied on the coarsest grid instead of the
                 Gauss-Seidel iteration. This should allow
                 larger coarsest grids, which may be useful
                 if one wants to parallelize the code.
\item{MGCGEO-FA: } Same as the previous algorithm, but using a CG
                   iteration on the coarsest grid
                   with even/odd preconditioning.
\item{CG-FA: } The inversion of the Laplacian
               is done using CG in single precision
               and in the stopping criterion we compute
               explicitly the
               residual defined in eq.\ \reff{eq:residual}.
\item{CGr-FA: } Same as the previous algorithm, but now in the
                stopping criterion we use the magnitude of the
                residual vector built by the CG method. Note
                that a CG method usually stores three vectors
                at each step: the approximate solution $x$,
                its residual $r$ and a search direction $p$.
\item{CGrEO-FA: } Same as the previous algorithm, but now the CG
                  is done with even/odd preconditioning.
\end{enumerate}

We used these six algorithms for numerical tests at $\beta = \infty$
in two, three and four dimensions. For the tuning parameter $\alpha$
we used the optimal choice obtained for the FFT-FA algorithm
(see Table \ref{Table.Fourier}). We found that the three methods
using MG are practically equivalent, with MG-FA and MGCG-FA
slightly faster than MGCGEO-FA. On the contrary, among the algorithms
using CG, the last one, namely CGrEO-FA, is always faster than the
other two, but still slower than MG-FA and MGCG-FA. If we
compare these algorithms to the original FFT-FA we see that
because of the use of FFT the algorithm gets progressively worse
as the lattice dimension $d$ increases. In particular, we get that
in three and in four dimensions the MG-FA algorithm is already
faster for lattice sizes $32^3$ and $16^4$ and the gain is larger
when considering the even/odd update.\footnote{~Note that
when using the Fourier acceleration method at finite $\beta$
one should avoid updating simultaneously all sites of the
lattice, since the resulting move in configuration space might
be too large and affect the convergence of the method.}
The CGrEO-FA algorithm is
essentially equivalent to FFT-FA in four dimensions for lattices
$16^4$ or larger.

\begin{figure}[t]
\vspace*{-3cm} \hspace*{-0cm}
\begin{center}
\epsfxsize = 6.0in
\leavevmode\epsffile{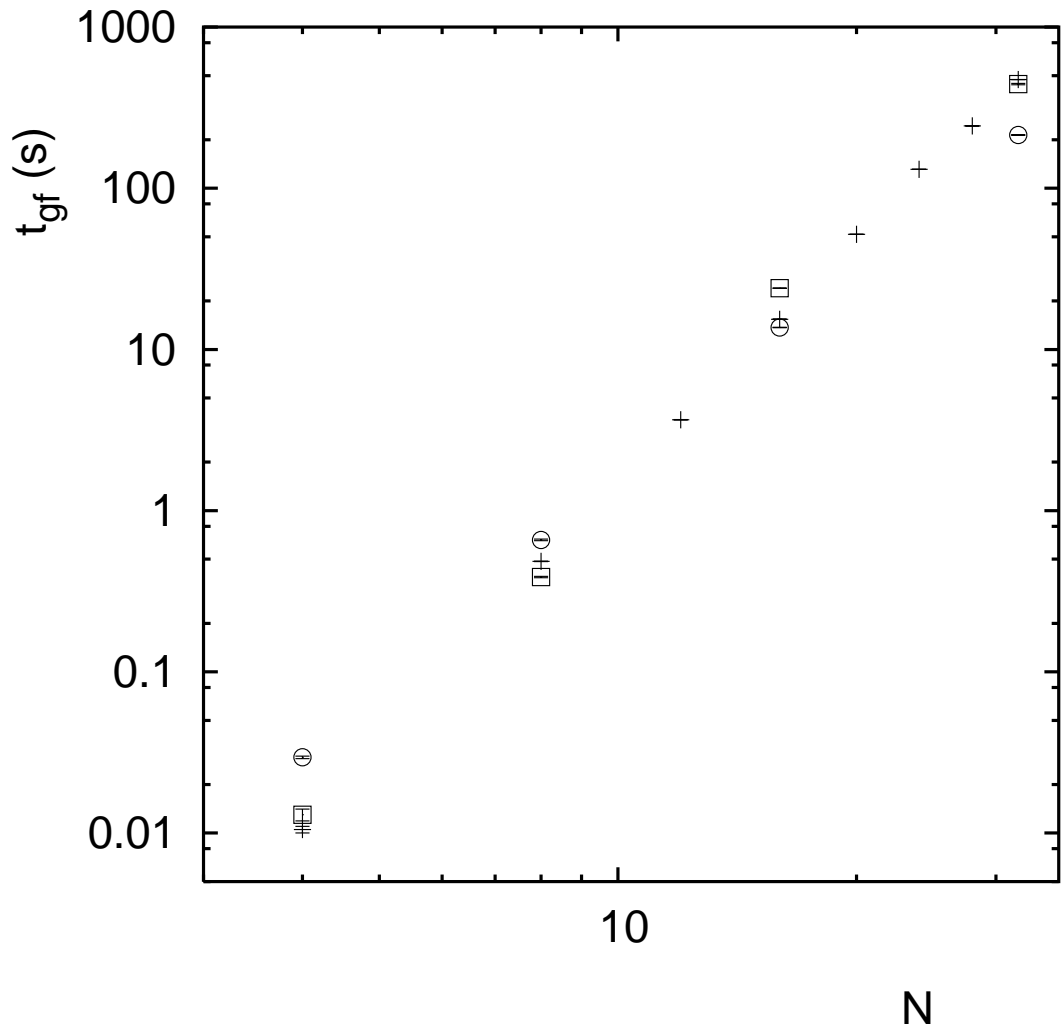} \\
\end{center}
\caption{~Plot of the time $t_{gf}$ (in seconds) used
         to complete the gauge fixing as a function of the
         lattice side $N$ for the Cornell method
         ($+$), the FFT-FA algorithm ($\Box$) and the
         MG-FA algorithm ($\bigcirc$) in
         four dimensions and considering even/odd update.}
\label{fig:tempo}
\end{figure}

Finally, we compared the computational cost of the
FFT-FA and the MG-FA algorithms with the best
among the improved local methods, namely the Cornell method. In
Fig.\ \ref{fig:tempo} we plot the CPU time needed to gauge-fix a
configuration using these three methods in the four-dimensional
case using even/odd update.
We can see that the MG-FA method is the fastest already
for the lattice volume $16^4$. This happens in three dimensions
at $32^3$.
We note that this analysis is very machine- and code-dependent,
and that the Fourier acceleration methods are
particularly well suited for the case $\beta = \infty$.
As we noted in Ref.\ \cite{CMLat96}, the performance of the
FA method is poor for small $\beta$, reaching $z = 1$ at 
$\beta = 0$. We are currently investigating this matter.


\section{$\Lambda$ Gauges}
\label{Lambda}

The analytic study presented in Sections \ref{sec:gf},
\ref{Infinito} and \ref{CSD} for the Landau-gauge
minimizing functional at $\beta = \infty$ --- 
namely when all the link variables $U_{\mu}(\bx)$ are
equal to the identity matrix $\1$ --- 
can be easily extended to the so-called
$\lambda$-gauges, which have been recently used in several
analytic \cite{BD} and numerical articles \cite{lambda,lambdaworks}.

To this end, let us recall that a general $\lambda$-gauge can be
defined by considering the minimizing functional
\be
{\cal E}_{U, \lambda}\left[ g \right] \, \equiv \,
   1 \,-\, \frac{1}{V\, \left(\sum_{\mu = 1}^d \lambda_{\mu} \right)}
         \sum_{\bx} \sum_{\mu = 1}^{d} \,
         \lambda_{\mu} \,
         \frac{\Tr}{2} \left[ \;
             g(\bx) \; U_{\mu}(\bx) \;
             g^{\dagger}(\bx + \bun_{\mu})
                        \; \right]
\label{eq:spinglambda}
\;\mbox{.}
\ee
Clearly, if $\lambda_{\mu} = 1$ for all $\mu$ we get back
the standard Landau-gauge minimizing functional given in
eq.\ \reff{eq:spingl}. Also, if $\lambda_i = 1$
for $i = 1\mbox{,}\, 2\mbox{,}\, \ldots\mbox{,}\, d-1$, then
we can interpolate \cite{BD}
between the Landau and the Coulomb gauge
by varying $\lambda_d$ between 1 and 0.

One can easily redo all the analysis in Sections \ref{sec:gf}
and \ref{Infinito} and observe that all the formulae are still
valid if we make the substitutions
\be
\sum_{\mu = 1}^{d} \,\to\, \sum_{\mu = 1}^{d} \lambda_{\mu}
\ee
and
\be
d \,\to\, \sum_{\mu = 1}^{d} \lambda_{\mu}
\;\mbox{.}
\ee
In particular we have
\be
h(\bx\mbox{,}\, \lambda) \equiv
        \sum_{\mu = 1}^{d} \lambda_{\mu}
            \left[ \; U_{\mu}(\bx) \, g^{\dagger}(\bx + \bun_{\mu}) +
        U_{\mu}^{\dagger}(\bx - \bun_{\mu}) \,
        g^{\dagger}(\bx - \bun_{\mu}) \; \right]
\label{eq:hdefilambda}
\ee
and, after setting $U_{\mu}(\bx) = \1$,
\be
{\cal E}_{U, \lambda}\left[ g \right] \, = \, 
\frac{1}{2\,V\,\left(\sum_{\mu = 1}^d \lambda_{\mu} \right)}
\sum_{\bx}\, \frac{\Tr}{2} \, 
\sum_{\mu = 1}^{d} \lambda_{\mu}
   \left\{ \left[ \, g(\bx)\,-\,g(\bx + \bun_{\mu})
                             \,\right]\,
\left[ \, g(\bx)\,-\,g(\bx + \bun_{\mu})\,\right]^{\dagger}
\right\} \label{E1lambda}
\;\mbox{.}
\ee
Also, the Laplacian $\Delta$ becomes a $\lambda$-Laplacian defined
by the relation
\be
\left(\,-\,\Delta_{\lambda} \,\bfv\,\right)(\bx) \equiv
\sum_{\mu = 1}^{d} \, \lambda_{\mu} \, \left[ \;2\,\bfv(\bx)\,-
\,\bfv(\bx + \bun_{\mu}) -
\bfv(\bx - \bun_{\mu}) \; \right]
\label{eq:laplacianolambda}
\;\mbox{,}
\ee
with eigenvalues in momentum space given by
\be
p^{2}_{\lambda}(\bk) \equiv 4 \, \sum_{\mu = 1}^{d} \,
\lambda_{\mu}\,
\sin^{2}\left( \, \pi \,k_{\mu} \, \right)
\label{eq:p2deflambda}
\;\mbox{.}
\ee
Thus, the Fourier acceleration method is now a $\lambda$-Laplacian
preconditioning, i.e.\
\be
\bvu(\bx) \,\equiv\, \left\{ {\widehat F}^{-1}\left[
         \, \frac{1}{p^{2}_{\lambda}(\bk)} \,
                {\widehat F} \bvw \right] \right\} (\bx)
\label{eq:bvulambda}
\; \mbox{.}
\ee
Finally, by using eq.\ \reff{eq:gI}, we can rewrite the minimizing
functional \reff{E1lambda} as 
\be
{\cal E}_{\lambda}\left[ f \right]
\, = \, \frac{\epsilon^2}{2\,V\,
\left(\sum_{\mu = 1}^d \lambda_{\mu} \right)}
\sum_{\bx}\,\bfv(\bx)\cdot
\left(\,- \, \Delta_{\lambda}\,\bfv\,\right)(\bx)\; +\,{\cal O}(\epsilon^3)
\;\mbox{,}
\label{E2terlambda}
\ee
which implies
\ba
\!\!\!\!\!\!\!h(\bx)\! &\!=\!&\!
        2\,\left(\sum_{\mu = 1}^d \lambda_{\mu} \right)
         \,\1\,-\,i\,\epsilon\,\bsig\cdot\,
        \sum_{\mu = 1}^{d} \lambda_{\mu}\,
            \left[ \; \bfv(\bx + \bun_{\mu}) +
        \bfv(\bx - \bun_{\mu}) \; \right]\,+\,{\cal O}(\epsilon^2)
\label{eq:happroxlambda} \\[0.2cm]
\!\!\!\!\!\!\!{\mathrm w}(\bx)\! &\!\!=\!\!&\!
       \, 2\,\left(\sum_{\mu = 1}^d \lambda_{\mu} \right)
         \,\1\,+\,i\,\epsilon\,\bsig\cdot\,
        \sum_{\mu = 1}^{d} \lambda_{\mu}\,
            \left[ \;2\,\bfv(\bx)\,-\, \bfv(\bx + \bun_{\mu})
                \,-\, \bfv(\bx - \bun_{\mu}) \;
                  \right]\,+\,{\cal O}(\epsilon^2)
\qquad
\ea
and
\ba
{\cal N}(\bx)\! & = &\! \, 2\,\left(\sum_{\mu = 1}^d \lambda_{\mu} \right)
          \,+\,{\cal O}(\epsilon^2)
\label{eq:Ndlambda} \\[0.3cm]
\bvw(\bx)\! & = & \! \,\epsilon\,\left(\,- \Delta_{\lambda}
            \,\bfv\,\right)(\bx) \,+\,{\cal O}(\epsilon^2)
\;\mbox{.}
\ea
Then, the update for the five gauge-fixing algorithms can be
written as
\ba
\bfv^{(LosAl.)}(\bx) &=& \frac{1}{2\,\left(\sum_{\mu = 1}^d
       \lambda_{\mu} \right)}\,
        \sum_{\mu = 1}^{d} \lambda_{\mu} \left[ \; \bfv(\bx + \bun_{\mu}) +
        \bfv(\bx - \bun_{\mu}) \; \right]
\label{eq:LOSbfvlambda}
\\[0.3cm] 
\bfv^{(cornell)}(\bx) &=& 2\,
           \left(\sum_{\mu = 1}^d \lambda_{\mu} \right)
                  \,\alpha\, \bfv^{(LosAl.)}(\bx)
             \,+\, \left[\,1\,-\,2\,
           \left(\sum_{\mu = 1}^d \lambda_{\mu} \right)\,\alpha\,\right]
        \,\bfv(\bx) \qquad
\label{eq:bfvCorlambda}
\\[0.3cm]
\bfv^{(over)}(\bx) &=& \omega\,\bfv^{(LosAl.)}(\bx) \,+\,
          \left(\,1\,-\,\omega\,\right) \,\bfv(\bx)
\label{eq:bfvOvelambda}
\\[0.3cm]
\bfv^{(stoc)}(\bx) &=&
   \left\{ \begin{array}{ll}
          2\,\bfv^{(LosAl.)}(\bx) \, - \, \bfv(\bx) &
                   \quad \mbox{with probability $p$} \\
      \phantom{ } & \phantom{ } \\ \bfv^{(LosAl.)}(\bx) &
                   \quad \mbox{with probability $1 - p$}
        \end{array} \right.
\label{eq:bfvstoclambda}
\\[0.3cm]
\bfv^{(Fourier)}(\bx) &=& \left(\,1\,-\,\alpha\,\right)\,\bfv(\bx)
\;\mbox{,}
\label{eq:FOUbflambdav}
\ea
and all the observations reported at the end of Section
\ref{Infinito} still apply, including the relation
\be
\omega\,=\,\alpha\,{\cal N}(\bx)\, + {\cal O}(\epsilon^2)
\label{eq:omegaalpha2}
\;\mbox{.}
\ee

The formulae in Section \ref{CSD} are also unchanged, with the
only exception of eq.\ \reff{eq:sdefini}, which now becomes
\be
c(\bk\mbox{,}\,\lambda)\equiv
\frac{1}{2} \left[ \frac{1}{\left(\sum_{\mu = 1}^d
\lambda_{\mu} \right)}\,
            \sum_{\mu = 1}^{d} \lambda_{\mu}\,
              \,\cos{(\,2\,\pi\,k_{\mu}\,)} \right]
\,=\,
\frac{1}{2} \left[\,1\,-\,\frac{p^2_{\lambda}(\bk)^2}{2
             \left(\sum_{\mu = 1}^d \lambda_{\mu} \right)}
\,\right]
\label{eq:sdefinilambda}
\;\mbox{.}
\ee
For the smallest non-zero momentum, in the limit
of large lattice side $N$, this gives\footnote{~One can check
that this relation holds also
for the largest momentum if one considers
the absolute value of $c(k(N)\mbox{,}\,\lambda)$.}
\be
c(k(N)\mbox{,}\,\lambda)\,\approx\,
   \frac{1}{2} \left[\,1\,-\,\frac{2 \,\lambda\,\pi^2}{
   \left(\sum_{\mu = 1}^d \lambda_{\mu} \right) \,N^2} \,\right]
\,\equiv\,\frac{1}{2} \left[\,1\,-\,\zeta(N\mbox{,}\,\lambda) \,\right]
\label{cdiNelambda2}
\;\mbox{,}
\ee
where $\lambda = \min_{\mu} \lambda_{\mu}$.
By using eq.\ \reff{eq:slargeN2} we can write
\be
\zeta(N\mbox{,}\,\lambda) \,=\, \frac{\zeta(N)}{f(\lambda)}
\ee
with
\be
f(\lambda) \equiv \frac{\sum_{\mu = 1}^d \lambda_{\mu}}{
      d \, \lambda}
\;\mbox{.}
\ee
Note that eq.\ \reff{cdiNelambda2}
is still valid in the case of asymmetric lattices if we set
\be
\frac{\lambda}{N^2} \,=\, \min_{\mu}\, \frac{\lambda_{\mu}}{
                             N_{\mu}^2}
\;\mbox{.}
\ee

It follows that the analysis of CSD and, when necessary, of
the tuning of the local algorithms considered here is modified in
the following way:
\ba
\tau_{Los Alamos}(\lambda)
  &\approx& \frac{d\,N^2}{4\,\pi^{2}}\,f(\lambda)
\,=\, \tau_{Los Alamos} \, f(\lambda)
\label{eq:tauLOSlambda}
\\[0.2cm]
\tau_{over}(\lambda) & \approx & \frac{1}{2\,\Omega} \,
       \approx \, \frac{\sqrt{d} \, N}{4\,\pi}
       \,\sqrt{f(\lambda)}
\,=\, \tau_{over} \, \sqrt{f(\lambda)}
\label{eq:tauoverlambda}
\\[0.2cm]
\tau_{stoc}(\lambda) & \approx & \frac{1}{P} \,
       \approx \,
    \frac{\sqrt{d}\,N}{2\,\pi}\, \sqrt{f(\lambda)}
\,=\, \tau_{stoc} \, \sqrt{f(\lambda)}
\;\mbox{.}
\label{eq:taustoclambda}
\ea
Clearly, the relation $\,p\,\approx\,\omega - 1\,$ is still valid.
Also, if we fix the lattice sides $N$, the quantity
$f(\lambda)$ --- and therefore the relaxation
times of these algorithms --- increases (respectively decreases)
by decreasing (respectively increasing) the value of $\lambda$.
This confirms the results obtained numerically
in Ref.\ \cite{lambda} for the cases $\lambda_i = 1$
for $i = 1\mbox{,}\, 2\mbox{,}\, 3$, and
$\lambda_4 = 1$ and $0.5$.

\vskip 3mm


In order to verify these results
we have done numerical tests in the two-dimensional case
with $\lambda_1 = 1$ and $\lambda_2 = 0.25$ for the
lattice sides $N = 16\mbox{,}\, 32\mbox{,}\, 48\mbox{,}\, \ldots
\mbox{,}\, 128$. This choice of $\lambda$'s gives
${\cal N}(\bx) \approx 1.25$ and
$f(\lambda) = 2.5$.
The data are reported in Tables
\ref{Table.LosAlamos2dlambda}--\ref{Table.Fourier2dlambda}.
[Again we don't show the statistical error since it is
usually very small.]
By comparison with the data obtained in Landau gauge and
reported in Tables
\ref{Table.LosAlamos}--\ref{Table.Fourier},
one can easily check the relations for the relaxation time
$\tau$ given in eqs.\ 
\reff{eq:tauLOSlambda}--\reff{eq:taustoclambda}
above. Also note that
eq.\ \reff{eq:omegaalpha2} and the relation
$\,p\,\approx\,\omega - 1\,$ are very well satisfied by
our data.


\section{Conclusions}
\label{Concl}
 
We studied numerically and analytically five gauge-fixing
algorithms in $SU(2)$ lattice gauge theory by considering the
case $\beta = \infty$, for Landau gauge and $\lambda$-gauges.
The analysis has been done for general dimension $d$ and
numerical checks were carried out at $d = 2\mbox{,}\,3$ and $4$.
Results are in agreement with those obtained
previously in Landau gauge
at finite $\beta$ in two dimensions \cite{CM}. In fact, we find that
the (local) Los Alamos method
has dynamic critical exponent $z \approx 2$, the three
improved local methods we considered --- the
overrelaxation method, the stochastic overrelaxation method
and the so-called Cornell method --- have critical exponent
$z \approx 1$, and the global method of Fourier acceleration
completely eliminates critical slowing-down. 

As said in the Introduction, if the system does not
undergo a phase transition going from $\beta = 0$ to
$\beta = \infty$, then the dynamic critical exponent $z$
{\bf should not} depend on the constant physics, i.e.\
it should be the same at finite $\beta$
and at $\beta = \infty$. On the contrary, the constant $c$
obtained from the fit $\,\tau = c\,N^z\,$ should be
different in the two cases and one expects
\be
c ( \beta = \infty )\; < \; c ( \,\mbox{finite}\, \beta)
\;\mbox{.}
\ee
To make this comparison simpler, we report
in Table \ref{Table.ccomparison}
the values obtained in Ref.\ \cite{CM} of the
constant $c$ for the five gauge-fixing algorithms at finite
$\beta$ and the results of the fits done for the same
algorithms at $\beta = \infty$. (In both cases we consider
the lexicographic update.) From the data it is clear that
the constant $c$ satisfies very well the
above inequality for the five algorithms.

Our numerical simulations show that the Cornell
method with even/odd
update is the best among the local algorithms.
It is very fast and at the same time
effective in relaxing the value of the quantity $\Sigma_Q$.
It would be interesting to check if this is true also
for finite $\beta$ and for the $SU(3)$ case.
As already observed in Ref.\ \cite{CM},
among the local algorithms one should choose
the stochastic overrelaxation method if the lexicographic
update is considered.
Finally, as expected, the Fourier acceleration method is extremely efficient
at $\beta = \infty$ and we checked that its implementation
can be improved by inverting the lattice Laplacian
using a MG algorithm.

The theoretical analysis, valid for any dimension $d$,
helped us clarify the tuning of these algorithms. In
particular, the relations between the parameter $\omega$
of the overrelaxation, the parameter $\alpha$ of the
Cornell method and the parameter $p$ of the
stochastic overrelaxation method simplify
the tuning and confirm nicely the expressions
obtained numerically in Ref.\ \cite{CM}.
For the Fourier acceleration
method we found analytically the tuning condition
$\alpha = 1$. This result is well verified numerically
at $\beta = \infty$ and at finite $\beta$ (see Ref.\
\cite{CM}).

We also studied generalizations of the overrelaxation
and of the stochastic overrelaxation algorithms.
In particular, following a
suggestion in \cite{BN}, we considered explicitly a local
algorithm (similar to overrelaxation) corresponding to
an updating matrix of size larger than $2 \times 2$
(i.e.\ $4 \times 4$). In all cases, we verified that one
cannot have a dynamic critical exponent $z$ smaller than 1 with
these local algorithms.

To sum up, in this work we have done a careful analysis
of CSD for the problem of numerical
gauge fixing in Landau gauge --- for the $SU(2)$ group and
in $d = 2, 3$ and $4$ dimensions ---
in a case that can be studied analytically,
i.e.\ $\beta = \infty$.
This study has provided several analytic predictions, which we
verified numerically. We note that at $\beta = \infty$ these
results clearly apply also to the problem of numerical gauge
fixing in Coulomb gauge.
We believe that these predictions can be very useful
in the investigation of
more realistic (i.e.\ finite) values of $\beta$ as
well as for $\beta = 0$ and in the extension of this analysis to
the general $SU(N)$ case.


\section*{Acknowledgments}

The authors' research is supported by FAPESP, Brazil
(Project No.\ 00/05047-5).



\clearpage

\clearpage

\begin{table}
\begin{center}
\addtolength{\tabcolsep}{3.0mm}
\vspace*{-0.9cm}
\protect\footnotesize
\begin{tabular}{|| c | c | c | c ||}
\hline
\hline
$ V=N^d $ & $ \tau $ & $n_{gf}$ & $ t_{gf} $ \\
\hline
\hline
$  16^2 $ & $  6.72 $ & $  177.3 $ & $  0.055 $ \\ \hline 
$  32^2 $ & $ 26.23 $ & $  587.7 $ & $  0.708 $ \\ \hline 
$  48^2 $ & $ 58.66 $ & $ 1149.3 $ & $  3.100 $ \\ \hline 
$  64^2 $ & $ 104.07 $ & $ 1868.2 $ & $ 11.200 $ \\ \hline 
$  80^2 $ & $ 162.43 $ & $ 2672.9 $ & $ 31.400 $ \\ \hline 
$  96^2 $ & $ 233.76 $ & $ 3661.2 $ & $ 61.800 $ \\ \hline 
$ 112^2 $ & $ 318.06 $ & $ 4665.9 $ & $ 106.000 $ \\ \hline 
$ 128^2 $ & $ 415.33 $ & $ 5792.1 $ & $ 170.000 $ \\ \hline 
\hline
$   8^3 $ & $  2.59 $ & $   75.5 $ & $  0.067 $ \\ \hline 
$  16^3 $ & $  9.95 $ & $  238.5 $ & $  2.580 $ \\ \hline 
$  24^3 $ & $ 22.13 $ & $  467.6 $ & $ 17.800 $ \\ \hline 
$  32^3 $ & $ 39.16 $ & $  750.7 $ & $ 68.200 $ \\ \hline 
$  40^3 $ & $ 61.05 $ & $ 1067.2 $ & $ 194.000 $ \\ \hline 
$  48^3 $ & $ 87.81 $ & $ 1420.8 $ & $ 467.000 $ \\ \hline 
$  56^3 $ & $ 119.42 $ & $ 1810.5 $ & $ 969.000 $ \\ \hline 
$  64^3 $ & $ 155.90 $ & $ 2214.9 $ & $ 1780.000 $ \\ \hline 
\hline
$   4^4 $ & $  0.92 $ & $   30.9 $ & $  0.018 $ \\ \hline 
$   8^4 $ & $  3.42 $ & $   93.1 $ & $  1.400 $ \\ \hline 
$  12^4 $ & $  7.50 $ & $  179.4 $ & $ 14.700 $ \\ \hline 
$  16^4 $ & $ 13.20 $ & $  283.4 $ & $ 74.300 $ \\ \hline 
$  20^4 $ & $ 20.51 $ & $  404.5 $ & $ 259.000 $ \\ \hline 
$  24^4 $ & $ 29.43 $ & $  538.0 $ & $ 725.000 $ \\ \hline 
$  28^4 $ & $ 39.97 $ & $  680.9 $ & $ 1670.000 $ \\ \hline 
$  32^4 $ & $ 52.14 $ & $  822.4 $ & $ 3560.000 $ \\ \hline 
\hline
$  16^2 $ & $  6.44 $ & $  178.4 $ & $  0.057 $ \\ \hline 
$  32^2 $ & $ 25.90 $ & $  609.9 $ & $  0.764 $ \\ \hline 
$  48^2 $ & $ 58.32 $ & $ 1224.9 $ & $  3.990 $ \\ \hline 
$  64^2 $ & $ 103.71 $ & $ 2009.4 $ & $ 13.100 $ \\ \hline 
$  80^2 $ & $ 162.07 $ & $ 2910.7 $ & $ 39.400 $ \\ \hline 
$  96^2 $ & $ 233.40 $ & $ 3978.9 $ & $ 83.100 $ \\ \hline 
$ 112^2 $ & $ 317.70 $ & $ 5080.8 $ & $ 148.000 $ \\ \hline 
$ 128^2 $ & $ 414.97 $ & $ 6348.2 $ & $ 240.000 $ \\ \hline 
\hline
$   8^3 $ & $  2.43 $ & $   73.6 $ & $  0.067 $ \\ \hline 
$  16^3 $ & $  9.73 $ & $  246.3 $ & $  3.080 $ \\ \hline 
$  24^3 $ & $ 21.89 $ & $  490.2 $ & $ 23.000 $ \\ \hline 
$  32^3 $ & $ 38.91 $ & $  799.9 $ & $ 88.700 $ \\ \hline 
$  40^3 $ & $ 60.79 $ & $ 1149.0 $ & $ 252.000 $ \\ \hline 
$  48^3 $ & $ 87.54 $ & $ 1555.1 $ & $ 599.000 $ \\ \hline 
$  56^3 $ & $ 119.15 $ & $ 1967.5 $ & $ 1220.000 $ \\ \hline 
$  64^3 $ & $ 155.63 $ & $ 2463.6 $ & $ 2290.000 $ \\ \hline 
\hline
$   4^4 $ & $  0.92 $ & $   30.1 $ & $  0.018 $ \\ \hline 
$   8^4 $ & $  3.29 $ & $   93.7 $ & $  1.700 $ \\ \hline 
$  12^4 $ & $  7.34 $ & $  185.3 $ & $ 18.300 $ \\ \hline 
$  16^4 $ & $ 13.01 $ & $  296.7 $ & $ 92.400 $ \\ \hline 
$  20^4 $ & $ 20.31 $ & $  427.6 $ & $ 325.000 $ \\ \hline 
$  24^4 $ & $ 29.22 $ & $  574.5 $ & $ 913.000 $ \\ \hline 
$  28^4 $ & $ 39.76 $ & $  732.5 $ & $ 2150.000 $ \\ \hline 
$  32^4 $ & $ 51.92 $ & $  892.4 $ & $ 4570.000 $ \\ \hline 
\hline 
\end{tabular}
\end{center}
\vspace*{-0.5cm}
\caption{~The relaxation time $\tau$, the number of sweeps $n_{gf}$
          and the time $t_{gf}$ (in seconds)
          for the Los Alamos method for $d = 2\mbox{,}\,
          3\mbox{,}$ and $4$ using
          lexicographic (upper rows) or even/odd (lower rows) update.}
\label{Table.LosAlamos}
\end{table}
\begin{table}
\begin{center}
\addtolength{\tabcolsep}{3.0mm}
\vspace*{-0.9cm}
\protect\footnotesize
\begin{tabular}{|| c | c | c | c | c ||}
\hline
\hline
$ V=N^d $ & $ \alpha $ & $ \tau $ & $n_{gf}$ & $ t_{gf} $ \\
\hline
\hline
$  16^2 $ & $ 0.410 $ & $  2.40 $ & $   73.5 $ & $  0.025 $ \\ \hline 
$  32^2 $ & $ 0.445 $ & $  4.27 $ & $  131.7 $ & $  0.175 $ \\ \hline 
$  48^2 $ & $ 0.460 $ & $  5.85 $ & $  186.7 $ & $  0.554 $ \\ \hline 
$  64^2 $ & $ 0.470 $ & $  7.70 $ & $  252.9 $ & $  1.800 $ \\ \hline 
$  80^2 $ & $ 0.475 $ & $  9.30 $ & $  306.4 $ & $  3.690 $ \\ \hline 
$  96^2 $ & $ 0.477 $ & $ 10.48 $ & $  342.4 $ & $  6.200 $ \\ \hline 
$ 112^2 $ & $ 0.479 $ & $ 11.70 $ & $  377.6 $ & $  9.200 $ \\ \hline 
$ 128^2 $ & $ 0.482 $ & $ 13.58 $ & $  443.3 $ & $ 13.800 $ \\ \hline 
\hline
$   8^3 $ & $ 0.235 $ & $  1.21 $ & $   38.4 $ & $  0.037 $ \\ \hline 
$  16^3 $ & $ 0.275 $ & $  2.30 $ & $   73.2 $ & $  0.835 $ \\ \hline 
$  24^3 $ & $ 0.290 $ & $  3.21 $ & $  105.0 $ & $  4.230 $ \\ \hline 
$  32^3 $ & $ 0.300 $ & $  4.26 $ & $  141.8 $ & $ 13.500 $ \\ \hline 
$  40^3 $ & $ 0.305 $ & $  5.12 $ & $  170.0 $ & $ 32.500 $ \\ \hline 
$  48^3 $ & $ 0.310 $ & $  6.33 $ & $  210.0 $ & $ 71.500 $ \\ \hline 
$  56^3 $ & $ 0.310 $ & $  6.51 $ & $  211.0 $ & $ 115.000 $ \\ \hline 
$  64^3 $ & $ 0.312 $ & $  7.30 $ & $  238.0 $ & $ 199.000 $ \\ \hline 
\hline
$   4^4 $ & $ 0.155 $ & $  0.56 $ & $   20.1 $ & $  0.013 $ \\ \hline 
$   8^4 $ & $ 0.185 $ & $  1.26 $ & $   41.0 $ & $  0.660 $ \\ \hline 
$  12^4 $ & $ 0.200 $ & $  1.82 $ & $   59.1 $ & $  5.120 $ \\ \hline 
$  16^4 $ & $ 0.210 $ & $  2.38 $ & $   79.0 $ & $ 21.900 $ \\ \hline 
$  20^4 $ & $ 0.215 $ & $  2.84 $ & $   93.0 $ & $ 62.800 $ \\ \hline 
$  24^4 $ & $ 0.220 $ & $  3.39 $ & $  112.0 $ & $ 158.000 $ \\ \hline 
$  28^4 $ & $ 0.225 $ & $  4.16 $ & $  138.0 $ & $ 348.000 $ \\ \hline 
$  32^4 $ & $ 0.225 $ & $  4.22 $ & $  139.0 $ & $ 622.000 $ \\ \hline 
\hline
$  16^2 $ & $ 0.405 $ & $  0.87 $ & $   36.0 $ & $  0.013 $ \\ \hline 
$  32^2 $ & $ 0.440 $ & $  1.89 $ & $   62.0 $ & $  0.086 $ \\ \hline 
$  48^2 $ & $ 0.460 $ & $  2.88 $ & $   97.0 $ & $  0.302 $ \\ \hline 
$  64^2 $ & $ 0.470 $ & $  3.91 $ & $  131.0 $ & $  0.906 $ \\ \hline 
$  80^2 $ & $ 0.474 $ & $  4.74 $ & $  154.0 $ & $  2.250 $ \\ \hline 
$  96^2 $ & $ 0.479 $ & $  5.88 $ & $  193.0 $ & $  4.470 $ \\ \hline 
$ 112^2 $ & $ 0.482 $ & $  6.69 $ & $  220.0 $ & $  6.760 $ \\ \hline 
$ 128^2 $ & $ 0.484 $ & $  7.79 $ & $  255.4 $ & $ 10.100 $ \\ \hline 
\hline
$   8^3 $ & $ 0.240 $ & $  0.55 $ & $   22.2 $ & $  0.022 $ \\ \hline 
$  16^3 $ & $ 0.275 $ & $  1.06 $ & $   41.0 $ & $  0.535 $ \\ \hline 
$  24^3 $ & $ 0.290 $ & $  1.67 $ & $   58.0 $ & $  2.860 $ \\ \hline 
$  32^3 $ & $ 0.300 $ & $  2.12 $ & $   78.0 $ & $  9.000 $ \\ \hline 
$  40^3 $ & $ 0.305 $ & $  3.04 $ & $   94.2 $ & $ 21.400 $ \\ \hline 
$  48^3 $ & $ 0.310 $ & $  3.23 $ & $  114.0 $ & $ 45.400 $ \\ \hline 
$  56^3 $ & $ 0.312 $ & $  3.94 $ & $  129.1 $ & $ 83.000 $ \\ \hline 
$  64^3 $ & $ 0.315 $ & $  4.42 $ & $  147.0 $ & $ 142.000 $ \\ \hline 
\hline
$   4^4 $ & $ 0.150 $ & $  0.29 $ & $   14.9 $ & $  0.011 $ \\ \hline 
$   8^4 $ & $ 0.185 $ & $  0.68 $ & $   25.0 $ & $  0.485 $ \\ \hline 
$  12^4 $ & $ 0.200 $ & $  0.96 $ & $   35.2 $ & $  3.660 $ \\ \hline 
$  16^4 $ & $ 0.210 $ & $  1.30 $ & $   46.5 $ & $ 15.400 $ \\ \hline 
$  20^4 $ & $ 0.220 $ & $  1.82 $ & $   65.0 $ & $ 51.800 $ \\ \hline 
$  24^4 $ & $ 0.225 $ & $  2.20 $ & $   79.0 $ & $ 131.000 $ \\ \hline 
$  28^4 $ & $ 0.225 $ & $  2.28 $ & $   79.0 $ & $ 244.000 $ \\ \hline 
$  32^4 $ & $ 0.228 $ & $  2.55 $ & $   89.0 $ & $ 473.000 $ \\ \hline 
\hline
\end{tabular}
\end{center}
\vspace*{-0.5cm}
\caption{~The tuning parameter $\alpha$, the relaxation time $\tau$,
          the number of sweeps $n_{gf}$
          and the time $t_{gf}$ (in seconds)
          for the Cornell method for $d = 2\mbox{,}\,
          3\mbox{,}$ and $4$ using
          lexicographic (upper rows) or even/odd (lower rows) update.}
\label{Table.Cornell}
\end{table}
\begin{table}
\begin{center}
\addtolength{\tabcolsep}{3.0mm}
\vspace*{-0.9cm}
\protect\footnotesize
\begin{tabular}{|| c | c | c | c | c ||}
\hline
\hline
$ V=N^d $ & $ \omega $ & $ \tau $ & $n_{gf}$ & $ t_{gf} $ \\
\hline
\hline
$  16^2 $ & $ 1.630 $ & $  2.27 $ & $   73.2 $ & $  0.026 $ \\ \hline 
$  32^2 $ & $ 1.780 $ & $  4.03 $ & $  135.8 $ & $  0.190 $ \\ \hline 
$  48^2 $ & $ 1.830 $ & $  5.69 $ & $  182.6 $ & $  0.642 $ \\ \hline 
$  64^2 $ & $ 1.870 $ & $  7.17 $ & $  242.0 $ & $  1.790 $ \\ \hline 
$  80^2 $ & $ 1.890 $ & $  8.59 $ & $  289.0 $ & $  3.680 $ \\ \hline 
$  96^2 $ & $ 1.900 $ & $ 10.24 $ & $  321.5 $ & $  5.990 $ \\ \hline 
$ 112^2 $ & $ 1.920 $ & $ 11.82 $ & $  401.7 $ & $ 10.100 $ \\ \hline 
$ 128^2 $ & $ 1.920 $ & $ 13.15 $ & $  405.4 $ & $ 13.100 $ \\ \hline 
\hline 
$   8^3 $ & $ 1.420 $ & $  1.17 $ & $   38.5 $ & $  0.037 $ \\ \hline 
$  16^3 $ & $ 1.640 $ & $  2.19 $ & $   73.5 $ & $  0.826 $ \\ \hline 
$  24^3 $ & $ 1.730 $ & $  3.14 $ & $  105.0 $ & $  4.290 $ \\ \hline 
$  32^3 $ & $ 1.780 $ & $  4.01 $ & $  133.0 $ & $ 13.000 $ \\ \hline 
$  40^3 $ & $ 1.820 $ & $  4.85 $ & $  166.0 $ & $ 32.100 $ \\ \hline 
$  48^3 $ & $ 1.840 $ & $  5.60 $ & $  189.2 $ & $ 65.200 $ \\ \hline 
$  56^3 $ & $ 1.860 $ & $  6.44 $ & $  219.0 $ & $ 124.000 $ \\ \hline 
$  64^3 $ & $ 1.870 $ & $  7.10 $ & $  237.3 $ & $ 200.000 $ \\ \hline 
\hline 
$   4^4 $ & $ 1.220 $ & $  0.53 $ & $   20.0 $ & $  0.013 $ \\ \hline 
$   8^4 $ & $ 1.470 $ & $  1.19 $ & $   41.1 $ & $  0.677 $ \\ \hline 
$  12^4 $ & $ 1.590 $ & $  1.75 $ & $   60.0 $ & $  5.270 $ \\ \hline 
$  16^4 $ & $ 1.660 $ & $  2.27 $ & $   77.0 $ & $ 21.400 $ \\ \hline 
$  20^4 $ & $ 1.710 $ & $  2.75 $ & $   93.1 $ & $ 63.300 $ \\ \hline 
$  24^4 $ & $ 1.740 $ & $  3.24 $ & $  107.0 $ & $ 152.000 $ \\ \hline 
$  28^4 $ & $ 1.770 $ & $  3.66 $ & $  123.0 $ & $ 319.000 $ \\ \hline 
$  32^4 $ & $ 1.790 $ & $  4.05 $ & $  137.0 $ & $ 625.000 $ \\ \hline 
\hline 
$  16^2 $ & $ 1.600 $ & $  0.89 $ & $   36.0 $ & $  0.013 $ \\ \hline 
$  32^2 $ & $ 1.780 $ & $  1.48 $ & $   72.0 $ & $  0.104 $ \\ \hline 
$  48^2 $ & $ 1.850 $ & $  2.82 $ & $  108.0 $ & $  0.396 $ \\ \hline 
$  64^2 $ & $ 1.870 $ & $  3.76 $ & $  128.2 $ & $  0.919 $ \\ \hline 
$  80^2 $ & $ 1.900 $ & $  4.57 $ & $  164.0 $ & $  2.460 $ \\ \hline 
$  96^2 $ & $ 1.920 $ & $  6.00 $ & $  206.0 $ & $  4.610 $ \\ \hline 
$ 112^2 $ & $ 1.930 $ & $  6.93 $ & $  236.0 $ & $  7.250 $ \\ \hline 
$ 128^2 $ & $ 1.940 $ & $  8.01 $ & $  275.0 $ & $ 11.200 $ \\ \hline 
\hline 
$   8^3 $ & $ 1.400 $ & $  0.52 $ & $   21.9 $ & $  0.023 $ \\ \hline 
$  16^3 $ & $ 1.650 $ & $  1.07 $ & $   43.0 $ & $  0.565 $ \\ \hline 
$  24^3 $ & $ 1.750 $ & $  1.64 $ & $   63.9 $ & $  3.210 $ \\ \hline 
$  32^3 $ & $ 1.800 $ & $  2.28 $ & $   81.0 $ & $  9.520 $ \\ \hline 
$  40^3 $ & $ 1.840 $ & $  2.87 $ & $  103.0 $ & $ 23.900 $ \\ \hline 
$  48^3 $ & $ 1.860 $ & $  3.25 $ & $  119.0 $ & $ 48.300 $ \\ \hline 
$  56^3 $ & $ 1.880 $ & $  3.94 $ & $  139.0 $ & $ 92.200 $ \\ \hline 
$  64^3 $ & $ 1.890 $ & $  4.72 $ & $  154.7 $ & $ 150.000 $ \\ \hline 
\hline 
$   4^4 $ & $ 1.220 $ & $  0.30 $ & $   13.0 $ & $  0.009 $ \\ \hline 
$   8^4 $ & $ 1.460 $ & $  0.67 $ & $   25.0 $ & $  0.496 $ \\ \hline 
$  12^4 $ & $ 1.600 $ & $  0.92 $ & $   37.0 $ & $  3.880 $ \\ \hline 
$  16^4 $ & $ 1.680 $ & $  1.30 $ & $   49.0 $ & $ 16.100 $ \\ \hline 
$  20^4 $ & $ 1.730 $ & $  1.59 $ & $   59.0 $ & $ 47.400 $ \\ \hline 
$  24^4 $ & $ 1.770 $ & $  1.93 $ & $   71.0 $ & $ 119.000 $ \\ \hline 
$  28^4 $ & $ 1.800 $ & $  2.28 $ & $   82.0 $ & $ 256.000 $ \\ \hline 
$  32^4 $ & $ 1.820 $ & $  2.51 $ & $   92.2 $ & $ 489.000 $ \\ \hline 
\hline 
\end{tabular}
\end{center}
\vspace*{-0.5cm}
\caption{~The tuning parameter $\omega$, the relaxation time $\tau$,
          the number of sweeps $n_{gf}$
          and the time $t_{gf}$ (in seconds)
          for the overrelaxation method for $d = 2\mbox{,}\,
          3\mbox{,}$ and $4$ using
          lexicographic (upper rows) or even/odd (lower rows) update.}
\label{Table.Overrel}
\end{table}
\begin{table}
\begin{center}
\addtolength{\tabcolsep}{3.0mm}
\vspace*{-0.9cm}
\protect\footnotesize
\begin{tabular}{|| c | c | c | c | c ||}
\hline
\hline
$ V=N^d $ & $ p $ & $ \tau $ & $n_{gf}$ & $ t_{gf} $ \\
\hline
\hline
$  16^2 $ & $ 0.580 $ & $  2.53 $ & $   89.7 $ & $  0.033 $ \\ \hline 
$  32^2 $ & $ 0.770 $ & $  5.05 $ & $  181.1 $ & $  0.261 $ \\ \hline 
$  48^2 $ & $ 0.830 $ & $  7.50 $ & $  258.1 $ & $  0.823 $ \\ \hline 
$  64^2 $ & $ 0.870 $ & $  9.94 $ & $  342.0 $ & $  2.400 $ \\ \hline 
$  80^2 $ & $ 0.900 $ & $ 12.29 $ & $  444.3 $ & $  5.640 $ \\ \hline 
$  96^2 $ & $ 0.910 $ & $ 14.65 $ & $  502.6 $ & $  9.540 $ \\ \hline 
$ 112^2 $ & $ 0.920 $ & $ 17.32 $ & $  573.2 $ & $ 14.600 $ \\ \hline 
$ 128^2 $ & $ 0.930 $ & $ 19.57 $ & $  656.2 $ & $ 21.800 $ \\ \hline 
\hline
$   8^3 $ & $ 0.400 $ & $  1.30 $ & $   48.4 $ & $  0.046 $ \\ \hline 
$  16^3 $ & $ 0.630 $ & $  2.61 $ & $   96.0 $ & $  1.120 $ \\ \hline 
$  24^3 $ & $ 0.730 $ & $  3.92 $ & $  141.1 $ & $  5.790 $ \\ \hline 
$  32^3 $ & $ 0.790 $ & $  5.12 $ & $  187.9 $ & $ 18.400 $ \\ \hline 
$  40^3 $ & $ 0.830 $ & $  6.38 $ & $  237.0 $ & $ 46.200 $ \\ \hline 
$  48^3 $ & $ 0.850 $ & $  7.53 $ & $  273.3 $ & $ 95.100 $ \\ \hline 
$  56^3 $ & $ 0.870 $ & $  8.66 $ & $  318.3 $ & $ 179.000 $ \\ \hline 
$  64^3 $ & $ 0.880 $ & $ 10.03 $ & $  349.7 $ & $ 298.000 $ \\ \hline 
\hline
$   4^4 $ & $ 0.160 $ & $  0.65 $ & $   25.1 $ & $  0.016 $ \\ \hline 
$   8^4 $ & $ 0.440 $ & $  1.38 $ & $   52.1 $ & $  0.868 $ \\ \hline 
$  12^4 $ & $ 0.580 $ & $  2.06 $ & $   78.4 $ & $  6.960 $ \\ \hline 
$  16^4 $ & $ 0.660 $ & $  2.73 $ & $  103.0 $ & $ 29.000 $ \\ \hline 
$  20^4 $ & $ 0.710 $ & $  3.38 $ & $  125.4 $ & $ 86.300 $ \\ \hline 
$  24^4 $ & $ 0.750 $ & $  4.00 $ & $  149.2 $ & $ 215.000 $ \\ \hline 
$  28^4 $ & $ 0.780 $ & $  4.62 $ & $  173.0 $ & $ 465.000 $ \\ \hline 
$  32^4 $ & $ 0.800 $ & $  5.22 $ & $  193.2 $ & $ 881.000 $ \\ \hline 
\hline
$  16^2 $ & $ 0.520 $ & $  1.73 $ & $   62.7 $ & $  0.024 $ \\ \hline 
$  32^2 $ & $ 0.730 $ & $  3.29 $ & $  120.6 $ & $  0.182 $ \\ \hline 
$  48^2 $ & $ 0.810 $ & $  4.88 $ & $  178.2 $ & $  0.601 $ \\ \hline 
$  64^2 $ & $ 0.850 $ & $  6.39 $ & $  232.4 $ & $  1.770 $ \\ \hline 
$  80^2 $ & $ 0.880 $ & $  7.95 $ & $  290.9 $ & $  4.550 $ \\ \hline 
$  96^2 $ & $ 0.900 $ & $  9.61 $ & $  352.8 $ & $  8.210 $ \\ \hline 
$ 112^2 $ & $ 0.910 $ & $ 11.25 $ & $  404.1 $ & $ 12.800 $ \\ \hline 
$ 128^2 $ & $ 0.920 $ & $ 13.15 $ & $  465.9 $ & $ 19.300 $ \\ \hline 
\hline
$   8^3 $ & $ 0.350 $ & $  1.03 $ & $   38.8 $ & $  0.040 $ \\ \hline 
$  16^3 $ & $ 0.590 $ & $  1.96 $ & $   73.4 $ & $  0.951 $ \\ \hline 
$  24^3 $ & $ 0.700 $ & $  2.89 $ & $  107.2 $ & $  5.450 $ \\ \hline 
$  32^3 $ & $ 0.770 $ & $  3.85 $ & $  144.7 $ & $ 17.300 $ \\ \hline 
$  40^3 $ & $ 0.810 $ & $  4.77 $ & $  179.1 $ & $ 42.300 $ \\ \hline 
$  48^3 $ & $ 0.840 $ & $  5.75 $ & $  216.0 $ & $ 88.800 $ \\ \hline 
$  56^3 $ & $ 0.860 $ & $  6.64 $ & $  249.7 $ & $ 165.000 $ \\ \hline 
$  64^3 $ & $ 0.870 $ & $  7.41 $ & $  271.6 $ & $ 267.000 $ \\ \hline 
\hline
$   4^4 $ & $ 0.160 $ & $  0.60 $ & $   23.2 $ & $  0.015 $ \\ \hline 
$   8^4 $ & $ 0.400 $ & $  1.13 $ & $   43.1 $ & $  0.824 $ \\ \hline 
$  12^4 $ & $ 0.540 $ & $  1.66 $ & $   63.1 $ & $  6.590 $ \\ \hline 
$  16^4 $ & $ 0.630 $ & $  2.19 $ & $   83.8 $ & $ 27.900 $ \\ \hline 
$  20^4 $ & $ 0.690 $ & $  2.71 $ & $  103.8 $ & $ 84.500 $ \\ \hline 
$  24^4 $ & $ 0.730 $ & $  3.21 $ & $  122.0 $ & $ 206.000 $ \\ \hline 
$  28^4 $ & $ 0.760 $ & $  3.71 $ & $  140.0 $ & $ 442.000 $ \\ \hline 
$  32^4 $ & $ 0.790 $ & $  4.26 $ & $  162.0 $ & $ 882.000 $ \\ \hline 
\hline
\end{tabular}
\end{center}
\vspace*{-0.5cm}
\caption{~The tuning parameter $p$, the relaxation time $\tau$,
          the number of sweeps $n_{gf}$
          and the time $t_{gf}$ (in seconds)
          for the stochastic overrelaxation method for $d = 2\mbox{,}\,
          3\mbox{,}$ and $4$ using
          lexicographic (upper rows) or even/odd (lower rows) update.}
\label{Table.Stochas}
\end{table}
\begin{table}
\begin{center}
\addtolength{\tabcolsep}{3.0mm}
\vspace*{-0.5cm}
\protect\footnotesize
\begin{tabular}{|| c | c | c | c | c ||}
\hline
\hline
$ V=N^d $ & $ \alpha $ & $ \tau $ & $n_{gf}$ & $ t_{gf} $ \\
\hline
\hline
$  16^2 $ & $ 1.000 $ & $  0.06 $ & $    5.1 $ & $  0.004 $ \\ \hline 
$  32^2 $ & $ 1.000 $ & $  0.06 $ & $    5.4 $ & $  0.015 $ \\ \hline 
$  64^2 $ & $ 1.000 $ & $  0.04 $ & $    6.0 $ & $  0.092 $ \\ \hline 
$ 128^2 $ & $ 1.000 $ & $  0.05 $ & $    6.0 $ & $  0.679 $ \\ \hline 
\hline
$   8^3 $ & $ 1.015 $ & $  0.01 $ & $    6.9 $ & $  0.011 $ \\ \hline 
$  16^3 $ & $ 1.000 $ & $  0.06 $ & $    5.0 $ & $  0.118 $ \\ \hline 
$  32^3 $ & $ 1.000 $ & $  0.06 $ & $    5.0 $ & $  2.590 $ \\ \hline 
$  64^3 $ & $ 1.000 $ & $  0.07 $ & $    5.1 $ & $ 30.500 $ \\ \hline 
\hline
$   4^4 $ & $ 1.000 $ & $  0.05 $ & $    5.0 $ & $  0.005 $ \\ \hline 
$   8^4 $ & $ 1.000 $ & $  0.05 $ & $    5.0 $ & $  0.145 $ \\ \hline 
$  16^4 $ & $ 1.000 $ & $  0.06 $ & $    5.0 $ & $  6.890 $ \\ \hline 
$  32^4 $ & $ 1.000 $ & $  0.06 $ & $    5.0 $ & $ 136.000 $ \\ \hline 
\hline
$  16^2 $ & $ 1.150 $ & $  0.27 $ & $   11.0 $ & $  0.009 $ \\ \hline 
$  32^2 $ & $ 1.140 $ & $  0.26 $ & $   11.0 $ & $  0.042 $ \\ \hline 
$  64^2 $ & $ 1.130 $ & $  0.25 $ & $   10.2 $ & $  0.232 $ \\ \hline 
$ 128^2 $ & $ 1.125 $ & $  0.25 $ & $   10.0 $ & $  2.580 $ \\ \hline 
\hline
$   8^3 $ & $ 1.150 $ & $  0.27 $ & $   11.0 $ & $  0.024 $ \\ \hline 
$  16^3 $ & $ 1.140 $ & $  0.26 $ & $   10.3 $ & $  0.324 $ \\ \hline 
$  32^3 $ & $ 1.130 $ & $  0.25 $ & $   10.0 $ & $  8.990 $ \\ \hline 
$  64^3 $ & $ 1.115 $ & $  0.24 $ & $   10.0 $ & $ 109.000 $ \\ \hline 
\hline
$   4^4 $ & $ 1.155 $ & $  0.27 $ & $   11.0 $ & $  0.013 $ \\ \hline 
$   8^4 $ & $ 1.140 $ & $  0.26 $ & $   10.4 $ & $  0.387 $ \\ \hline 
$  16^4 $ & $ 1.125 $ & $  0.25 $ & $   10.0 $ & $ 24.000 $ \\ \hline 
$  32^4 $ & $ 1.105 $ & $  0.23 $ & $    9.2 $ & $ 443.000 $ \\ \hline 
\hline
\end{tabular}
\end{center}
\caption{~The tuning parameter $\alpha$, the relaxation time $\tau$,
          the number of sweeps $n_{gf}$
          and the time $t_{gf}$ (in seconds)
          for the Fourier acceleration method for $d = 2\mbox{,}\,
          3\mbox{,}$ and $4$ using for the 
          Laplacian preconditioning the whole lattice
          (upper rows) or even/odd sublattices (lower rows).}
\label{Table.Fourier}
\end{table}
%


\clearpage

\begin{table}
\begin{center}
\begin{tabular}{|| c | c | c | c ||}
\hline
\hline
$ V = N^2 $ & $ \tau $ & $n_{gf}$ & $ t_{gf} $ \\
\hline
\hline
$  16^2 $ & $ 16.46 $ & $  378.1 $ & $  0.115 $ \\ \hline 
$  32^2 $ & $ 65.13 $ & $ 1217.7 $ & $  1.460 $ \\ \hline 
$  48^2 $ & $ 146.19 $ & $ 2343.3 $ & $  7.310 $ \\ \hline 
$  64^2 $ & $ 259.68 $ & $ 3750.3 $ & $ 23.300 $ \\ \hline 
$  80^2 $ & $ 405.58 $ & $ 5209.9 $ & $ 61.700 $ \\ \hline 
$  96^2 $ & $ 583.91 $ & $ 7033.1 $ & $ 119.000 $ \\ \hline 
$ 112^2 $ & $ 794.66 $ & $ 8754.2 $ & $ 198.000 $ \\ \hline 
$ 128^2 $ & $ 1037.80 $ & $ 9996.4 $ & $ 292.000 $ \\ \hline 
\hline
$  16^2 $ & $ 16.30 $ & $  392.3 $ & $  0.125 $ \\ \hline 
$  32^2 $ & $ 64.93 $ & $ 1279.7 $ & $  1.590 $ \\ \hline 
$  48^2 $ & $ 145.99 $ & $ 2527.7 $ & $  7.060 $ \\ \hline 
$  64^2 $ & $ 259.47 $ & $ 4095.9 $ & $ 26.400 $ \\ \hline 
$  80^2 $ & $ 405.37 $ & $ 5799.6 $ & $ 80.000 $ \\ \hline 
$  96^2 $ & $ 583.69 $ & $ 7714.4 $ & $ 162.000 $ \\ \hline 
$ 112^2 $ & $ 794.44 $ & $ 9727.4 $ & $ 285.000 $ \\ \hline 
$ 128^2 $ & $ 1037.60 $ & $ 10001.0 $ & $ 383.000 $ \\ \hline 
\hline
\end{tabular}
\end{center}
\caption{~The relaxation time $\tau$, the number of sweeps $n_{gf}$
          and the time $t_{gf}$ (in seconds)
          for the Los Alamos method for $d = 2$ using
          lexicographic (upper rows) or even/odd (lower rows) update
          with $\lambda_1 = 1$ and $\lambda_2 = 0.25$.}
\label{Table.LosAlamos2dlambda}
\end{table}
\begin{table}
\begin{center}
\begin{tabular}{|| c | c | c | c | c ||}
\hline
\hline
$ V = N^2 $ & $ \alpha $ & $ \tau $ & $n_{gf}$ & $ t_{gf} $ \\
\hline
\hline
$  16^2 $ & $ 0.680 $ & $  2.71 $ & $   87.1 $ & $  0.029 $ \\ \hline 
$  32^2 $ & $ 0.730 $ & $  5.14 $ & $  163.6 $ & $  0.217 $ \\ \hline 
$  48^2 $ & $ 0.750 $ & $  7.42 $ & $  235.6 $ & $  0.708 $ \\ \hline 
$  64^2 $ & $ 0.760 $ & $  9.61 $ & $  298.9 $ & $  1.920 $ \\ \hline 
$  80^2 $ & $ 0.767 $ & $ 11.67 $ & $  369.9 $ & $  4.670 $ \\ \hline 
$  96^2 $ & $ 0.772 $ & $ 13.86 $ & $  439.0 $ & $  7.940 $ \\ \hline 
$ 112^2 $ & $ 0.775 $ & $ 15.57 $ & $  484.5 $ & $ 11.700 $ \\ \hline 
$ 128^2 $ & $ 0.777 $ & $ 17.44 $ & $  539.4 $ & $ 16.700 $ \\ \hline 
\hline
$  16^2 $ & $ 0.685 $ & $  1.60 $ & $   50.0 $ & $  0.018 $ \\ \hline 
$  32^2 $ & $ 0.740 $ & $  2.95 $ & $  101.0 $ & $  0.140 $ \\ \hline 
$  48^2 $ & $ 0.755 $ & $  4.68 $ & $  139.4 $ & $  0.437 $ \\ \hline 
$  64^2 $ & $ 0.767 $ & $  5.82 $ & $  189.0 $ & $  1.280 $ \\ \hline 
$  80^2 $ & $ 0.772 $ & $  7.22 $ & $  225.3 $ & $  3.380 $ \\ \hline 
$  96^2 $ & $ 0.777 $ & $  8.64 $ & $  274.0 $ & $  6.130 $ \\ \hline 
$ 112^2 $ & $ 0.780 $ & $ 10.05 $ & $  309.6 $ & $  9.550 $ \\ \hline 
$ 128^2 $ & $ 0.782 $ & $ 11.41 $ & $  352.9 $ & $ 14.100 $ \\ \hline 
\hline
\end{tabular}
\end{center}
\caption{~The tuning parameter $\alpha$, the
          relaxation time $\tau$, the number of sweeps $n_{gf}$
          and the time $t_{gf}$ (in seconds)
          for the Cornell method for $d = 2$ using
          lexicographic (upper rows) or even/odd (lower rows) update
          with $\lambda_1 = 1$ and $\lambda_2 = 0.25$.}
\label{Table.Cornell2dlambda}
\end{table}
\begin{table}
\begin{center}
\begin{tabular}{|| c | c | c | c | c ||}
\hline
\hline
$ V = N^2 $ & $ \omega $ & $ \tau $ & $n_{gf}$ & $ t_{gf} $ \\
\hline
\hline
$  16^2 $ & $ 1.700 $ & $  2.67 $ & $   90.3 $ & $  0.033 $ \\ \hline 
$  32^2 $ & $ 1.820 $ & $  5.08 $ & $  165.5 $ & $  0.231 $ \\ \hline 
$  48^2 $ & $ 1.870 $ & $  7.48 $ & $  236.7 $ & $  0.747 $ \\ \hline 
$  64^2 $ & $ 1.900 $ & $  9.46 $ & $  310.9 $ & $  2.310 $ \\ \hline 
$  80^2 $ & $ 1.920 $ & $ 11.77 $ & $  391.2 $ & $  4.890 $ \\ \hline 
$  96^2 $ & $ 1.930 $ & $ 13.72 $ & $  449.2 $ & $  8.380 $ \\ \hline 
$ 112^2 $ & $ 1.940 $ & $ 15.91 $ & $  525.1 $ & $ 13.400 $ \\ \hline 
$ 128^2 $ & $ 1.940 $ & $ 18.69 $ & $  531.9 $ & $ 17.100 $ \\ \hline 
\hline
$  16^2 $ & $ 1.690 $ & $  0.64 $ & $   62.6 $ & $  0.024 $ \\ \hline 
$  32^2 $ & $ 1.840 $ & $  2.83 $ & $   98.9 $ & $  0.145 $ \\ \hline 
$  48^2 $ & $ 1.890 $ & $  4.35 $ & $  145.9 $ & $  0.474 $ \\ \hline 
$  64^2 $ & $ 1.920 $ & $  5.91 $ & $  200.3 $ & $  1.490 $ \\ \hline 
$  80^2 $ & $ 1.940 $ & $  7.99 $ & $  267.8 $ & $  3.980 $ \\ \hline 
$  96^2 $ & $ 1.950 $ & $  9.74 $ & $  321.0 $ & $  7.390 $ \\ \hline 
$ 112^2 $ & $ 1.950 $ & $ 10.90 $ & $  329.1 $ & $ 10.500 $ \\ \hline 
$ 128^2 $ & $ 1.960 $ & $ 12.24 $ & $  400.0 $ & $ 16.400 $ \\ \hline 
\hline
\end{tabular}
\end{center}
\caption{~The tuning parameter $\omega$, the
          relaxation time $\tau$, the number of sweeps $n_{gf}$
          and the time $t_{gf}$ (in seconds)
          for the overrelaxation method for $d = 2$ using
          lexicographic (upper rows) or even/odd (lower rows) update
          with $\lambda_1 = 1$ and $\lambda_2 = 0.25$.}
\label{Table.Overr2dlambda}
\end{table}
\begin{table}
\begin{center}
\begin{tabular}{|| c | c | c | c | c ||}
\hline
\hline
$ V = N^2 $ & $ p $ & $ \tau $ & $n_{gf}$ & $ t_{gf} $ \\
\hline
\hline
$  16^2 $ & $ 0.680 $ & $  3.42 $ & $  121.5 $ & $  0.044 $ \\ \hline 
$  32^2 $ & $ 0.820 $ & $  6.99 $ & $  239.2 $ & $  0.342 $ \\ \hline 
$  48^2 $ & $ 0.870 $ & $ 10.26 $ & $  346.1 $ & $  1.260 $ \\ \hline 
$  64^2 $ & $ 0.900 $ & $ 13.69 $ & $  455.1 $ & $  3.230 $ \\ \hline 
$  80^2 $ & $ 0.920 $ & $ 16.44 $ & $  569.9 $ & $  7.530 $ \\ \hline 
$  96^2 $ & $ 0.930 $ & $ 20.98 $ & $  668.9 $ & $ 12.700 $ \\ \hline 
$ 112^2 $ & $ 0.940 $ & $ 23.92 $ & $  779.6 $ & $ 19.800 $ \\ \hline 
$ 128^2 $ & $ 0.950 $ & $ 26.08 $ & $  923.8 $ & $ 30.400 $ \\ \hline 
\hline
$  16^2 $ & $ 0.670 $ & $  2.67 $ & $   95.2 $ & $  0.036 $ \\ \hline 
$  32^2 $ & $ 0.820 $ & $  5.16 $ & $  185.3 $ & $  0.278 $ \\ \hline 
$  48^2 $ & $ 0.870 $ & $  7.79 $ & $  277.9 $ & $  1.060 $ \\ \hline 
$  64^2 $ & $ 0.900 $ & $ 10.57 $ & $  375.7 $ & $  3.040 $ \\ \hline 
$  80^2 $ & $ 0.920 $ & $ 12.62 $ & $  446.9 $ & $  6.910 $ \\ \hline 
$  96^2 $ & $ 0.930 $ & $ 16.16 $ & $  525.6 $ & $ 12.200 $ \\ \hline 
$ 112^2 $ & $ 0.940 $ & $ 18.18 $ & $  603.0 $ & $ 19.100 $ \\ \hline 
$ 128^2 $ & $ 0.950 $ & $ 19.75 $ & $  703.5 $ & $ 29.300 $ \\ \hline 
\hline
\end{tabular}
\end{center}
\caption{~The tuning parameter $p$, the
          relaxation time $\tau$, the number of sweeps $n_{gf}$
          and the time $t_{gf}$ (in seconds)
          for the stochastic overrelaxation method for $d = 2$ using
          lexicographic (upper rows) or even/odd (lower rows) update
          with $\lambda_1 = 1$ and $\lambda_2 = 0.25$.}
\label{Table.Stoch2dlambda}
\end{table}
\begin{table}
\begin{center}
\begin{tabular}{|| c | c | c | c | c ||}
\hline
\hline
$ V = N^2 $ & $ \alpha $ & $ \tau $ & $n_{gf}$ & $ t_{gf} $ \\
\hline
\hline
$  16^2 $ & $ 1.000 $ & $  0.06 $ & $    5.4 $ & $  0.004 $ \\ \hline 
$  32^2 $ & $ 1.000 $ & $  0.05 $ & $    6.0 $ & $  0.017 $ \\ \hline 
$  64^2 $ & $ 1.000 $ & $  0.05 $ & $    6.0 $ & $  0.093 $ \\ \hline 
$ 128^2 $ & $ 1.000 $ & $  0.06 $ & $    6.0 $ & $  0.748 $ \\ \hline 
\hline
$  16^2 $ & $ 1.150 $ & $  0.27 $ & $   11.0 $ & $  0.009 $ \\ \hline 
$  32^2 $ & $ 1.140 $ & $  0.26 $ & $   10.2 $ & $  0.040 $ \\ \hline 
$  64^2 $ & $ 1.130 $ & $  0.25 $ & $   10.0 $ & $  0.230 $ \\ \hline 
$ 128^2 $ & $ 1.120 $ & $  0.24 $ & $   10.0 $ & $  1.840 $ \\ \hline 
\hline
\end{tabular}
\end{center}
\caption{~The tuning parameter $\alpha$, the
          relaxation time $\tau$, the number of sweeps $n_{gf}$
          and the time $t_{gf}$ (in seconds)
          for the Fourier acceleration method for $d = 2$ using
          for the Laplacian preconditioning the whole lattice
          (upper rows) or even/odd sublattices (lower rows)
          with $\lambda_1 = 1$ and $\lambda_2 = 0.25$.}
\label{Table.Fourier2dlambda}
\end{table}
%


%
\begin{table}
\begin{center}
\begin{tabular}{|| c | c | c ||}
\hline
\hline
$ \mbox{Algorithm} $ & $ c\; \mbox{at finite} \;\beta $ & $
  c\; \mbox{at infinite}\; \beta $ \\
\hline
\hline
$\mbox{Los Alamos}$   & $0.217 \pm 0.028$ & $0.026 \pm 0.000$ \\ \hline
$\mbox{Cornell}$      & $0.609 \pm 0.183$ & $0.164 \pm 0.004$ \\ \hline
$\mbox{overrelax.}$   & $0.220 \pm 0.049$ & $0.197 \pm 0.040$ \\ \hline
$\mbox{stoc. overr.}$ & $0.300 \pm 0.047$ & $0.152 \pm 0.021$ \\ \hline
$\mbox{Fourier}$      & $2.851 \pm 0.606$ & $0.042 \pm 0.004$ \\ \hline
\hline
\end{tabular}
\end{center}
\caption{~The coefficients $c$ obtained by
          fits of the relaxation time $\tau$ for the five
          gauge-fixing algorithms in two dimensions (with
          lexicographic update) at finite $\beta$ (from
          Ref.\ \cite{CM}) and at infinite $\beta$.}
\label{Table.ccomparison}
\end{table}
%


\end{document}